\documentclass[12pt]{article}
\usepackage[english]{babel}
\usepackage{dsfont}
\usepackage[utf8]{inputenc}
\usepackage[letterpaper,top=2cm,bottom=2cm,left=3cm,right=3cm,marginparwidth=1.75cm]{geometry}
\usepackage{physics}
\usepackage{amsmath}
\usepackage{amssymb}
\usepackage{mathtools}
\usepackage{graphicx,booktabs,array}
\usepackage[colorlinks=true, allcolors=blue]{hyperref}
\usepackage{biblatex}
\usepackage{tabularx}
\usepackage{geometry}  
\usepackage{lscape}    
\usepackage{amsmath}   
\usepackage{caption}
\usepackage{subcaption}
\usepackage[normalem]{ulem}
\newif\ifrlxshow\rlxshowtrue

\usepackage{booktabs}
\usepackage{tikz}
\usepackage{adjustbox}
\usepackage{rotating}
\usepackage{algorithm}
\usepackage{algpseudocode}
\usepackage{algorithmicx}
\usepackage[T1]{fontenc}
\usepackage{adjustbox}  
\usepackage{authblk}
\usepackage[utf8]{inputenc}
\usepackage{hyperref}

\renewcommand{\arraystretch}{1.5}

\newtheorem{theorem}{Theorem}
\newcommand{\sdcom}[1]{\textit{\color{orange} [SD: #1]}}

 \algdef{SE}[DOWHILE]{Do}{doWhile}{\algorithmicdo}[1]{\algorithmicwhile\ #1}%

\newcommand{\dsrc}{\includegraphics[width=8em]{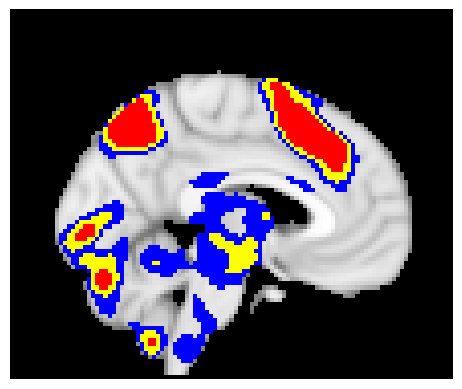}}
\newcommand{\dsrcc}{\includegraphics[width=8em]{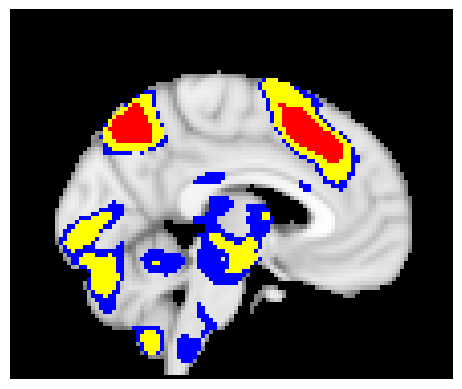}}
\newcommand{\dsrccc}{\includegraphics[width=8em]{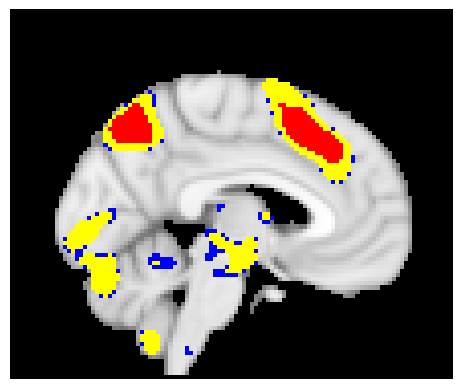}}
\newcommand{\dsrcccc}{\includegraphics[width=8em]{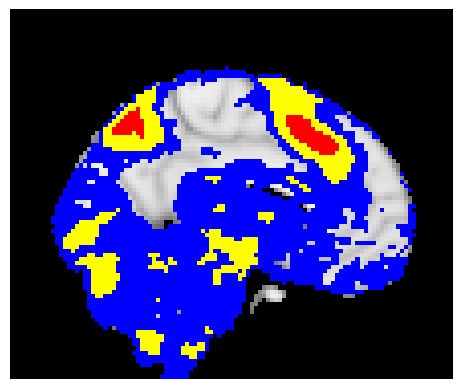}}
\newcommand{\dsrrc}{\includegraphics[width=8em]{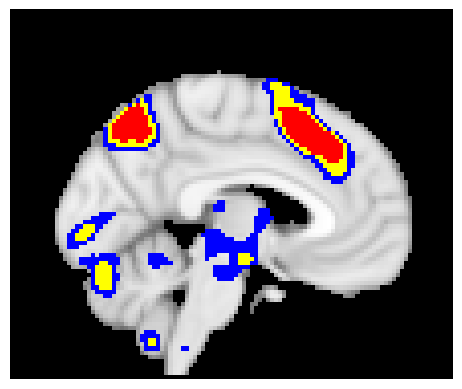}}
\newcommand{\dsrrcc}{\includegraphics[width=8em]{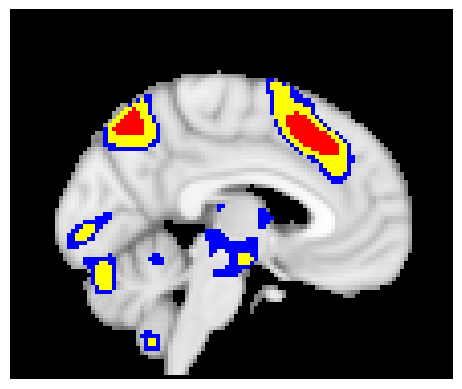}}
\newcommand{\dsrrccc}{\includegraphics[width=8em]{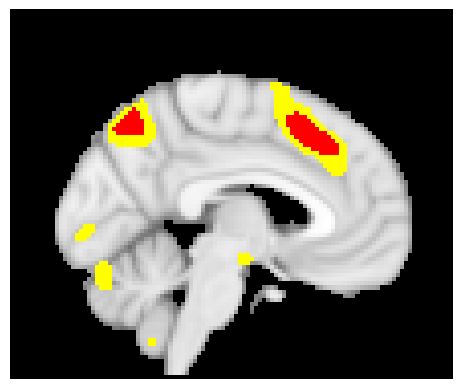}}
\newcommand{\dsrrcccc}{\includegraphics[width=8em]{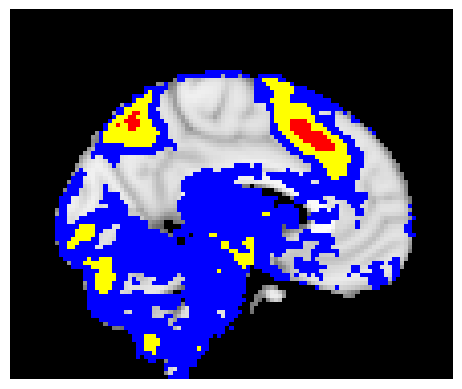}}
\newcommand{\dsrrrc}{\includegraphics[width=8em]{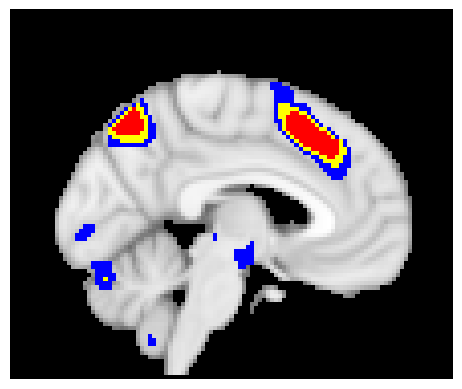}}
\newcommand{\dsrrrcc}{\includegraphics[width=8em]{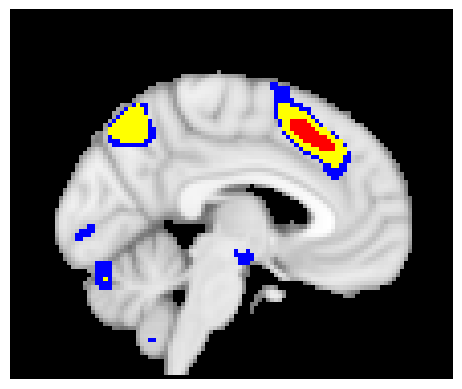}}
\newcommand{\dsrrrccc}{\includegraphics[width=8em]{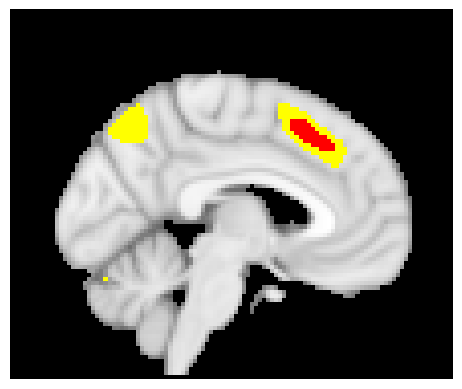}}
\newcommand{\dsrrrcccc}{\includegraphics[width=8em]{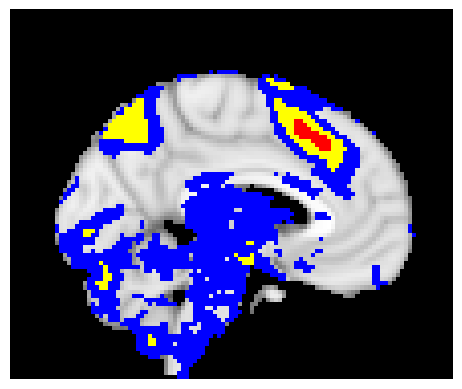}}

\newcommand{\dcrc}{\includegraphics[width=8em]{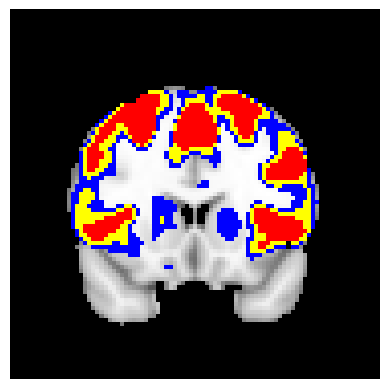}}
\newcommand{\dcrcc}{\includegraphics[width=8em]{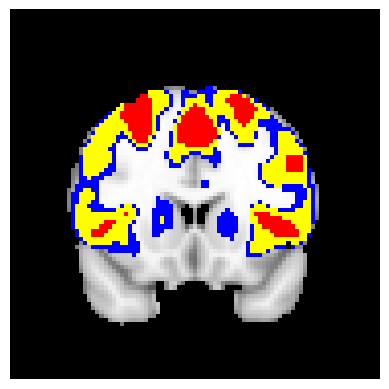}}
\newcommand{\dcrccc}{\includegraphics[width=8em]{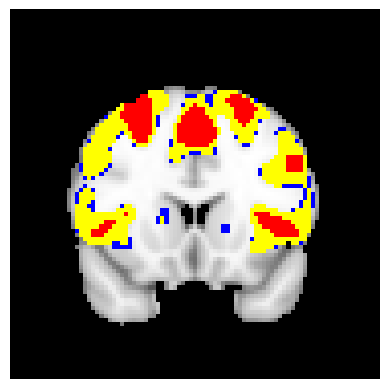}}
\newcommand{\dcrcccc}{\includegraphics[width=8em]{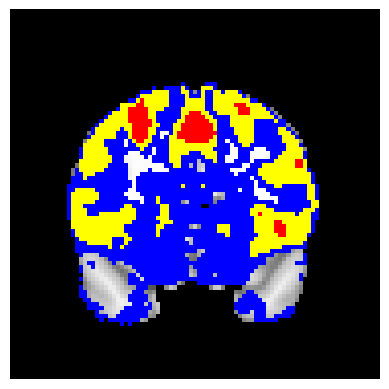}}
\newcommand{\dcrrc}{\includegraphics[width=8em]{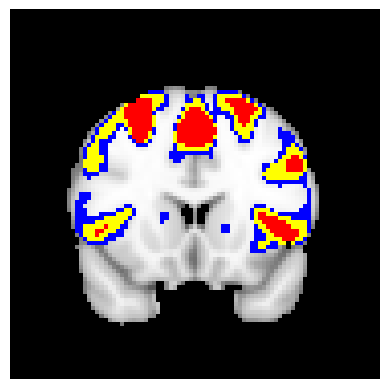}}
\newcommand{\dcrrcc}{\includegraphics[width=8em]{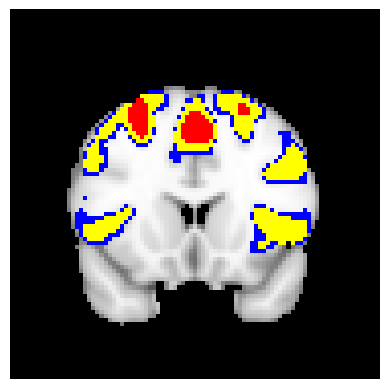}}
\newcommand{\dcrrccc}{\includegraphics[width=8em]{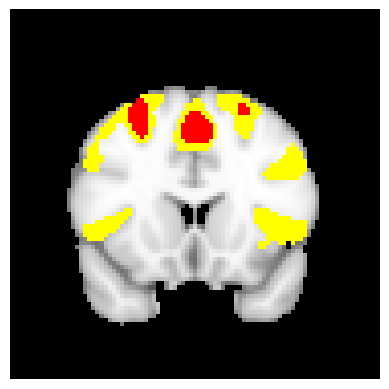}}
\newcommand{\dcrrcccc}{\includegraphics[width=8em]{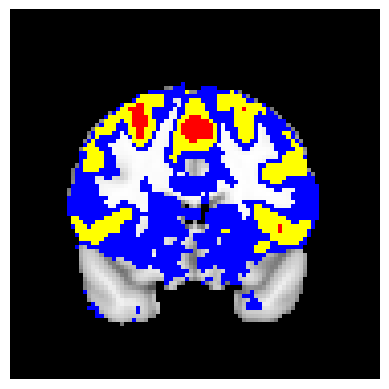}}
\newcommand{\dcrrrc}{\includegraphics[width=8em]{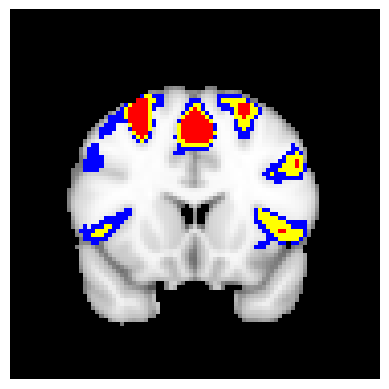}}
\newcommand{\dcrrrcc}{\includegraphics[width=8em]{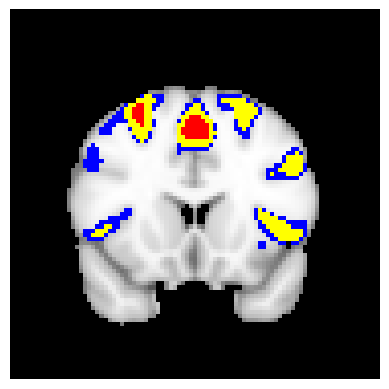}}
\newcommand{\dcrrrccc}{\includegraphics[width=8em]{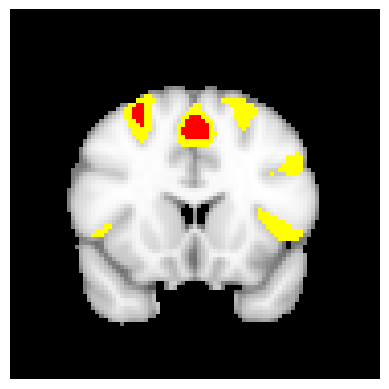}}
\newcommand{\dcrrrcccc}{\includegraphics[width=8em]{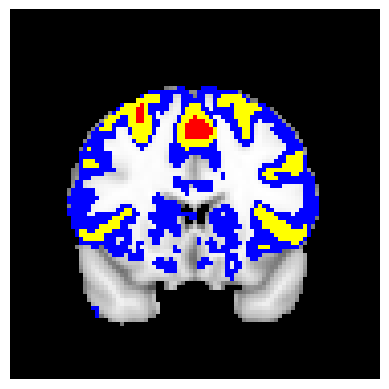}}

\newcommand{\darc}{\includegraphics[width=8em]{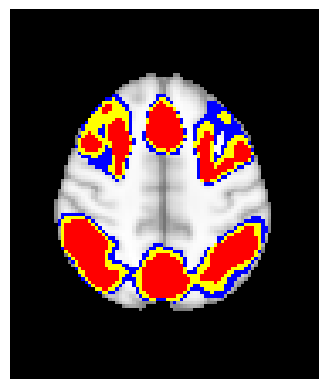}}
\newcommand{\darcc}{\includegraphics[width=8em]{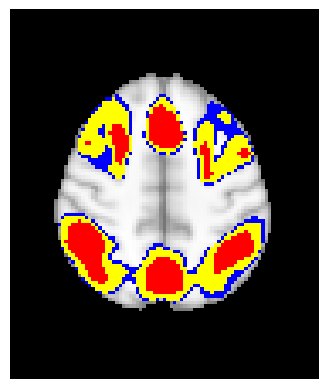}}
\newcommand{\darccc}{\includegraphics[width=8em]{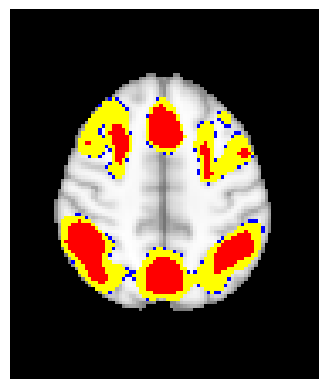}}
\newcommand{\darcccc}{\includegraphics[width=8em]{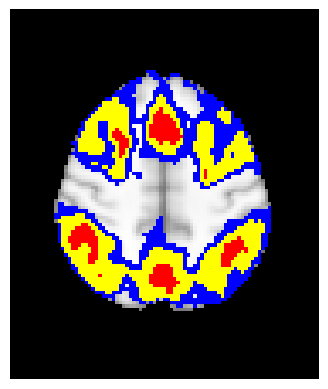}}
\newcommand{\darrc}{\includegraphics[width=8em]{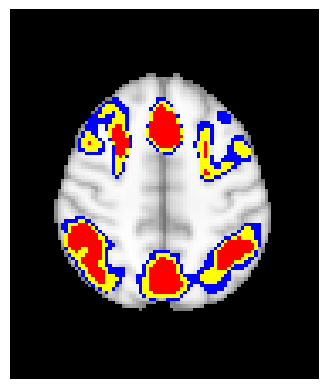}}
\newcommand{\darrcc}{\includegraphics[width=8em]{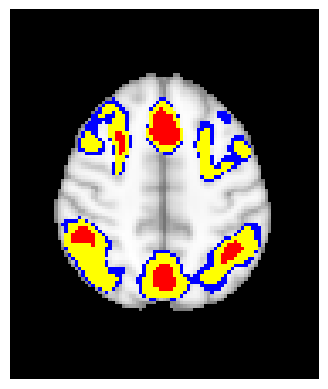}}
\newcommand{\darrccc}{\includegraphics[width=8em]{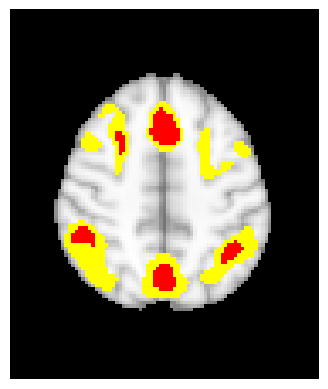}}
\newcommand{\darrcccc}{\includegraphics[width=8em]{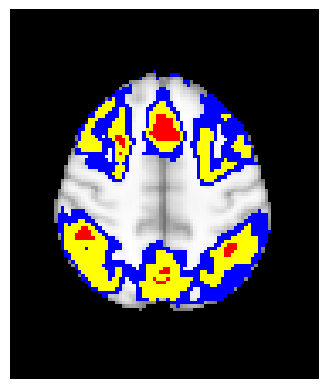}}
\newcommand{\darrrc}{\includegraphics[width=8em]{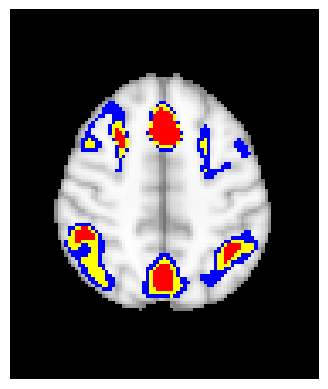}}
\newcommand{\darrrcc}{\includegraphics[width=8em]{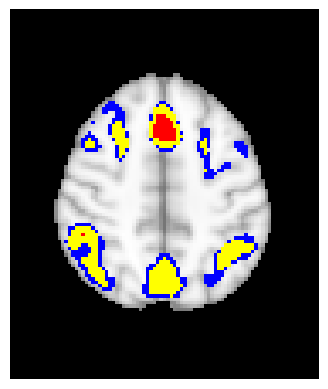}}
\newcommand{\darrrccc}{\includegraphics[width=8em]{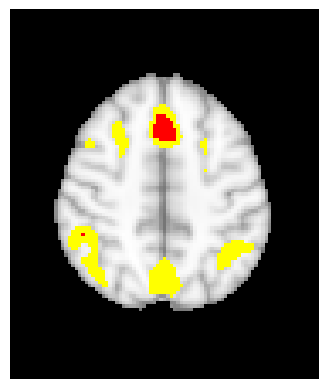}}
\newcommand{\darrrcccc}{\includegraphics[width=8em]{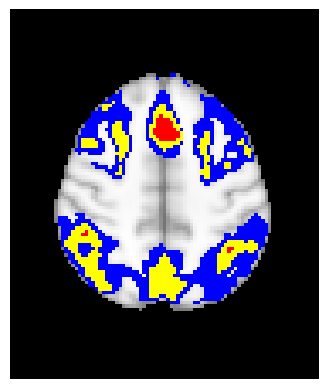}}
\algrenewcommand\algorithmicrequire{\textbf{Input:}}
\algrenewcommand\algorithmicensure{\textbf{Output:}}

\newpage
\title{False Discovery Controlled Regions for Excursion Sets}
\author[1,*]{Howon Ryu}
\author[2,3]{Thomas Maullin-Sapey}
\author[1,4]{Armin Schwartzman}
\author[1]{Samuel Davenport}
\affil[1]{Division of Biostatistics and Bioinformatics, University of California, San Diego, San Diego, CA, USA}
\affil[2]{Big Data Institute, Nuffield Department of Population Health, University of Oxford, Oxford, United Kingdom}
\affil[3]{School of Mathematics, University of Bristol, Bristol, United Kingdom}
\affil[4]{Hal{\i}c{\i}o\u{g}lu Data Science Institute, University of California, San Diego, San Diego, CA, USA}
\affil[*]{\small{Corresponding author: Division of Biostatistics and Bioinformatics, 9500 Gilman Dr, La Jolla, CA 92093, United States. Email: howonryu@ucsd.edu}}
\begin{document}
\maketitle

\begin{abstract}
    Identifying areas where the signal is prominent is an important task in image analysis, with particular applications in brain mapping. In this work, we develop False Discovery Controlled Regions (FDCRs) for excursion sets, the set of locations where the values are above or below a given level. We achieve this by treating the confidence procedure as a testing problem at the given level, allowing control of the False Discovery Rate (FDR). Methods are developed to control the FDR, separately for positive and negative excursions, as well as jointly over both. Furthermore, power is increased by incorporating a two-stage adaptive procedure. Simulation results in a variety of different settings show that the FDCRs successfully control the FDR under the nominal $\alpha$ level. We showcase our methods with an application to functional magnetic resonance imaging (fMRI) data from the Human Connectome Project illustrating the improvement in statistical power over existing approaches.
\end{abstract}

\section{Introduction \label{intro}}

Spatial inference is an important topic in image analysis, especially in neuroimaging applications. The goal of spatial inference is to provide localization of signals of interest, that is, to statistically identify regions within the image where the effect of interest is thought to be the strongest or exceeds a certain level. A common example of this problem is to identify regions of the brain that show activation in response to a stimulus in a task-based functional magnetic resonance imaging (fMRI) experiment.

A traditional approach to solving this problem is via a mass univariate voxel-wise testing procedure, which attempts to identify the locations of voxels where the activation is non-zero. Mass univariate inference typically involves performing hundreds of thousands of tests throughout the brain, leading to a substantial multiple testing problem \cite{bennett2009neural, worsley1992three}. In order to address this, standard analyses control the family-wise error rate (FWER) \cite{nichols2003controlling} or the false discovery rate (FDR) \cite{genovese2002thresholding, perone2004false}. FWER control can be achieved using permutation- or bootstrap-based inference \cite{hayasaka2004nonstationary, hayasaka2003validating, hayasaka2004combining, ZHANG200951}, or using random field theory \cite{flandin2019analysis, worsley1992three, Davenport2023conv}. These procedures test individual voxels, with contiguous clusters of surviving voxels reported as activated regions. Other alternatives include thresholding based on topological features such as peaks \cite{davenport2020selective, chumbley2010topological} or clusters \cite{friston1994functional, smith2009threshold}. Although these procedures test individual voxels, the surviving voxels are conventionally grouped into contiguous clusters and reported as the activated regions.

The above approaches seek to identify brain regions where the activation is non-zero. With the increasing size of fMRI datasets, however, this type of spatial inference suffers from the “null hypothesis fallacy”. Due to large sample sizes, even small effects are picked up by the model as significant \cite{rozeboom1960fallacy, eklund2016cluster}, and commonly used methods identify large regions of the brain as active (see \cite{BOWRING2019116187} and Figure 1 of \cite{Davenport2022peakCRs}) \cite{gonzalez2012whole}.

As an alternative spatial inference approach, to avoid the null hypothesis fallacy, Sommerfeld et al. \cite{SSS} (which from hereon will be referred to as SSS after the names of the authors) proposed the construction of confidence regions at a level $c \not = 0$, in the setting of 2D geospatial climate data.. Confidence regions, in their general sense, are designed to provide spatial uncertainty for the {\em excursion set}, $\mathcal{A}_c = \lbrace v \in \mathcal{V}: \mu(v) > c \rbrace$, corresponding to the locations where $\mu(v)$, the true mean at each element $v$ in the image space $\mathcal{V}$, is greater than the threshold $c \in \mathbb{R}$. (Throughout we will refer to the elements of $\mathcal{V}$ as locations. For images defined on a 2D/3D lattice, the locations are pixels/voxels respectively and for images defined on a surface the locations are vertices.) In the neuroimaging context, Bowring et al. \cite{BOWRING2019116187} and \cite{BOWRING2021117477} extended this work to raw effect size and standardized effect size of 3D fMRI data and \cite{maullin2023spatial} extended the SSS approach to account for conjunction and disjunction effects.

The confidence regions constructed in SSS and Bowring et al. \cite{BOWRING2019116187} consist of an upper region $\hat{\mathcal{A}}_c^+$, and a lower region $\hat{\mathcal{A}}^-_c$ aim to control a rather strict error rate, ensuring complete inclusion of the confidence regions with high probability, i.e., $\mathds{P}(\hat{\mathcal{A}}_c^+ \subseteq \mathcal{A}_c \subseteq \hat{\mathcal{A}}^-_c) \ge 1-\alpha$ for some $\alpha \in(0,1)$. The inclusion statement is violated if even a single location belongs to $\hat{\mathcal{A}_c^{+}}$ but not $\mathcal{A}_c$, or to $\mathcal{A}_c$ but not $\hat{\mathcal{A}_c^{-}}$. In this sense, this type of coverage is analogous to FWER, which is the probability of making any error at all when using a multiple testing procedure. Such a strict criterion is conservative and may produce large confidence regions with the correct coverage rate but low spatial precision. Moreover, the theory behind SSS requires the data to be defined on a continuous domain and not the lattice structures which often arise in neuroimaging.

As opposed to SSS, we develop methods that control another error criterion based on the false discovery rate (FDR). FDR is known to be less conservative and therefore targetting it as a criterion can increase precision. FDR control in neuroimaging was first introduced by \cite{genovese2002thresholding}, and since then it has been widely used as an alternative measure for error in spatial inference in fMRI \cite{benjamini2007false, chumbley2010topological, chumbley2009false}. Targeting the FDR provides more powerful inference than targeting FWER, resulting in a greater number of discoveries. This occurs because it aims to control only the false positives over all the rejected tests, rather than over all the tests. Various FDR corrections include Benjamini-Hochberg (BH) procedure \cite{benjamini1995controlling}, the Benjamini--Yekutieli (BY) procedure for use under dependency \cite{benjamini2001control}, and others \cite{benjamini2006adaptive, yekutieli1999resampling, benjamini2000adaptive}.

Our goal in this work is to develop testing-based methods that quantify the uncertainty in the location and the extent of excursion sets representing effects above a level $c$, while taking advantage of the increase in statistical power brought by adopting FDR control. To this end, we propose a location-wise testing method that searches for signals above (or below) a level $c \ne 0$ (which can be positive or negative) with FDR control over the entire image space. The key here is that the level being tested is not 0 but $c$, thus avoiding the null hypothesis fallacy which occurs at 0. The regions are then constructed from the outcomes of the location-wise tests.

Confidence regions make directional inclusion statements. Directional errors occur in two-sided hypothesis testing when the null hypothesis is rejected in the wrong direction.
Some early works in directional error control in two-tailed test involve \cite{finner1999stepwise}, and \cite{weinstein2013selection, benjamini2005false} for multiple testing confidence intervals. Whether BH continues to control the error rate once the direction of each rejection is declared --- a directional, or type III, error --- was conjectured by \cite{williams1999controlling, benjamini2000adaptive, shaffer2002multiplicity}; see also \cite{shaffer1995multiple}. \textcite{benjamini2005false} study this within the FDR framework, and consider an approach where one-sided BH procedures applied to $p_i$ and $1-p_i$, with the mixed directional FDR of each bounded. Different correctional methods for directional FDR control are presented in \cite{winkler2024false} for neuroimaging and \cite{guo2010controlling} for genetics data applications, while \cite{guo2015stepwise} presents error control under different correlational structures, or under independence \cite{heller2023simultaneous, leung2024adaptive}.

\begin{figure}[htbp]
    \centering
    \begin{subfigure}{0.49\linewidth}
        \includegraphics[width=\linewidth]{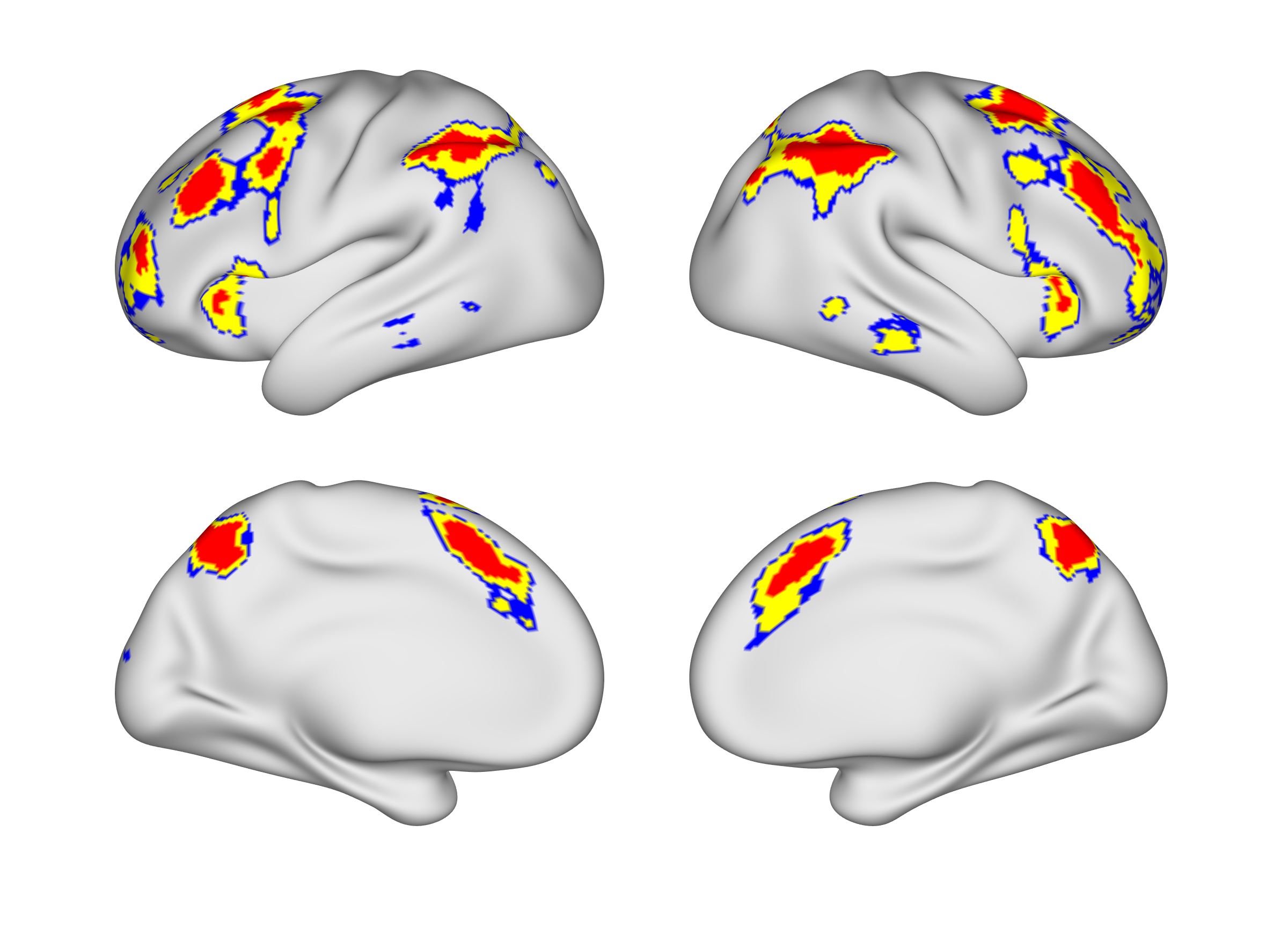}
        \caption{Separate Method, $c=1.5$}
        \label{fig:sub1}
    \end{subfigure}
    \begin{subfigure}{0.49\linewidth}
        \includegraphics[width=\linewidth]{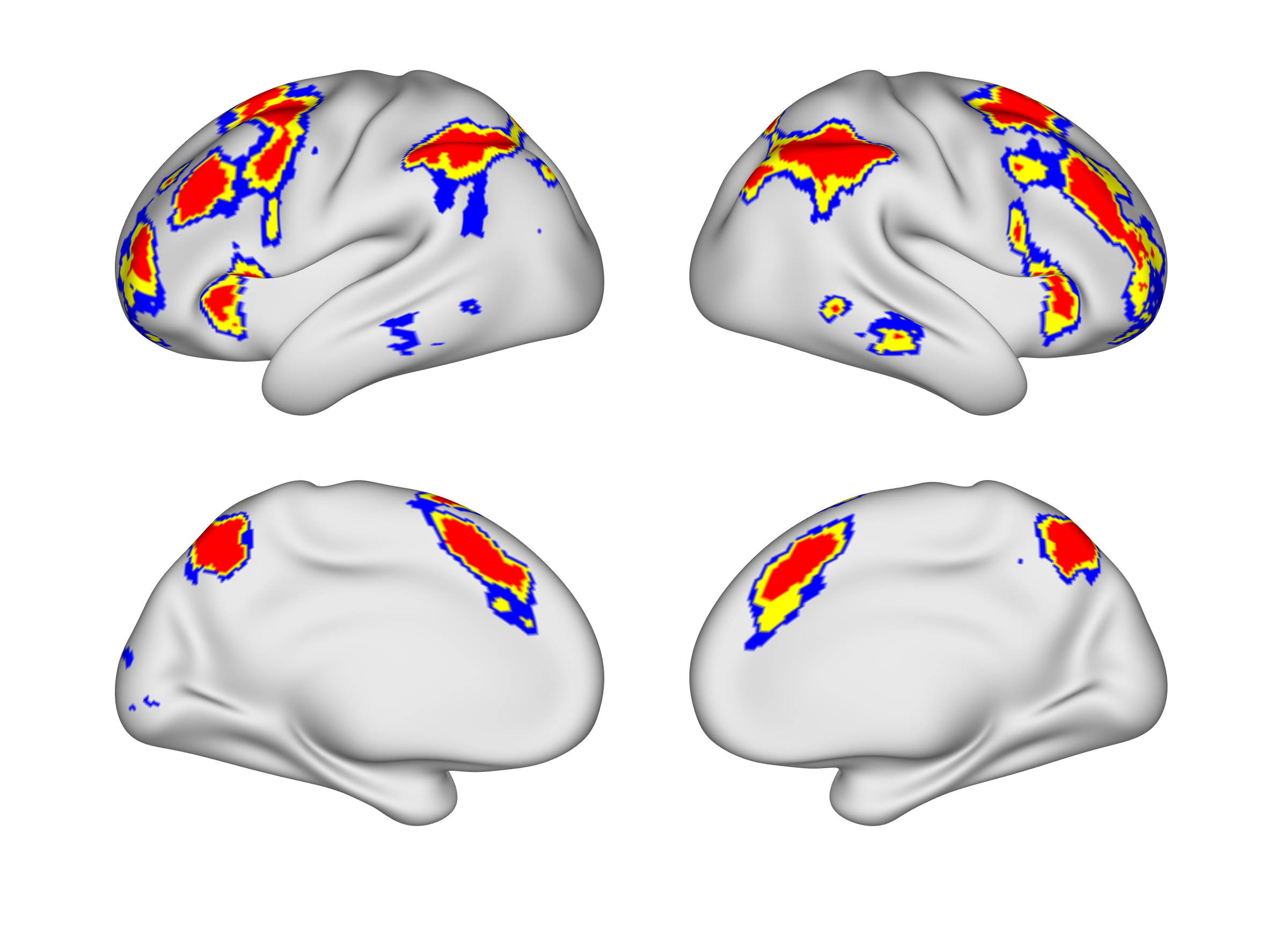}
        \caption{Joint Method, $c=1.5$}
        \label{fig:sub2}
    \end{subfigure}
    \\
    \begin{subfigure}{0.49\linewidth}
        \includegraphics[width=\linewidth]{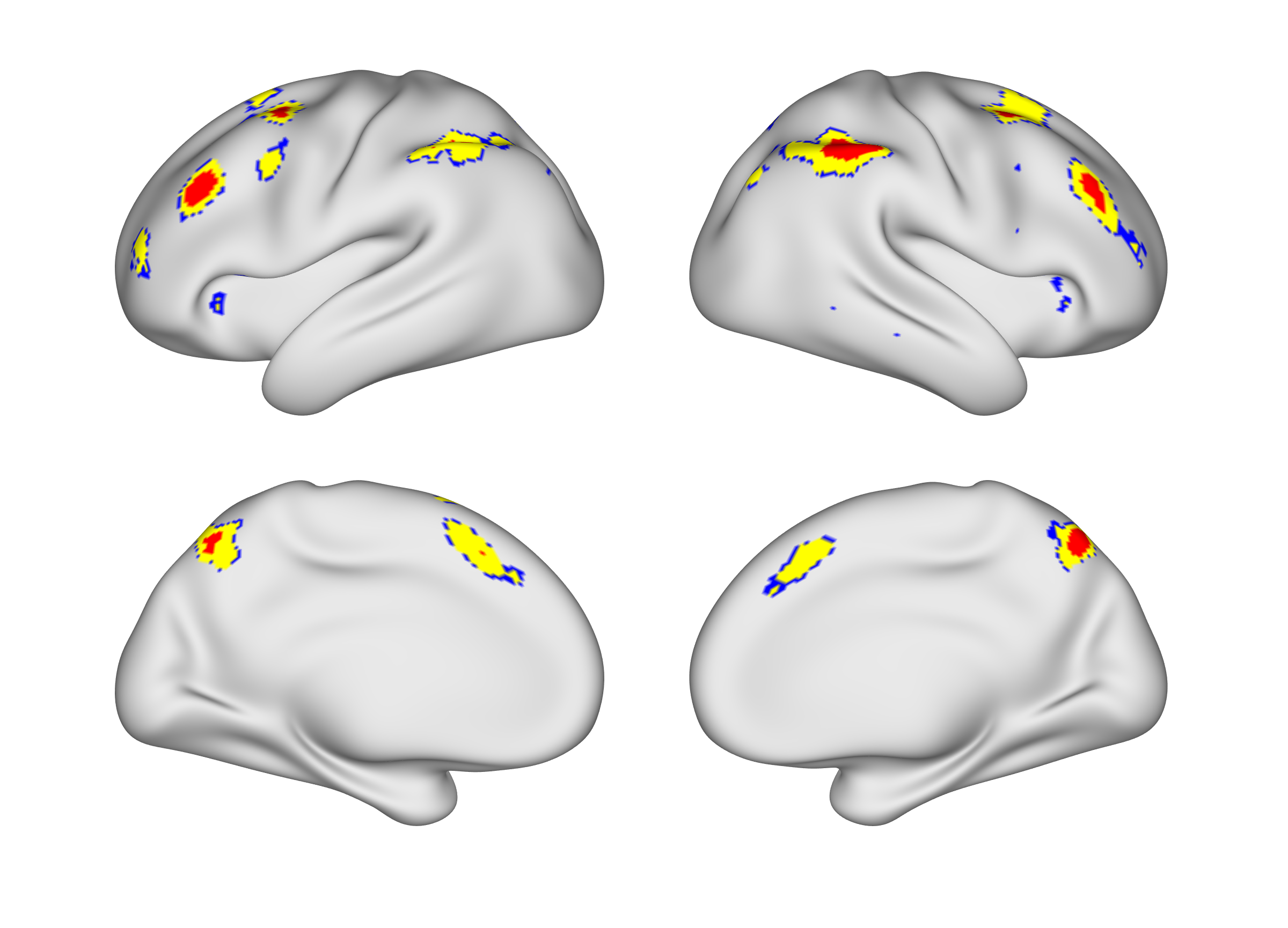}
        \caption{Separate Method, $c=2.5$}
        \label{fig:sub3}
    \end{subfigure}
    \begin{subfigure}{0.49\linewidth}
        \includegraphics[width=\linewidth]{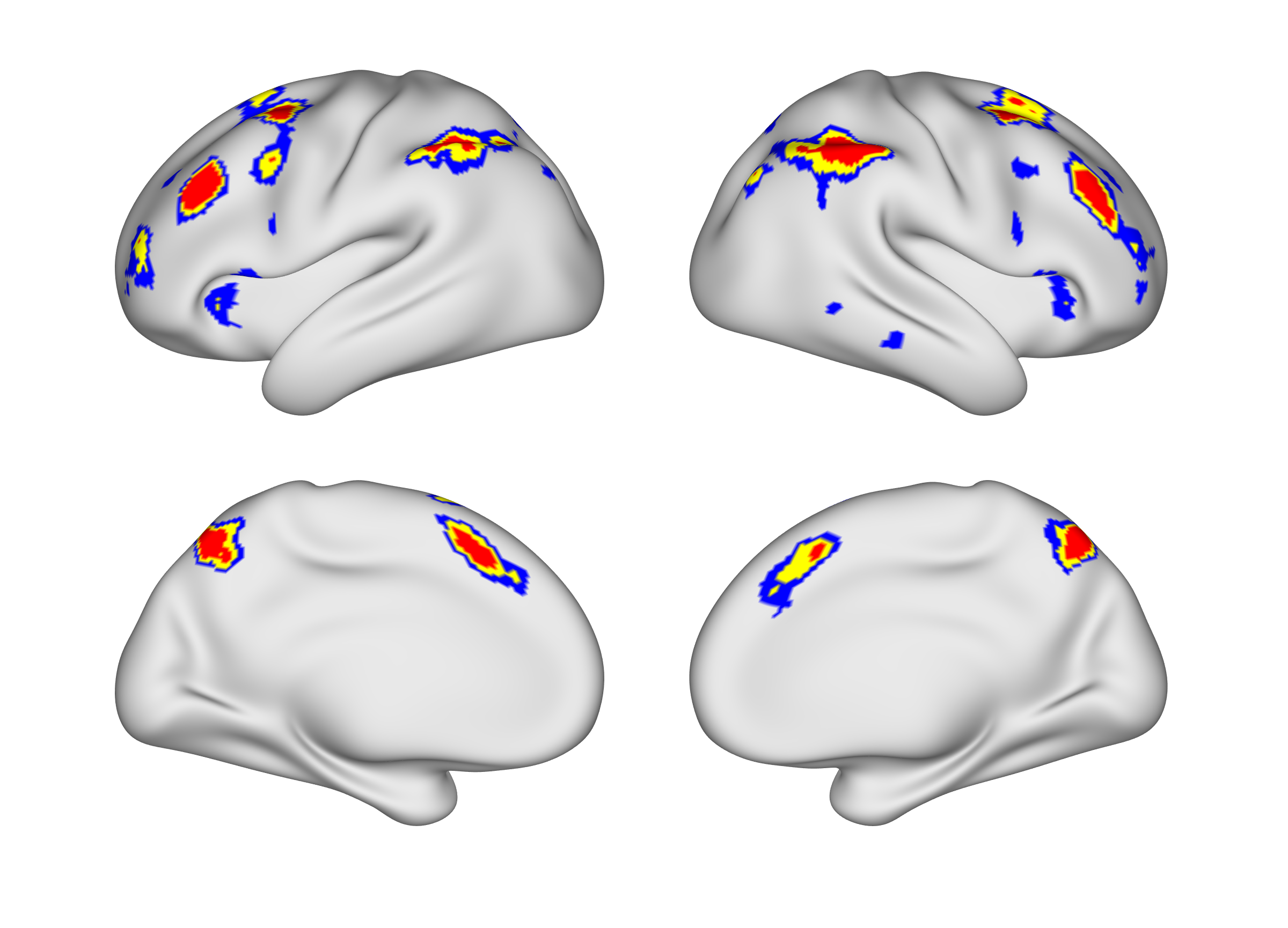}
        \caption{Joint Method, $c=2.5$}
        \label{fig:sub4}
    \end{subfigure}
    \caption{An illustration of false discovery controlled regions (FDCRs), developed in this work, applied to task-based fMRI: red area denotes upper false discovery controlled region $\hat{\mathcal{A}_c^+}$, blue area denotes lower false discovery controlled region $\hat{\mathcal{A}_c^-}$, and the yellow area denotes the point estimate for the excursion set $\hat{\mathcal{A}_c}$. The rows differ in threshold $c$ in the \% blood-oxygen-level-dependent (BOLD) change, and the columns differ in false discovery controlled region construction methods. }
\label{fig-intro_CR}
\end{figure}

In this paper, we introduce testing-based regions $\hat{\mathcal{A}}_c^+$ and $\hat{\mathcal{A}}_c^-$, called false discovery controlled regions (FDCRs) which control the FDR over space in the upper and lower directions respectively, with respect to the true excursion set $\mathcal{A}_c$. Our proposed approach uses one-sided testing at the level $c$, instead of the usual two-sided testing, which allows us to control directional errors.  We investigate two modes of multiple testing for FDCR construction: one using the Benjamini-Hochberg procedure, and the other using a two-stage adaptive procedure proposed in \cite{blanchard2009adaptive}. This two-stage procedure is useful in increasing statistical power when the targeted excursion set is small, which occurs frequently for higher levels $c$. Testing in the positive and negative directions separately we obtain FDCRs for each direction. We continue by constructing FDCRs with joint error control over the two directions, in which the tests for the positive and negative directions are considered together in a single multiple testing procedure. An illustration of the separate and joint FDCRs in practice is presented in Figure \ref{fig-intro_CR}. We evaluate the performance of the proposed methods in terms of FDR, for type-I error, and false non-discovery rate (FNDR), for type-II error, defined with respect to the excursion set.

The rest of the paper is organized as follows. In Section \ref{separate}, we present the problem formally, define the error rates for separate error control, and present the algorithms for controlling them. In Section \ref{joint}, this approach is extended to the joint error control over the two directions. In Section \ref{simulation} we evaluate the performance of the methods via simulation. In Section \ref{dataapplication}, a real data application to task fMRI data from the Human Connectome Project working memory task, and the realistic noise simulation with UK Biobank noise is presented.

\section{False Discovery Controlled Regions with Separate Error Control\label{separate}}
In this section, we introduce a spatial inference framework that controls the FDR separately through two one-sided tests. Upper and lower FDCRs are defined via positive and negative one-sided hypothesis testing where the spatial FDR is controlled at level $\alpha$ separately for each test.


\subsection{Upper False Discovery Controlled Regions}
\subsubsection{Construction}
Given locations $v$ within a finite lattice $\mathcal{V} \subset \mathbb{R}^d$, we can construct an upper FDCR by testing in the positive direction at the level $c$. For each $v \in \mathcal{V}$, we consider the null and alternative hypotheses given by
\begin{equation}
    H_0^U(v): \mu(v) \le c \quad \mbox{vs.} \quad H_A^U(v): \mu(v) > c
\label{eq:upper_hypothesis}
\end{equation}
with corresponding null set $(\mathcal{A}_c)^C \subseteq \mathcal{V}$, the complement of $\mathcal{A}_c$. To do so, let
\begin{equation}
t(v) = \frac{\hat{\mu}(v)-c}{\hat{\sigma}(v)/\sqrt{n}}
\label{eq:tstat}
\end{equation}
be the $t$-statistic for testing against the level $c$ as in (\ref{eq:upper_hypothesis}) where $\hat{\mu}(v)$ and $\hat{\sigma}(v)$ are the point estimates of the true mean and standard deviation at each location $v \in \mathcal{V}$ respectively from $n$ i.i.d. samples. We can transform these test-statistics into a set of $p$-values, $\mathbf{p}^U = \left( p^U(1), \ldots, p^{U}(m) \right)$, where $m = |\mathcal{V}|$ is the number of locations, to obtain an upper FDCR, at a given level $\alpha \in (0,1)$, by applying the Benjamini-Hochberg procedure \cite{benjamini1995controlling, benjamini2001control} as demonstrated in Algorithm \ref{alg_upper}.

\begin{algorithm}[H]    
    \caption{Upper False Discovery Controlled Regions}
    \label{alg_upper}
    \begin{algorithmic}[1]
        \Require{$(t(v), \alpha)$}
        \Ensure{ $\hat{\mathcal{A}}_c^+$}
        \State Obtain $m$ $p$-values $p^U{(v)} = 1- F_{T_{n-1}}\left( t(v) \right), v=1, \ldots, m,$ where $F_{T_{n-1}}$ is the cdf of $t$-distribution with $n-1$ degrees of freedom from the positive one-sided test.
        \State Apply the Benjamini-Hochberg procedure to the $p$-values  $\mathbf{p}^U = \left( p^U(1), \ldots, p^{U}(m) \right)$ to get $q$, the threshold for rejection.
        \State Define $\hat{\mathcal{A}}_c^+ = \{v \in \mathcal{V} :  p^U(v) \leq q\}$, the set of locations rejected.
        \State \textbf{Return} the upper false discovery controlled regions $\hat{\mathcal{A}}_c^+$.
\end{algorithmic}
\end{algorithm}

Here, we assume that the p-values satisfy PRDS (Positive Regression Dependence on a subset), a necessary condition for application of the Benjamini-Hochberg procedure \cite{benjamini2001control}. As the one-sided $p$-values are an order-preserving transformation of the test statistics, PRDS is inherited whenever the underlying statistics are themselves PRDS \cite[\S6.3]{benjamini2005false}. This is a reasonable assumption for fMRI images as this is induced by the smoothness of the image \cite{genovese2002thresholding}. 

\begin{figure}[H]
    \centering
    \includegraphics[width=1\linewidth]{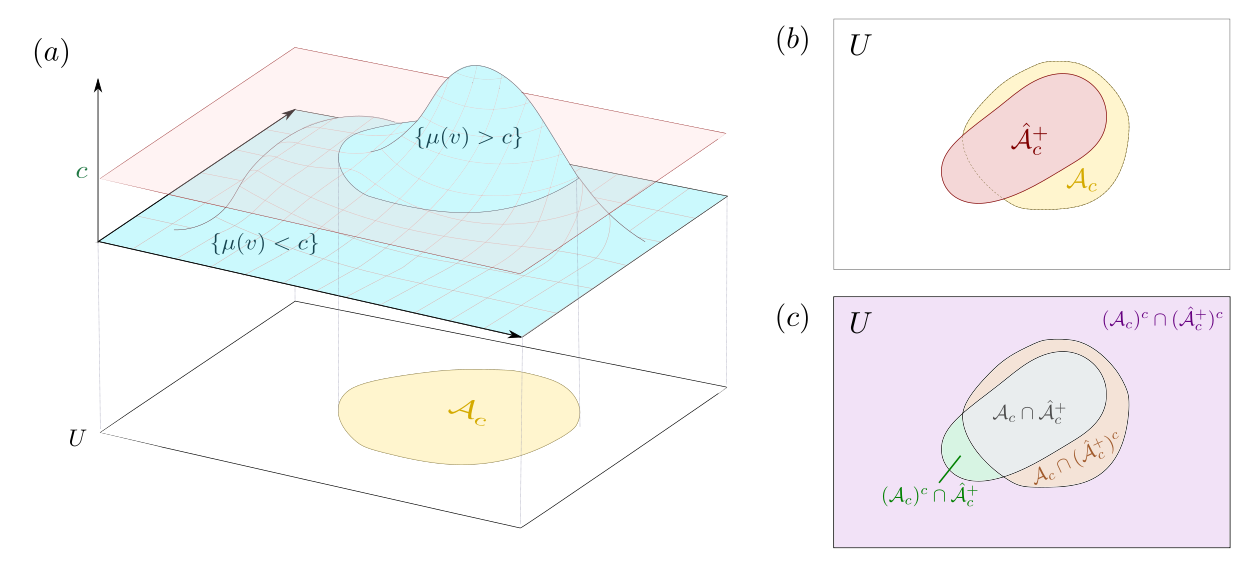}
\caption{Upper FDCR schematic: (a) Example signal and the projection of the excursion set above $c$ ($\mathcal{A}_c$). (b) A hypothetical upper FDCR $\hat{\mathcal{A}}_c^+$ superimposed on the ground truth $\mathcal{A}_c$. (c) The division of (b) into regions corresponding to those specified in Table \ref{table-upper_division} as a visual representation of false positives (green) and false non-positives (orange).}
    \label{figure-upper_division}
\end{figure}

The image space $\mathcal{V}$ can be naturally partitioned according to the rejection regions and the true non-null regions, see Table \ref{table-upper_division}. For a given level $c$, the test and the corresponding regions can be visualized as shown in Figure \ref{figure-upper_division}.

\begin{table}[H]
    \centering
    \renewcommand{\arraystretch}{1.2} \setlength{\extrarowheight}{2pt}
    \setlength{\tabcolsep}{4pt}
    \begin{tabular}{|>{\centering\arraybackslash}p{3cm}|>{\centering\arraybackslash}p{4cm}|>{\centering\arraybackslash}p{4cm}|>{\centering\arraybackslash}p{3cm}|}
    \hline
         Hypothesis & Not rejected & Rejected &  Total  \\ \hline
         $H_0^U$ & \textcolor{violet}{$(\mathcal{A}_c)^C \cap (\hat{A}_c^+)^C$} & \textcolor{green}{$(\mathcal{A}_c)^C \cap \hat{\mathcal{A}}_c^+$} & $(\mathcal{A}_c)^C$ \\

         $H_A^U$ & \textcolor{brown}{$\mathcal{A}_c \cap (\hat{\mathcal{A}}_c^+)^C$} & \textcolor{gray}{$\mathcal{A}_c \cap \hat{\mathcal{A}}_c^+$} & $\mathcal{A}_c$ \\ \hline

        Total & $(\hat{\mathcal{A}}_c^+)^C$ & $\hat{\mathcal{A}}_c^+$ & $\mathcal{V}$  \\
    \hline
    \end{tabular}
    \caption{Partition of the image space $\mathcal{V}$ according to the ground truth and the rejection status (positive direction, one-sided). The color of the text in each cell corresponds to the color of the area in Figure \ref{figure-upper_division} (b) and (c).}
    \label{table-upper_division}
\end{table}

\subsubsection{Spatial FDR and FNDR}
The false discovery proportion is given by the ratio of the number of locations at which the upper null $H_0^U$ is true (that is, $\mu(v) \leq c$) that are rejected ($|(\mathcal{A}_c)^C \cap \hat{\mathcal{A}}_c^+| =|\hat{\mathcal{A}}^+_c\setminus \mathcal{A}_c|$) to the number of all rejected locations ($|\hat{\mathcal{A}}^+_c|$)). The construction of $\hat{\mathcal{A}}_c^+$ ensures control of the upper spatial false discovery rate (${\rm FDR}_U$), which is defined as the expectation of the false discovery proportion, namely
\begin{equation}
{\rm FDR}_U = \mathbb{E}\left[\frac{ |\hat{\mathcal{A}}^+_c\setminus \mathcal{A}_c|}{|\hat{\mathcal{A}}^+_c| \vee 1} \right].
\end{equation}
Algorithm \ref{alg_upper} ensures that this is controlled to a level $\alpha$ under PRDS, see Proposition S6.1 of the supplementary material.

In order to measure a proxy for statistical power in Section \ref{simulation}, we shall use the false non-discovery proportion, which measures the ratio between the number of false non-discoveries and the number of total non-discoveries. Then, the upper false non-discovery rate (FNDR$_U$) is defined as
\begin{equation}
{\rm FNDR}_U = \mathbb{E}\left[  \frac{ |\mathcal{A}_c  \setminus \hat{\mathcal{A}}^+_c|}{|(\hat{\mathcal{A}}^+_c)^C| \vee 1} \right],
\label{FNDR_U}
\end{equation}
where FNDR$_U$ is the expected value of the false non-discovery proportion.

\subsection{Lower False Discovery Controlled Regions}
\label{method_lower}

\subsubsection{Construction}
In order to construct a lower FDCR $\hat{\mathcal{A}_c^-}$ we consider testing against the negative one-sided null hypothesis at level $c$, defined as
\begin{equation}
    H_0^L(v): \mu(v) > c \quad \mbox{vs.} \quad H_A^L(v): \mu(v) \le c.
\label{eq:lower_hypothesis}
\end{equation}
Testing the two directions as two separate one-sided families in this way is the approach advocated by \textcite[\S6.3]{benjamini2005false}. A lower FDCR can be constructed in the same way as for Algorithm 1. However, motivated by high thresholds $c$ in which only a small region may lie above the threshold we instead construct it using an adaptive procedure. 

Construction of the lower FDCR is then formally given by Algorithm \ref{alg_lower}, where as in the previous section we assume that the obtained $p$-values satisfy PRDS. Algorithm \ref{alg_lower} differs from Algorithm \ref{alg_upper} in that it incorporates a two-stage adaptive procedure rather than the BH procedure, the details of which are explained in Section \ref{adapt}.

\begin{algorithm}[H]    
    \caption{Lower False Discovery Controlled Regions}
    \label{alg_lower}
    \begin{algorithmic}[1]
        \Require{$(t(v), \alpha)$}
        \Ensure{ $\hat{\mathcal{A}}_c^-$ }
        \State Obtain $m$ $p$-values $p^L{(v)} = F_{T_{n-1}}\left( t(v)  \right)$, where $t(v)$ is the $t$-statistic defined in (\ref{eq:tstat}) and  $F_{T_{n-1}}$ is the cdf of $t$-distribution with $n-1$ degrees of freedom from the negative one-sided test, so that small values of $p^L(v)$ correspond to the one-sided alternative $H_A^L(v): \mu(v) \le c$.
        \State Apply the two-stage adaptive procedure (Blanchard et al. \cite{blanchard2009adaptive}) to the $p$-values  $\mathbf{p}^L = \left( p^L(1), \ldots, p^{L}(m) \right)$ to get $q$, the threshold for rejection.
        \State Define $\hat{\mathcal{A}}_c^- = \{v \in \mathcal{V}:  p^L(v) \leq q\}^C$, the complement of the set of locations rejected.
        \State \textbf{Return} the lower false discovery controlled regions $\hat{\mathcal{A}}_c^-$.
    \end{algorithmic}
\end{algorithm}


Then, the null set corresponding to the hypothesis (\ref{eq:lower_hypothesis}) is  $\mathcal{A}_c$, the excursion set itself. Note here that unlike the case of the upper FDCR $\hat{\mathcal{A}_c^+}$, the lower FDCR $\hat{\mathcal{A}_c^-}$ now represents the non-significant set of locations. This asymmetry in the definition is by choice, stemming from the fact that while the null hypothesis in the negative direction is used in constructing $\hat{\mathcal{A}_c^-}$, the excursion set is still in the positive direction. Algorithm \ref{alg_lower} may also be run with the BH procedure in place of the two-stage adaptive procedure at step 2, which we refer to below as the BH variant of Algorithm \ref{alg_lower}. We compare these two versions of Algorithm 2 in simulations of Section \ref{simulation} for comparison.

\begin{figure}[H]
    \centering
    \includegraphics[width=1\linewidth]{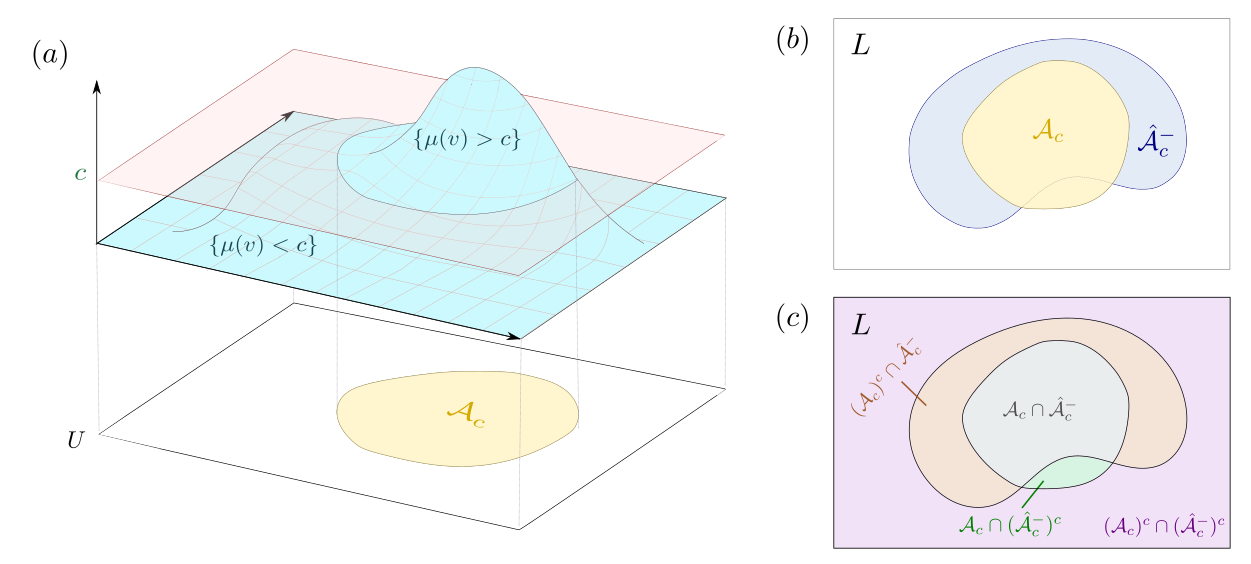}

\caption{Lower FDCR schematic: (a) Example signal and the projection of the excursion set above $c$ ($\mathcal{A}_c$). (b) A hypothetical lower FDCR $\hat{\mathcal{A}}_c^-$ superimposed on the ground truth $\mathcal{A}_c$. (c) The division of (b) into regions corresponding to those specified in Table \ref{table-lower_division} as a visual representation of false positives (green) and false non-positives (orange).}
    \label{figure-lower_division}
\end{figure}

The negative one-sided test partitions the image space $\mathcal{V}$ in the following way (Table \ref{table-lower_division}). For a given level $c$, the test and the corresponding regions are visualized as shown in Figure \ref{figure-lower_division}.

\begin{table}[H]
    \centering
    \renewcommand{\arraystretch}{1.2} \setlength{\extrarowheight}{2pt}
    \setlength{\tabcolsep}{4pt}
    \begin{tabular}{|>{\centering\arraybackslash}p{3cm}|>{\centering\arraybackslash}p{4cm}|>{\centering\arraybackslash}p{4cm}|>{\centering\arraybackslash}p{3cm}|}
    \hline
          Hypothesis & Not rejected & Rejected & Total   \\ \hline
          
         $H_0^L$ & \textcolor{violet}{$\mathcal{A}_c \cap \hat{\mathcal{A}}_c^-$} & \textcolor{green}{$\mathcal{A}_c \cap (\hat{\mathcal{A}}_c^-)^C$} & $\mathcal{A}_c$ \\
         
         $H_A^L$ & \textcolor{brown}{$(\mathcal{A}_c)^C \cap \hat{\mathcal{A}}_c^-$} & \textcolor{gray}{$(\mathcal{A}_c)^C\cap (\hat{\mathcal{A}}_c^-)^C$} & $(\mathcal{A}_c)^C$ \\ \hline
         
        Total & $\hat{\mathcal{A}}_c^-$ & $(\hat{\mathcal{A}}_c^-)^C$ & $\mathcal{V}$  \\
    \hline
    \end{tabular}
    \caption{Partition of the image space $\mathcal{V}$ according to the ground truth and the rejection status (negative direction, one-sided). The color of the text in each cell corresponds to the color of the area in Figure \ref{figure-lower_division} (b) and (c).}
    \label{table-lower_division}
\end{table}

\subsubsection{Spatial FDR and FNDR}
Similar to the upper regions, the lower false discovery proportion is measured by the number of locations at which the lower null $H_0^L$ is true (that is, $\mu(v) > c$) that are rejected ($ |\mathcal{A}_c \cap (\hat{\mathcal{A}}_c^-)^C| = |\mathcal{A}_c \setminus\hat{\mathcal{A}}^-_c|$) divided by the number of rejected locations  ($|(\hat{\mathcal{A}}^-_c)^c|$) from the negative one-sided test.
The false discovery rate for the lower FDCR (${\rm FDR}_L$) is defined as the expectation of the false discovery proportion, namely

\begin{equation}
{\rm FDR}_L = \mathbb{E}\left[ \frac{ |\mathcal{A}_c\setminus\hat{\mathcal{A}}^-_c|}{|(\hat{\mathcal{A}}^-_c)^C| \vee 1} \right].
\label{FDR_lower}
\end{equation}

Algorithm \ref{alg_lower} ensures that this is controlled to a level $\alpha$ under PRDS, see Proposition S6.2 of the supplementary material.

The false non-discovery proportion is measured by the number of locations at which the lower null $H_0^L$ is false that are not rejected in the negative testing, $ |(\mathcal{A}_c)^C \cap \hat{\mathcal{A}}_c^-| = |\hat{\mathcal{A}}^-_c  \setminus \mathcal{A}_c |$, divided by the number of all the non-rejected locations in the negative one-sided test, $|\hat{\mathcal{A}}^-_c|$. Again, the FNDR for the lower FDCR (${\rm FNDR}_L$) is defined as the expectation of the false non-discovery proportion,

\begin{equation}
{\rm FNDR}_L = \mathbb{E}\left[ \frac{ |\hat{\mathcal{A}}^-_c  \setminus \mathcal{A}_c |}{|\hat{\mathcal{A}}^-_c| \vee 1} \right].
\label{FNDR_L}
\end{equation}

\subsection{Increasing Statistical Power via the Two-stage Adaptive Procedure}\label{adapt}

While having a similar construction procedure, the biggest difference between Algorithms \ref{alg_upper} and \ref{alg_lower} is that the latter uses an adaptive procedure to perform multiple testing. The BH procedure controls the FDR at $\alpha \cdot \pi_0$ level where $\pi_0 \coloneqq \frac{m_0}{m}$ with $m$\ being the total number of all the hypotheses, and $m_0$ being the number of true null hypotheses. When the fraction of true null hypotheses $\pi_0$ is small, as is typically the case in the lower-direction test (test \ref{eq:lower_hypothesis}) for high values of $c$, the control of the FDR can be conservative. This warrants an adaptive procedure that entails a correction on $\alpha \cdot \pi_0$ for the lower direction so that the actual error rate is kept at a higher (less conservative) level.

We have chosen to state Algorithm \ref{alg_lower} with high levels of $c$ in mind. However the problem is completely symmetric (in the direction) and so the adaptive procedure could also be applied to the upper set. This could for instance be obtained directly by negating the test-statistics before feeding them into Algorithm \ref{alg_lower}. Similarly the lower FDCR can be constructred using an analogue to Algorithm \ref{alg_upper}. To demonstrate this point, we provide simmulation results comparing the lower FDCR constructed from the BH procedure against the lower region from the two-stage adaptive procedure (See Section \ref{simulation}).

Various adaptive FDR control measures have been suggested, including Benjamini and Hochberg's adaptive modification on the BH procedure \cite{benjamini2000adaptive} and others that followed \cite{storey2002direct, storey2004strong, genovese2004stochastic,  liang2012adaptive, gavrilov2009adaptive, leung2024adaptive} under different assumptions on the data. We employ the two-stage adaptive procedure by Blanchard and Roquain (2009) \cite{blanchard2009adaptive} to form the lower FDCR. We choose this procedure because most adaptive procedures establish FDR control only under independence of the $p$-values, whereas Blanchard and Roquain \cite{blanchard2009adaptive} are one of the few to provide guarantees under dependence, and in particular under the PRDS assumption that we make here. This is essential in the imaging setting, where the spatial smoothness of the data induces strong dependence between neighbouring $p$-values. This works by estimating the number of null hypotheses in the first stage and using this information in the second stage to perform inference.

Blanchard and Roquain express step-up procedures through a threshold collection written in the normalized form $\Delta_i = \pi^{-1}_0 \frac{\alpha \beta(i)}{m}$, where $\beta$ is termed the \emph{shape function}. Under PRDS they establish FDR control only for $\beta$ the identity, and so we take $\beta(i) = i$ throughout, which gives the ordinary linear step-up weights. By multiplying the threshold collection by a factor of $\pi^{-1}_0$, we expect to have a less conservative procedure. In practice, we use $\Delta_i = \hat{\pi}_0^{-1} \frac{\alpha i}{m}$ where $\hat{\pi}_0^{-1} = \frac{m}{\hat{m_0}}$ is an estimator of the true $\pi^{-1}_0$.

The two-stage adaptive procedure by Blanchard and Roquain \cite{blanchard2009adaptive} provably controls the FDR at the $\alpha$ level under PRDS, by their Theorem 23. In the first stage, a step-up procedure is conducted with threshold collection $\Delta_i = \frac{\alpha_0 i}{m}$ to get an estimate $\hat{m}_0 = m - |R_0|$ where $|R_0|$ is the number of rejected hypotheses. In the second stage, an adaptive procedure is conducted with $\Delta_i = \hat{\pi}_0^{-1} \frac{\alpha_1 i}{m}$ where $ \hat{\pi}_0^{-1} = F_{\kappa}(\frac{\hat{m_0}}{m})$. $F_{\kappa}(x)$ for $x \in [0,1]$ is given as follows:

\[  F_{\kappa}(x) = 
\begin{cases}
    1, & \text{if } x \leq \kappa^{-1}, \\
    \frac{2\kappa^{-1}}{1- \sqrt{1-4(1-x)\kappa^{-1}}}, & \text{otherwise}.
\end{cases}
\]

Blanchard and Roquain state that the two-stage procedure with $\alpha_0=\alpha/4, \alpha_1=\alpha/2$ and $\kappa =2$ produces less conservative FDR control compared to the linear step-up procedure alone (BH procedure) when $F_{\kappa=2}(\hat{m_0}/m) \ge 2$, which translates to when we expect at least $F_2^{-1}(2)=62.5\%$ of the tests to be rejected \cite{blanchard2009adaptive}. In the FDCR setting, the rejection corresponds to $(\hat{\mathcal{A}}^-_c)^C$. When $c$ is high enough so that $(\hat{\mathcal{A}}^-_c)^C$ is expected to be larger than $\hat{\mathcal{A}}^-_c$ (which is typically the case in neuroimaging inference), the two-stage adaptive procedure produces more powerful lower FDCRs.

\subsection{Nesting of the upper and lower FDCRs}
\label{sec:nesting-sep}

The upper and lower FDCRs are estimates of the same excursion set from
opposite directions, so it is natural to ask whether the upper region is
always contained within the lower one, i.e.\ whether
$\hat{\mathcal{A}}_c^+ \subseteq \hat{\mathcal{A}}_c^-$. This nesting
property is implicitly presumed whenever the pair is interpreted as
bracketing the excursion set. When both directions are tested using the BH
procedure it holds automatically.

\begin{theorem}
\label{thm:nesting-sep}
Suppose $\hat{\mathcal{A}}_c^+$ is given by Algorithm \ref{alg_upper} and
$\hat{\mathcal{A}}_c^-$ by the BH variant of Algorithm \ref{alg_lower}, both
at level $\alpha$. If $\alpha < 1/2$ then
$\hat{\mathcal{A}}_c^+ \subseteq \hat{\mathcal{A}}_c^-$.
\end{theorem}

The argument is short: a location lying in $\hat{\mathcal{A}}_c^+$ but not in
$\hat{\mathcal{A}}_c^-$ is rejected in both directions, so that $p^U(v)$ and
$p^L(v)$ are both at most $\alpha$; since $p^U(v) + p^L(v) = 1$ this gives
$\alpha \geq 1/2$. The proof is given in Proposition S2.1 of the
supplementary material, and Remark S2.1 records the corresponding counts of
rejections.

The conclusion does not extend to Algorithm \ref{alg_lower} itself. There the
null-proportion correction can increase the threshold applied to the lower
$p$-values sufficiently that a location is rejected in both directions, and
nesting can fail; this is shown in Proposition S5.1 of the supplementary
material. The configurations required are however highly asymmetric, of a
kind that the spatial smoothness of neuroimaging data makes implausible; see
Remark S5.3.

\section{False Discovery Controlled Regions with Joint Error Control\label{joint}}
In this section we develop upper and lower FDCRs which have joint (instead of separate) error rate control. To do so, we shall combine the directional $p$-values from the upper and lower direction tests into a single BH algorithm.

\subsection{Hypothesis Testing}
We now propose a FDCR construction procedure for jointly testing two null hypotheses:
\begin{equation}
    H^U_0(v): \mu(v) \leq c \quad \mbox{vs.} \quad H^U_A(v): \mu(v) > c,
\label{hypothesis-joint1}
\end{equation}
\begin{equation}
    H^L_0(v): \mu(v) > c \quad \mbox{vs.} \quad H^L_A(v): \mu(v) \le c,
\label{hypothesis-joint2}
\end{equation}
where the superscripts $U$ and $L$ denote upper and lower respectively. Define $\mathcal{L} = \mathcal{V} \times \{-1\}$ and $\mathcal{U} = \mathcal{V} \times \{1\}$ to be two homeomorphic copies of $\mathcal{V}$. For sets $A, B \subseteq \mathcal{V}$ we write $A \sqcup B := A \times \{-1\} \cup B \times \{1\}$ for their disjoint union, in which $A$ is placed in the lower copy $\mathcal{L}$ and $B$ in the upper copy $\mathcal{U}$, following the notation of \cite{lee2010introduction}. Given this notation we can define the joint hypothesis space $\mathcal{H} =\mathcal{L}\sqcup \mathcal{U}$. The joint null set for the hypotheses is then $\mathcal{A}_c^C \sqcup \mathcal{A}_c$. Considering the disjoint union will allow us to be able to provide joint confidence statements which hold for the upper and lower spaces simultaneously. 

\begin{figure}[H]
    \centering
    \includegraphics[width=1\linewidth]{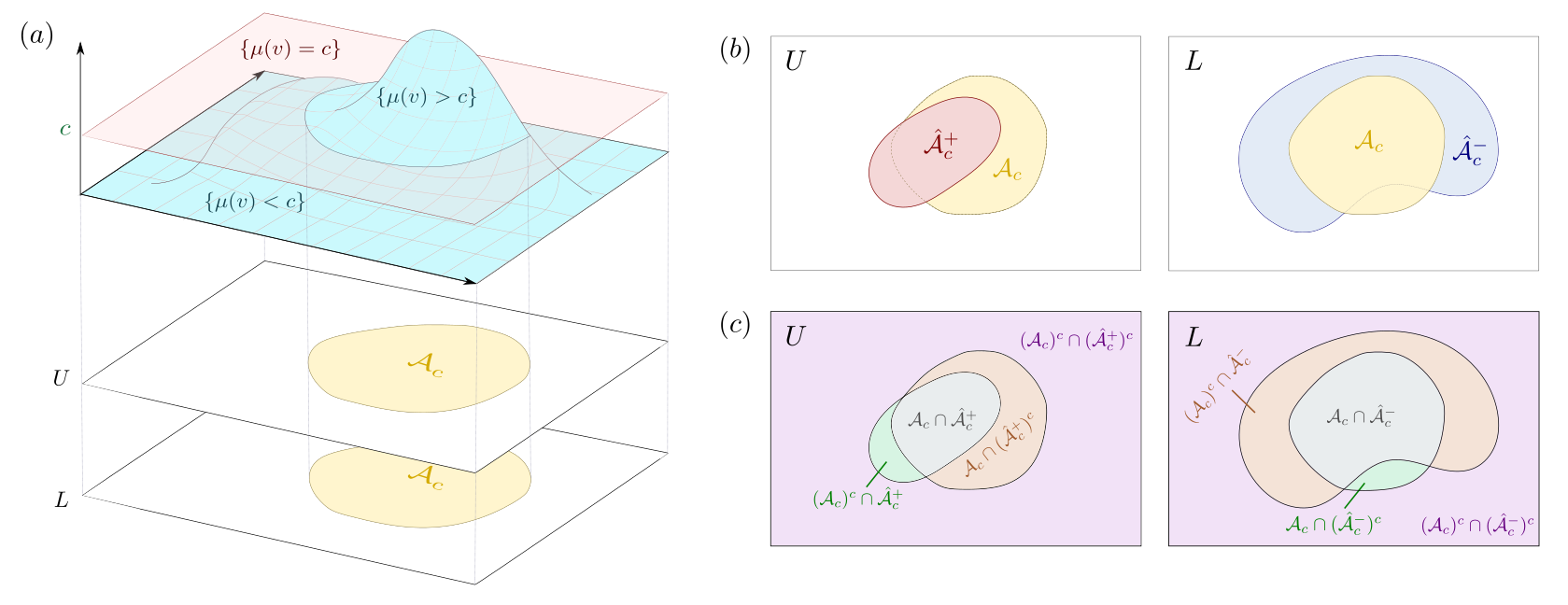}
\caption{Upper and lower FDCRs from joint control method schematic: (a) Example signal and the projection of the excursion set above $c$ ($\mathcal{A}_c$). (b) Hypothetical upper lower FDCRs $\hat{\mathcal{A}}_c^+$ and $\hat{\mathcal{A}}_c^-$ superimposed on the ground truth $\mathcal{A}_c$. (c) The division of (b) into regions corresponding to those specified in Table \ref{table-joint_division} as a visual representation of false positives (green) and false non-positives (orange).}
    \label{figure-joint_division}
\end{figure}

The space $\mathcal{V}$ is partitioned by the joint error control hypothesis testing framework as in Table \ref{table-joint_division}. The corresponding regions are visualized in Figure \ref{figure-joint_division}.

\begin{table}[H]
    \centering
    \renewcommand{\arraystretch}{1.2} \setlength{\extrarowheight}{5pt}
    \setlength{\tabcolsep}{4pt}
    \begin{tabular}{|>{\centering\arraybackslash}p{3cm}|>{\centering\arraybackslash}p{4cm}|>{\centering\arraybackslash}p{4cm}|>{\centering\arraybackslash}p{3cm}|}
    \hline
         Hypothesis  &  Not Rejected & Rejected & Total   \\ \hline
         $H_0$ & \textcolor{violet}{$\{ (\mathcal{A}_c)^C \cap (\mathcal{\hat{A}}_c^+)^C\}  \sqcup \newline \{ \mathcal{A}_c \cap \mathcal{\hat{A}}_c^- \} $} & \textcolor{green}{$\{ (\mathcal{A}_c)^C \cap \mathcal{\hat{A}}_c^+\} \sqcup \newline \{ \mathcal{A}_c \cap (\mathcal{\hat{A}}_c^-)^C \} $} & $  (\mathcal{A}_c)^C     \sqcup  \mathcal{A}_c $ \\

    $H_A$ & \textcolor{brown}{$\{ \mathcal{A}_c \cap (\mathcal{\hat{A}}_c^+)^C  \} \sqcup \newline \{ (\mathcal{A}_c)^C \cap \mathcal{\hat{A}}_c^-  \} $} & \textcolor{gray}{$\{ \mathcal{A}_c \cap \mathcal{\hat{A}}_c^+  \} \sqcup \newline \{ (\mathcal{A}_c)^C \cap (\mathcal{\hat{A}}_c^-)^C 
  \}$} & $ \mathcal{A}_c \sqcup (\mathcal{A}_c)^C $ \\ \hline

       Total   & $(\mathcal{\hat{A}}_c^+)^C  \sqcup  \mathcal{\hat{A}}_c^-  $ & $\mathcal{\hat{A}}_c^+  \sqcup  (\mathcal{\hat{A}}_c^-)^C $ & $\mathcal{V} \times \{ L , U\}$  \\
    \hline
    \end{tabular}
\caption{Table demonstrating the partition of $\mathcal{V}$ according to true hypothesis and rejection. Here the rows refer to the hypotheses of both one-sided families at once, regarded as hypotheses on the joint space $\mathcal{H} = \mathcal{L} \sqcup \mathcal{U}$: the $H_0$ row collects the locations at which the relevant one-sided null holds and likewise for $H_A$.}
\label{table-joint_division}
\end{table}

\subsection{Construction}
Given the same setting of image space $\mathcal{V}$ consisting of $m$ locations, we construct the upper and lower FDCRs for the area above the threshold $c$, $\mathcal{A}_c$. 
Let $p^U(v)$ be the $p$-value calculated from the $t$-statistics corresponding to the positive test (\ref{hypothesis-joint1}), and $p^L(v) = 1-p^U(v)$ be the $p$-value calculated from the t-statistics corresponding to the negative test (\ref{hypothesis-joint2}). The two $p$-values are exact complements because both are obtained from the same $t$-statistic $t(v)$ under the same reference distribution: $p^U(v) = 1 - F_{T_{n-1}}(t(v))$ and $p^L(v) = F_{T_{n-1}}(t(v))$, so that their sum is $1$ at every location.
Subsequently, $\mathbf{p} = \left( p^L(1), \ldots, p^L(m), p^U(1), \ldots, \ p^U(m) \right)$ collectively undergoes the BH procedure for joint FDR control through which the voxels are identified as rejected or not rejected.

\begin{algorithm}[H]    
    \caption{Upper and Lower False Discovery Controlled Regions with Joint Error Control}
    \label{jointerrorcontrol}
    \begin{algorithmic}[1]
        \Require{$(t(v), \alpha)$}
        \Ensure{ $(\hat{\mathcal{A}}_c^+, \hat{\mathcal{A}}_c^-)$ }
        \State Obtain $m$ $p$-values $p^U(v) =  1-F_{T_{n-1}}\left( t(v) \right)$, where $t(v)$ is the $t$-statistic defined in (\ref{eq:tstat}) and $F_{T_{n-1}}$ is the cdf of $t$-distribution with $n-1$ degrees of freedom from the positive one-sided test, so that small values of $p^U(v)$ correspond to the one-sided alternative $H^U_A(v): \mu(v) > c$ of (\ref{hypothesis-joint1}). 
        \State Obtain $p^L(v) = 1-p^U(v)$.
        \State Apply the Benjamini-Hochberg procedure to $\mathbf{p} = \left( p^L(1), \ \ldots, \ p^L(m), \ p^U(1), \ \ldots, \ p^U(m) \right)$ to get $q$, the biggest p-value for rejection.
        \State Define $\hat{\mathcal{A}}_c^+  = \{v \in \mathcal{V} :  p^U(v) \leq q\}$, the set of voxels rejected from $U$.
        \State Define $\hat{\mathcal{A}}_c^- = \{v \in \mathcal{V} :  p^L(v) \leq q\}^C$, the complement of the set of voxels rejected from $L$.
        
        \State \textbf{return} the upper FDCR $\hat{\mathcal{A}}_c^+$ and the lower FDCR $\hat{\mathcal{A}}_c^-$.

    \end{algorithmic}
\end{algorithm}

Importantly the BH procedure controls the FDR to a level $\frac{m_0}{m}\cdot \alpha$ where $m$ denotes the number of locations tested, and $m_0$ denotes the true non-null locations \textcolor{blue}{\cite{benjamini2001control}}. With the joint error control hypothesis testing, the locations in the image are tested twice with each location being the true non-null hypothesis for at least one of the directions considered. This means that the FDR is effectively controlled at a level $\frac{m}{2m}\cdot \alpha$. We can take advantage of this and use a nominal level of $2\alpha$ instead of $\alpha$ while still providing FDR control at a level $\alpha$, leading to a substantial power improvement.

Technically, PRDS does not hold for the jointly defined $p$-values due to a limited amount of negative correlation between lower and upper $p$-values at the $\mathcal{A}_c$, however in practice this effect is very small and does not affect error control, as shown in Section \ref{simulation_FDR}. In particular Section S6.4 of the supplementary material shows that this negative dependence is confined to pairs of null hypotheses lying close to the boundary of the excursion set.

If the $t$-statistics are instead assumed assumed to be independent across locations then Algorithm \ref{jointerrorcontrol} run at a nominal level $\alpha' < 1/2$ satisfies ${\rm FDR}_J \le \alpha'/2$, so that the nominal level $2\alpha$ delivers the target level $\alpha$ with no slack lost to the doubling, provided $\alpha < 1/4$; see Proposition S6.3 of the supplementary material. Notably this requires no assumption on the dependence between the two directions at a given location: exactly one of the two hypotheses at each location is null, and the non-null partner is neutralised before any probabilistic step is taken.

\subsection{Joint False Discovery Controlled Regions: spatial FDR and FNDR}

The false discovery proportion in FDCRs with joint FDR control is defined by the number of true null hypotheses that are rejected (with $H_0$ as defined in Table \ref{table-joint_division}, that is $H_0^U$ on the upper copy and $H_0^L$ on the lower), $ |\{(\mathcal{A}_c)^C \cap \mathcal{\hat{A}}_c^+\} \sqcup  \{ \mathcal{A}_c \cap (\mathcal{\hat{A}}_c^-)^C\}|= |(\hat{\mathcal{A}}^+_c\setminus \mathcal{A}_c) \sqcup (\mathcal{A}_c\setminus\hat{\mathcal{A}}^-_c)|$, divided by the total number of rejections across both one-sided tests, $| \hat{\mathcal{A}}^+_c \sqcup(\hat{\mathcal{A}}_c^{-})^C|$. The FDR for FDCRs with joint error rate control (${\rm FDR}_J$) is thus defined as the expected false discovery proportion,

\begin{equation}
    {\rm FDR}_J = \mathbb{E}\left[ \frac{ |\{(\mathcal{A}_c)^C \cap \mathcal{\hat{A}}_c^+\} \sqcup  \{ \mathcal{A}_c \cap (\mathcal{\hat{A}}_c^-)^C\}|}{| \hat{\mathcal{A}}^+_c \sqcup (\hat{\mathcal{A}}_c^{-})^C| \vee 1}\right] = \mathbb{E}\left[ \frac{ |\{(\mathcal{A}_c)^C \cap \mathcal{\hat{A}}_c^+\}| +  |\{ \mathcal{A}_c \cap (\mathcal{\hat{A}}_c^-)^C\}|}{(|\hat{\mathcal{A}}^+_c| + |(\hat{\mathcal{A}}_c^{-})^C|) \vee 1}\right].
\label{FDR_J}
\end{equation}

Similarly, the false non-discovery proportion for joint FDR control is defined by the number of hypotheses that are non-null --- $H_A^U$ on the upper copy, $H_A^L$ on the lower --- but are not rejected, over the upper and lower sets $(| \{ \mathcal{A}_c \cap (\mathcal{\hat{A}}_c^+)^C  \} \sqcup \{ (\mathcal{A}_c)^C \cap \mathcal{\hat{A}}_c^-  \} | = |(\mathcal{A}_c \setminus \hat{\mathcal{A}}^+_c)  \sqcup (\hat{\mathcal{A}}^-_c \setminus \mathcal{A}_c)|)$  and the number of total non-discoveries $(|(\mathcal{\hat{A}}_c^+)^C  \sqcup  \mathcal{\hat{A}}_c^- | )$.  Therefore, the FNDR for joint error control (${\rm FNDR}_J$) is defined as the expected false non-discovery proportion,

\begin{equation}
{\rm FNDR}_J = \mathbb{E}\left[\frac{ |(\mathcal{A}_c \setminus \hat{\mathcal{A}}^+_c)  \sqcup (\hat{\mathcal{A}}^-_c \setminus \mathcal{A}_c)|}{ |(\mathcal{\hat{A}}_c^+)^C  \sqcup  \mathcal{\hat{A}}_c^- | \vee 1 }\right] =  \mathbb{E}\left[\frac{ |\mathcal{A}_c \setminus \hat{\mathcal{A}}^+_c| +  |\hat{\mathcal{A}}^-_c \setminus \mathcal{A}_c|}{(|(\hat{\mathcal{A}}^+_c)^C| + |\hat{\mathcal{A}}^-_c|)  \vee 1 }\right].
\label{FNDR_J}
\end{equation}

Table \ref{tab:FDRSummary} summarizes the spatial FDR and FNDR for the upper, lower and joint methods.

\begin{table}[H]
    \centering
    \renewcommand{\arraystretch}{1.2} \setlength{\extrarowheight}{6pt}
    \setlength{\tabcolsep}{6pt}
    \begin{tabular}{|>{\centering\arraybackslash}p{4cm}|>{\centering\arraybackslash}p{5cm}|>{\centering\arraybackslash}p{5cm}|}
    \hline
         & FDR & FNDR \\ \hline
     Separate Upper   &  $\mathbb{E}\left[\frac{ |\hat{\mathcal{A}}^+_c\setminus \mathcal{A}_c|}{|\hat{\mathcal{A}}^+_c|  \vee 1 }\right]$ &  $\mathbb{E}\left[ \frac{ |    \mathcal{A}_c  \setminus \hat{\mathcal{A}}^+_c|}{|(\hat{\mathcal{A}}^+_c)^c|  \vee 1} \right]$\\ \hline
     Separate Lower &  $\mathbb{E}\left[\frac{ |\mathcal{A}_c\setminus\hat{\mathcal{A}}^-_c|)}{|(\hat{\mathcal{A}}^-_c)^C|  \vee 1}\right]$ & $\mathbb{E}\left[\frac{ |\hat{\mathcal{A}}^-_c  \setminus \mathcal{A}_c |}{|\hat{\mathcal{A}}^-_c|  \vee 1} \right]$ \\ \hline
     Joint & $\mathbb{E}\left[ \frac{ |\hat{\mathcal{A}}^+_c\setminus \mathcal{A}_c| + |\mathcal{A}_c\setminus\hat{\mathcal{A}}^-_c|}{(|\hat{\mathcal{A}}^+_c| + |(\hat{\mathcal{A}}_c^{-})^C|) \vee 1}\right]$  & $\mathbb{E}\left[\frac{ |\mathcal{A}_c \setminus \hat{\mathcal{A}}^+_c| +  |\hat{\mathcal{A}}^-_c \setminus \mathcal{A}_c|}{(|(\hat{\mathcal{A}}^+_c)^C| + |\hat{\mathcal{A}}^-_c|)  \vee 1 }\right]$ \\
     \hline
    \end{tabular}
    \caption{Summary of the spatial FDR and FNDR for FDCRs produced by different methods.}
    \label{tab:FDRSummary}
\end{table}

\subsection{Nesting of the upper and lower FDCRs}
\label{sec:nesting-joint}

As in the separate case of Section \ref{sec:nesting-sep}, we may ask whether
the upper FDCR produced by Algorithm \ref{jointerrorcontrol} is contained
within the lower one. It is, and the argument is in fact shorter, as the
rejections are counted on a single hypothesis space of size $2m$.

\begin{theorem}
\label{thm:nesting-joint}
Suppose $\hat{\mathcal{A}}_c^+$ and $\hat{\mathcal{A}}_c^-$ are constructed
jointly, as in Algorithm \ref{jointerrorcontrol}, by applying the BH
procedure at level $\alpha$ to the $2m$ joint $p$-values. If $\alpha < 1/2$
then $\hat{\mathcal{A}}_c^+ \subseteq \hat{\mathcal{A}}_c^-$. In particular,
when Algorithm \ref{jointerrorcontrol} is run at the nominal level $2\alpha$,
nesting is guaranteed whenever $\alpha < 1/4$, which covers the conventional
choices of $\alpha$ used in practice.
\end{theorem}

A location lying in $\hat{\mathcal{A}}_c^+$ but not in
$\hat{\mathcal{A}}_c^-$ has both its upper and its lower copy rejected, so
that $p^U(v)$ and $p^L(v)$ are both at most $K\alpha/(2m)$, where $K$ is the
total number of rejections. Adding the two bounds and using
$p^U(v) + p^L(v) = 1$ gives $K \geq m/\alpha$, which together with
$K \leq 2m$ forces $\alpha \geq 1/2$. The proof is given in Proposition
S3.1 of the supplementary material.

\section{Simulations\label{simulation}}
\subsection{Error Rate Simulation Setting}
\begin{figure}[H]
    \centering
    \includegraphics[width=1\linewidth]{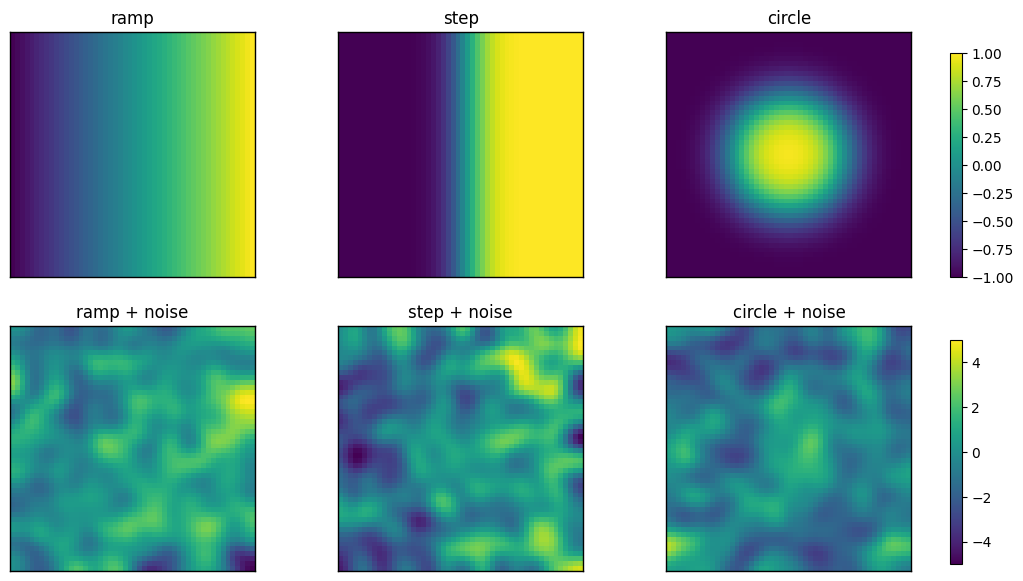}
    \caption{An example synthetic image used for the simulation. The first row shows ramp, step, and circle signals with smoothing (signal smoothing with full width at half maximum 10 pixels). The second row shows signal images with added smoothed Gaussian noise. The pixels in the noise field follow $\mathcal{N}(0,1)$. The noise field is smoothed with full width at half maximum 5 pixels.}
    \label{illustration_signal}
\end{figure}

The error rate simulation was conducted to demonstrate the performance of the proposed FDCR methods in terms of empirical FDR and empirical FNDR depending on different $c$ levels. The simulation was repeated 1,000 times where in each simulation instance, a set of 80 synthetic 2D images were generated with each image consisting of $50 \times 50 = 2500$ pixels.

Within each signal, we denote each image as $y_i: \mathcal{V} \rightarrow  \mathbb{R}, i = 1, \ldots, 80$. The images $y_i$ are generated from signal and noise field such that $y_i(v) = \Theta_\mu(\mu(v)) + \Theta_\epsilon(\epsilon_i(v))$ where $\mu(v)$ denotes the underlying signal function across $v \in \mathcal{V}$, and $\epsilon_i(v) \sim \mathcal{N}(0, \sigma^2)$ denotes the Gaussian random noise. Given $\mu(v)$ and $\epsilon_i(v)$, $\Theta_\mu$ and $\Theta_\epsilon$ perform smoothing with Gaussian kernel with differing degrees of full width at half maximum (FWHM).

Then, the FDCRs were constructed via:
\begin{enumerate}
    \item joint error control for upper and lower FDCRs at $\alpha = 0.1$,
    \item separate error control for upper FDCR (BH procedure) at $\alpha = 0.05$,
    \item separate error control for lower FDCR (BH procedure) at $\alpha = 0.05$, and 
    \item separate error control for lower FDCR (two-stage adaptive procedure) at $\alpha = 0.05$. 
\end{enumerate}

Based on these four sets of FDCRs, false discovery and false non-discovery proportions were calculated at $c$ levels ranging from $-2$ to $2$ with 0.2 increments, resulting in 21 different $c$ levels in total. Empirical FDR and FNDR were then measured as the mean of the false discovery and false non-discovery proportions over 1,000 simulations at each $c$ level.

This process is repeated across different simulation settings generated using the combinations of (i) three signal types (ramp, step, and circle), (ii) three noise smoothing levels (${\rm FWHM}=0, 5, 10$) for $\Theta_\epsilon$ to emulate correlated noise field assumption which is realistic and commonly observed in real world fMRI data, and (iii) three signal smoothing levels (${\rm FWHM}=5, 10, 15$) for $\Theta_\mu$.

We consider three signals: ramp, circle, and step. The ramp signal increases linearly from -1 to 1 over 50 pixels horizontally from left to right. There is no signal smoothing involved in the ramp signals and thus ramp signals are unaffected by different signal smoothing levels, as it is defined as linearly gradual increase across the grid.
The step signal is generated by smoothing over a 2D signal where the left half of the image is of value -1, and the right half of the image is of the value 1.
The circle signal is generated by smoothing over a circle of radius 12 pixels which has magnitude 1 on the background of field of value -1.

There are some benefits for using ramp, step and circle signals for error simulation. Ramp signal is effective in demonstrating how the error rates behave depending on different $c$ levels when there is linear change in slope in signal. Step signals illustrate the effect of flat areas in the image on error rates. Finally, the circle signal is a variation on the step signal where the flat region of magnitude 1 is sloped out by Gaussian kernel smoothing over the background of value -1, creating a blob of flat region that resembles signals observed from fMRI scans.


\begin{figure}[h]
    \centering
    \includegraphics[width=1\linewidth]{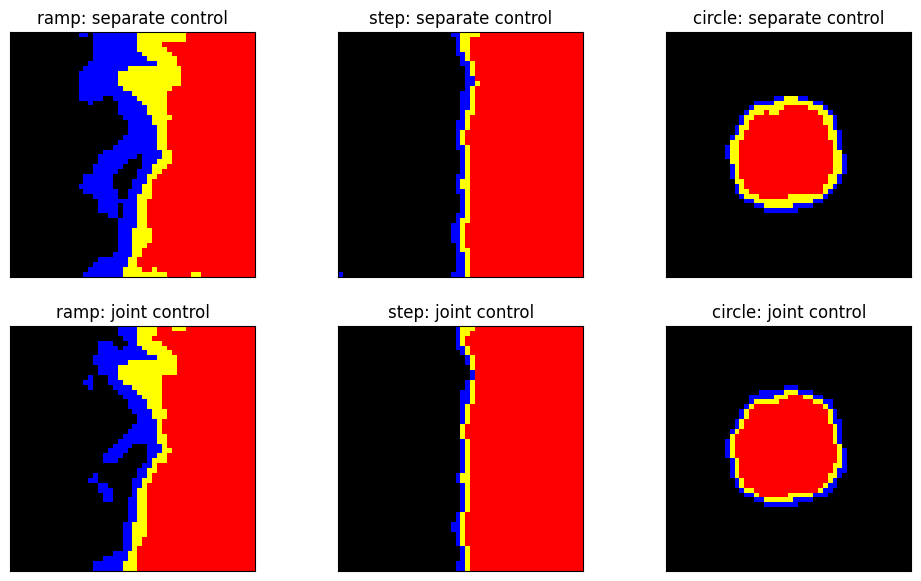}
    \caption{An illustration of FDCRs for the 2D synthetic images ($n=80$) with $\alpha = 0.05$ for separate control (the upper region with BH procedure, and the lower region with the two-stage adaptive procedure), and $\alpha = 0.1$ for the joint control; the red area denotes the upper FDCR $\hat{\mathcal{A}}_c^+$, the yellow area including the red area denotes the excursion set $\hat{\mathcal{A}}_c$, and the blue area including the yellow area denotes the lower FDCR $\hat{\mathcal{A}}_c^-$. The smoothing and noise field settings are replicated from Figure \ref{illustration_signal}.}
    \label{illustration_confset}
\end{figure}

Figure \ref{illustration_signal} shows example synthetic images per signal type with signal smoothing at FWHM 10 pixels, noise smoothing at FWHM 5 pixels, and noise field following $\mathcal{N}(0,1)$. With the same smoothing and noise field setting, an example FDCR inference result from a set of 80 synthetic images is displayed in Figure \ref{illustration_confset}. Figure \ref{illustration_confset} shows the upper and lower FDCRs superimposed on $\hat{\mathcal{A}}_c = \{v \in \mathcal{V}: \hat{\mu}(v) > c\}$, where $\hat{\mu}(v)$ is the sample mean of the images. The first row shows FDCRs with separate control method (upper BH and lower two-stage adaptive), while the second row shows FDCRs with joint control method.

In summary, all the combinations of the settings resulted in 25 different simulation setting combinations, as only one signal smoothing level (no smoothing) was considered for the ramp signal (due to ramp indifference to signal smoothing). Among noise smoothing settings, only ${\rm FWHM}=5$ is reported, as the other two settings showed similar results.

\subsection{FDR Simulation Results}
\label{simulation_FDR}

\newcommand{\simFDRramp}{\includegraphics[width=0.23\textwidth]{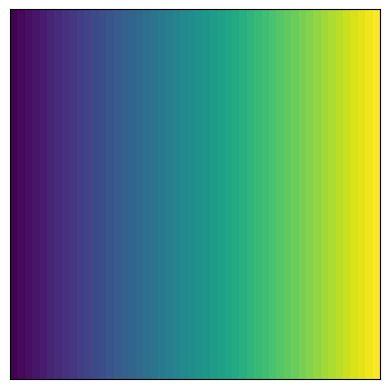}}
\newcommand{\simFDRstep}{\includegraphics[width=0.23\textwidth]{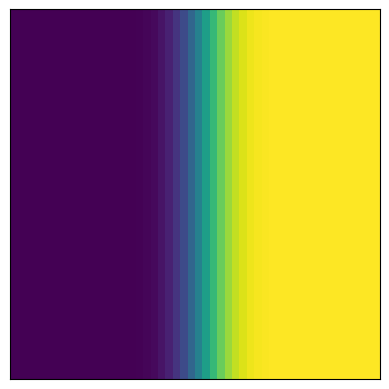}}
\newcommand{\simFDRcircle}{\includegraphics[width=0.23\textwidth]{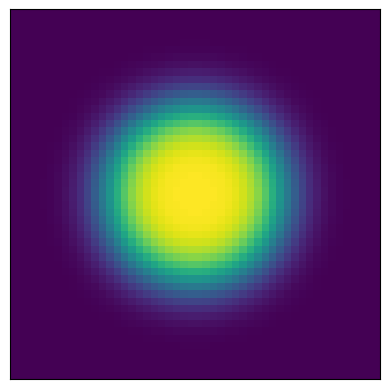}}
\newcommand{\simFDRoo}{\raisebox{1em}{\adjustbox{valign=m, max width=\dimexpr0.33\textwidth-2\tabcolsep}{\includegraphics{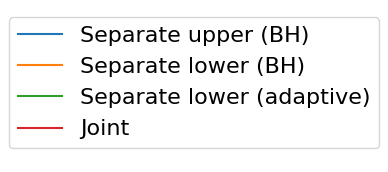}}}}
\newcommand{\simFDRot}{\includegraphics[width=0.23\textwidth]{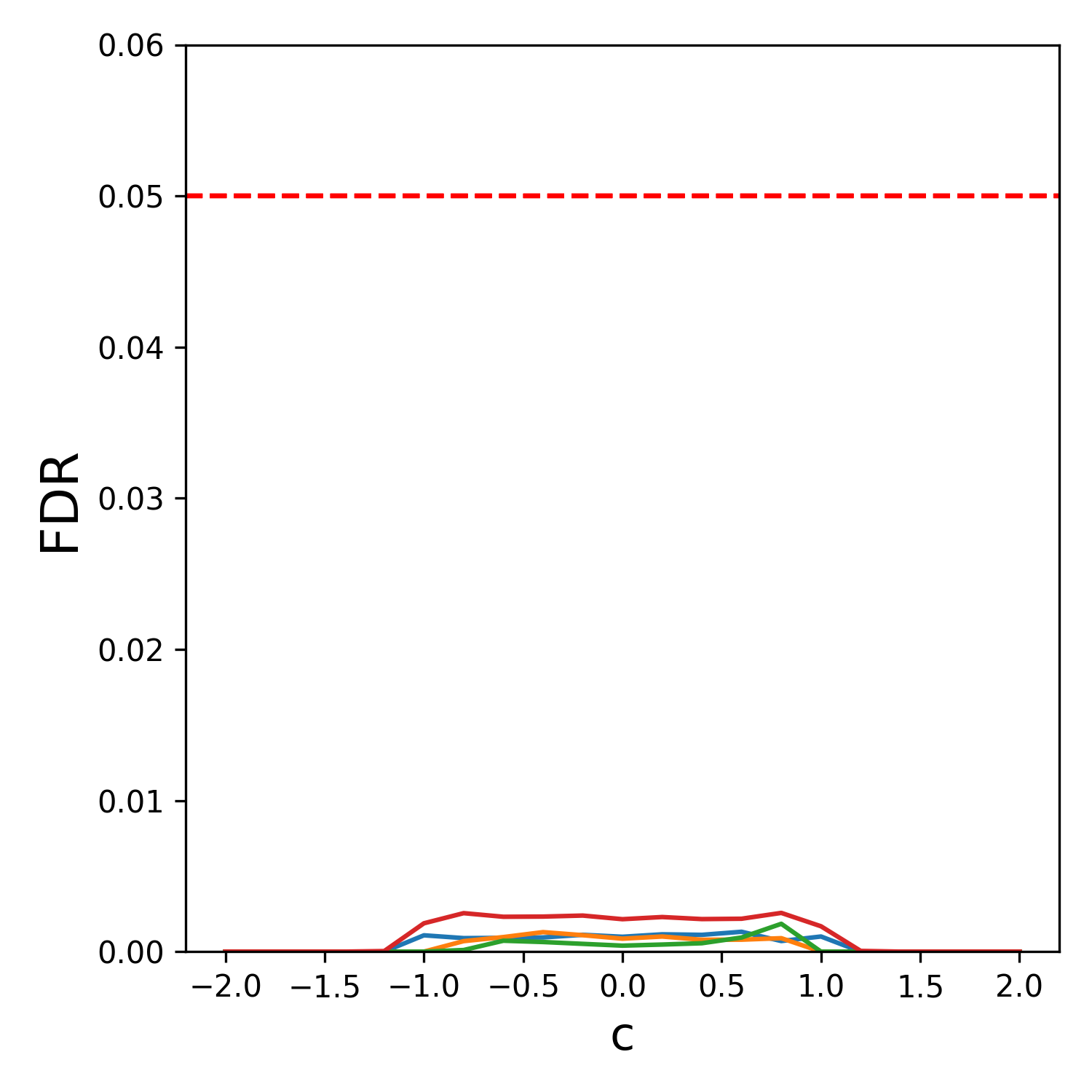}}
\newcommand{\simFDRoth}{\includegraphics[width=0.23\textwidth]{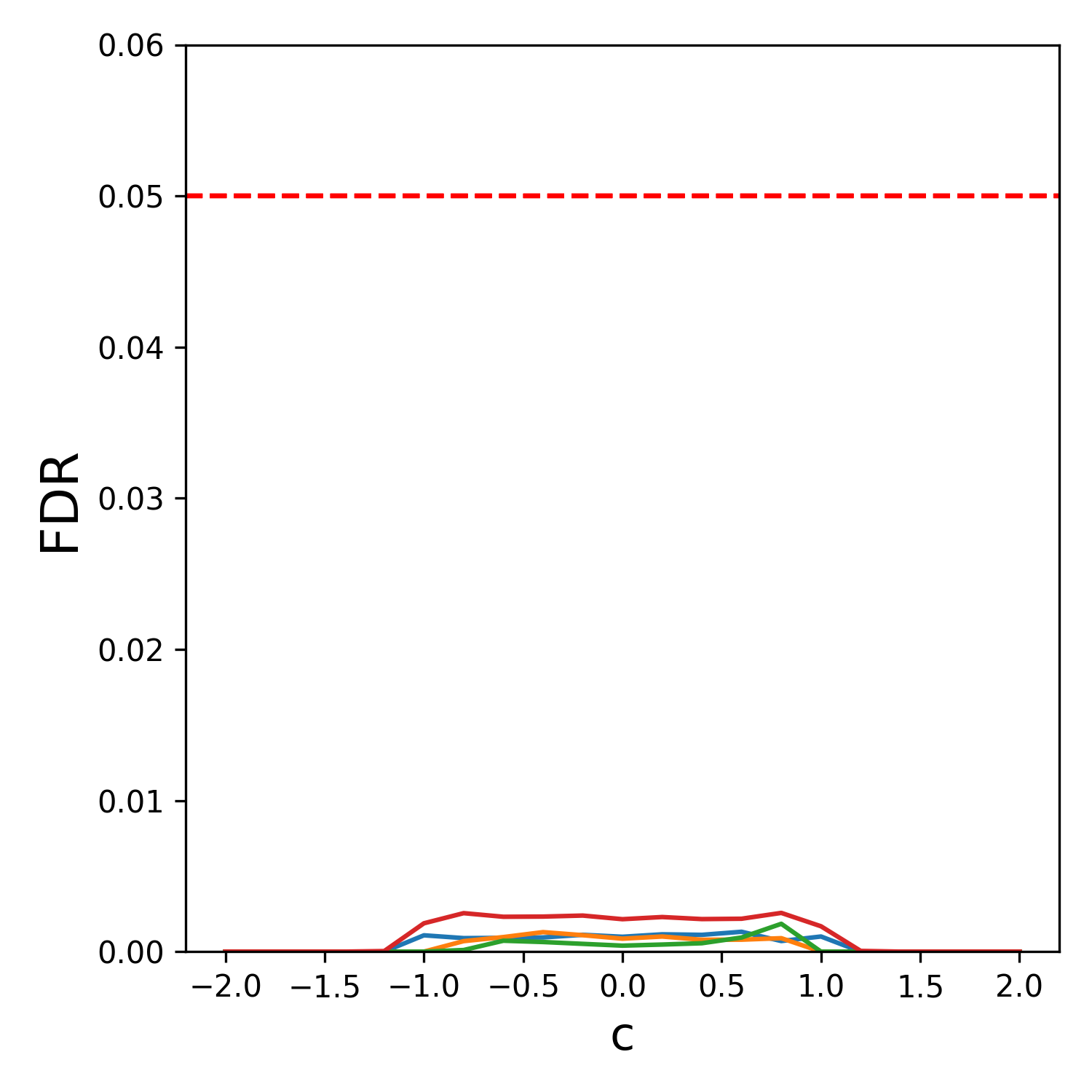}}
\newcommand{\simFDRto}{\includegraphics[width=0.23\textwidth]{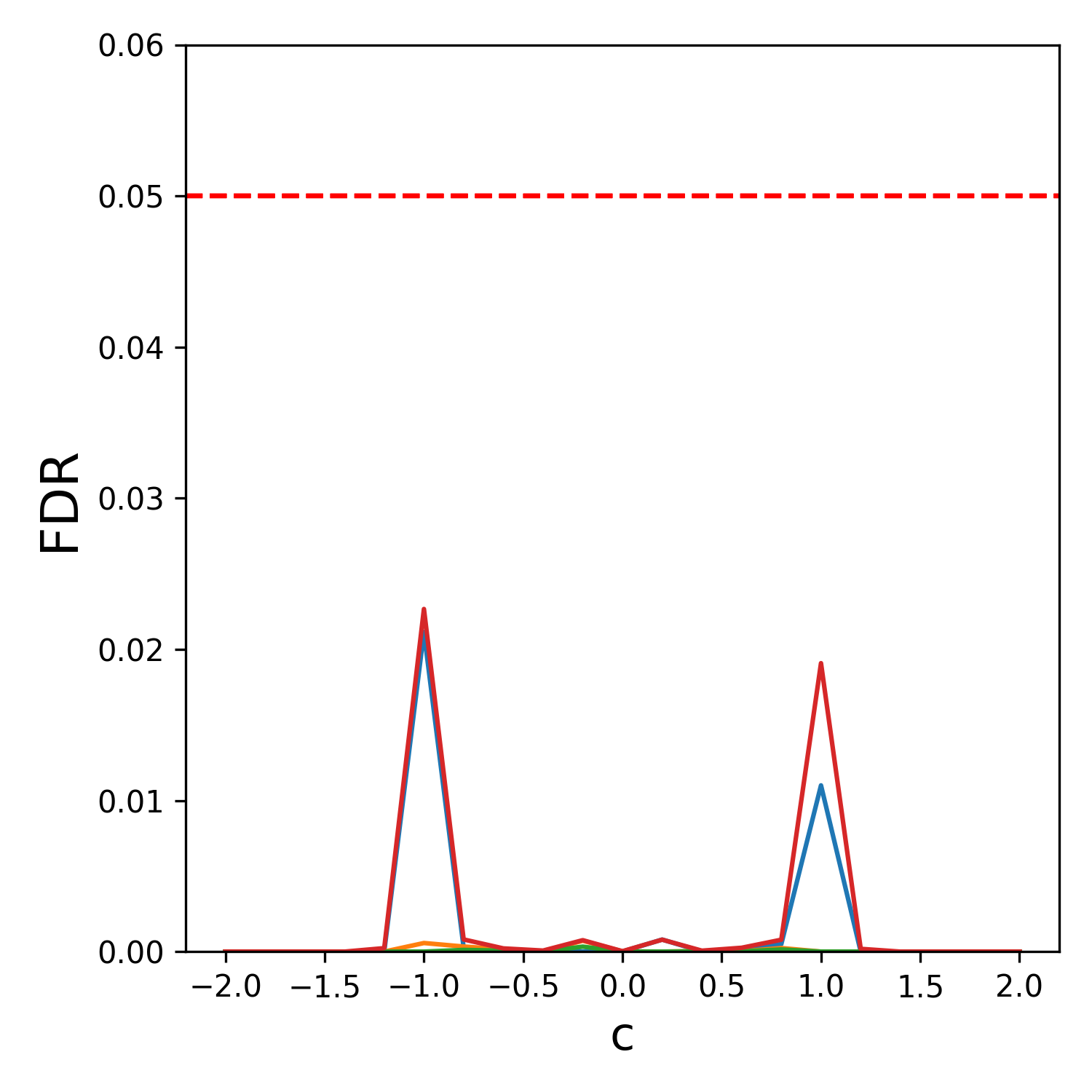}}
\newcommand{\simFDRtt}{\includegraphics[width=0.23\textwidth]{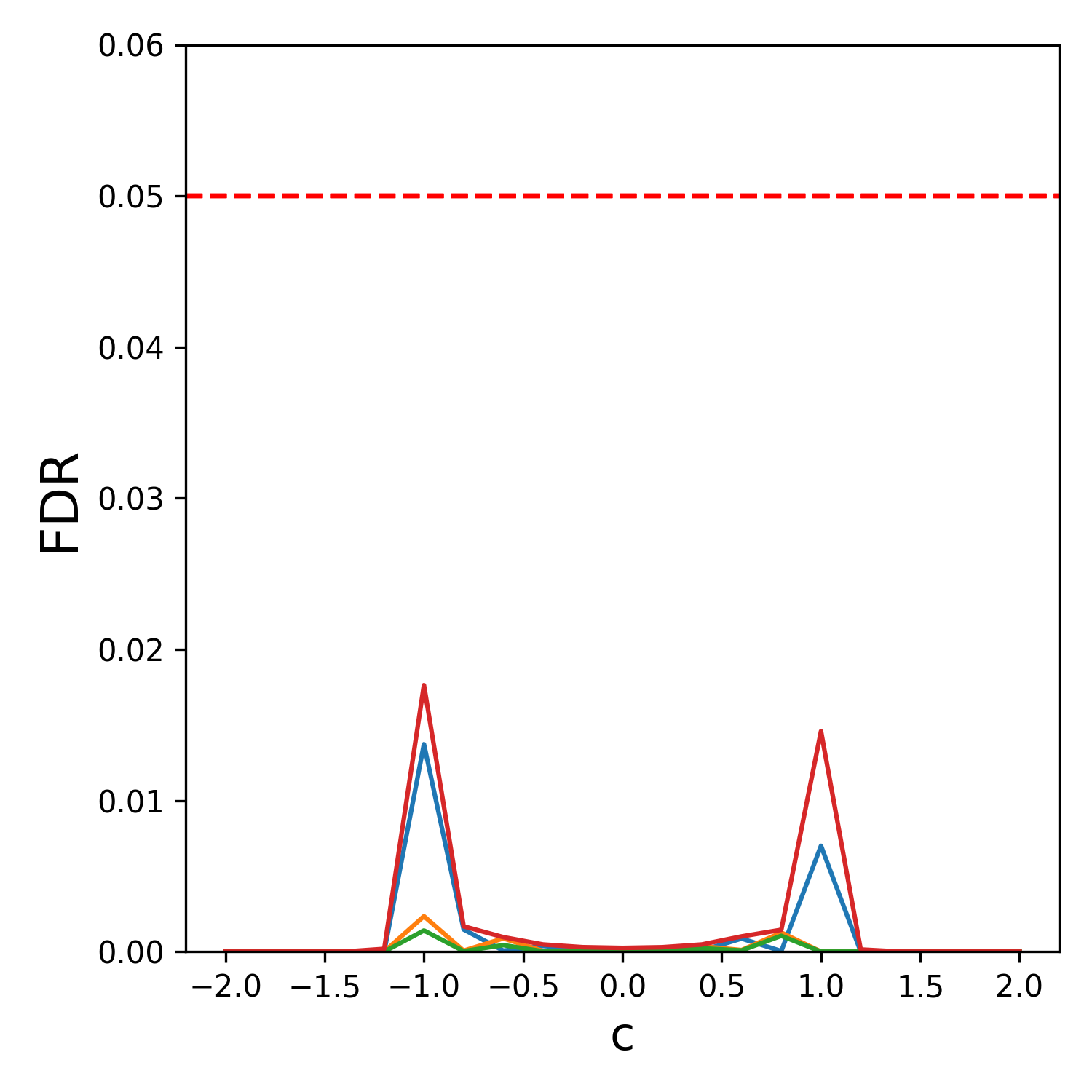}}
\newcommand{\simFDRtth}{\includegraphics[width=0.23\textwidth]{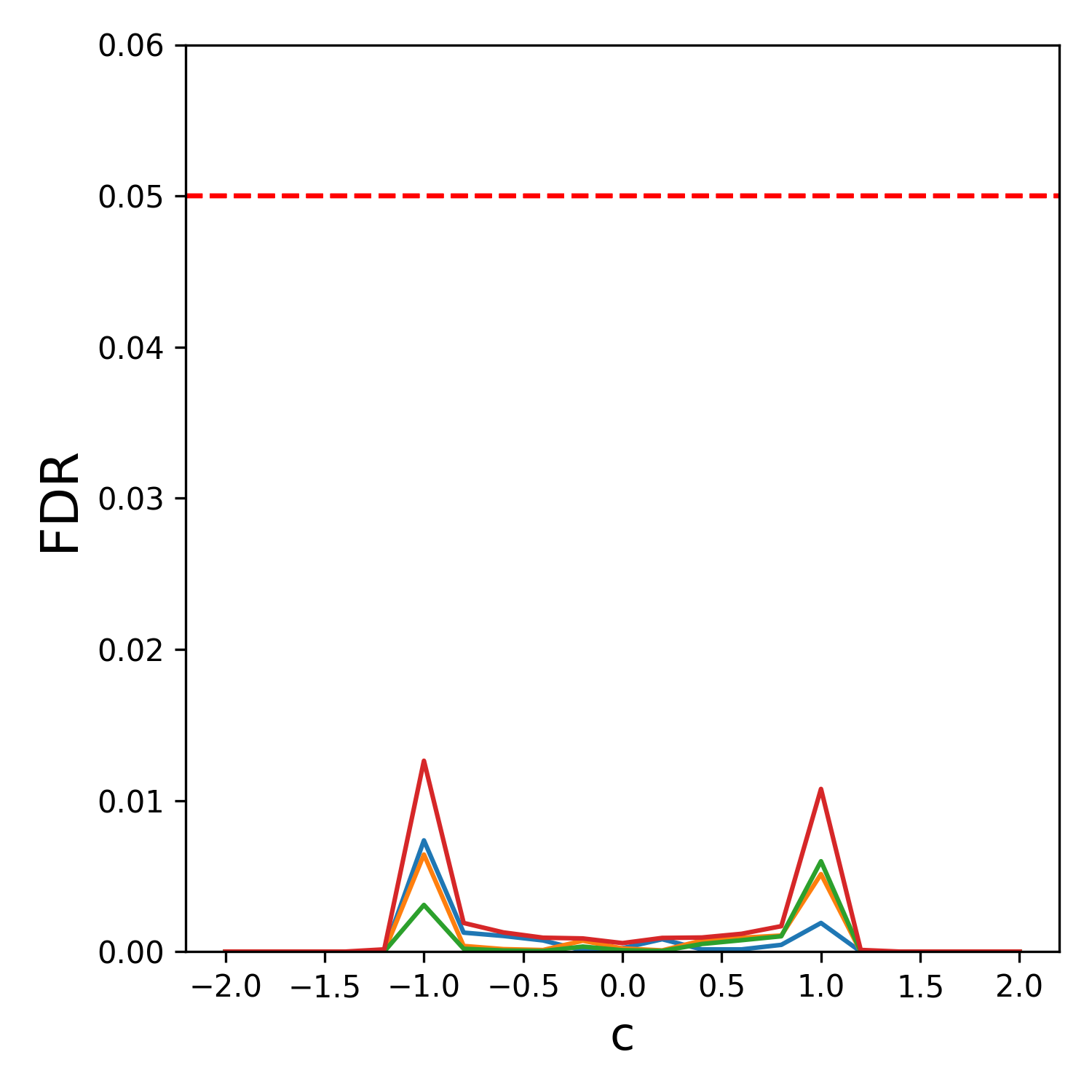}}
\newcommand{\simFDRtho}{\includegraphics[width=0.23\textwidth]{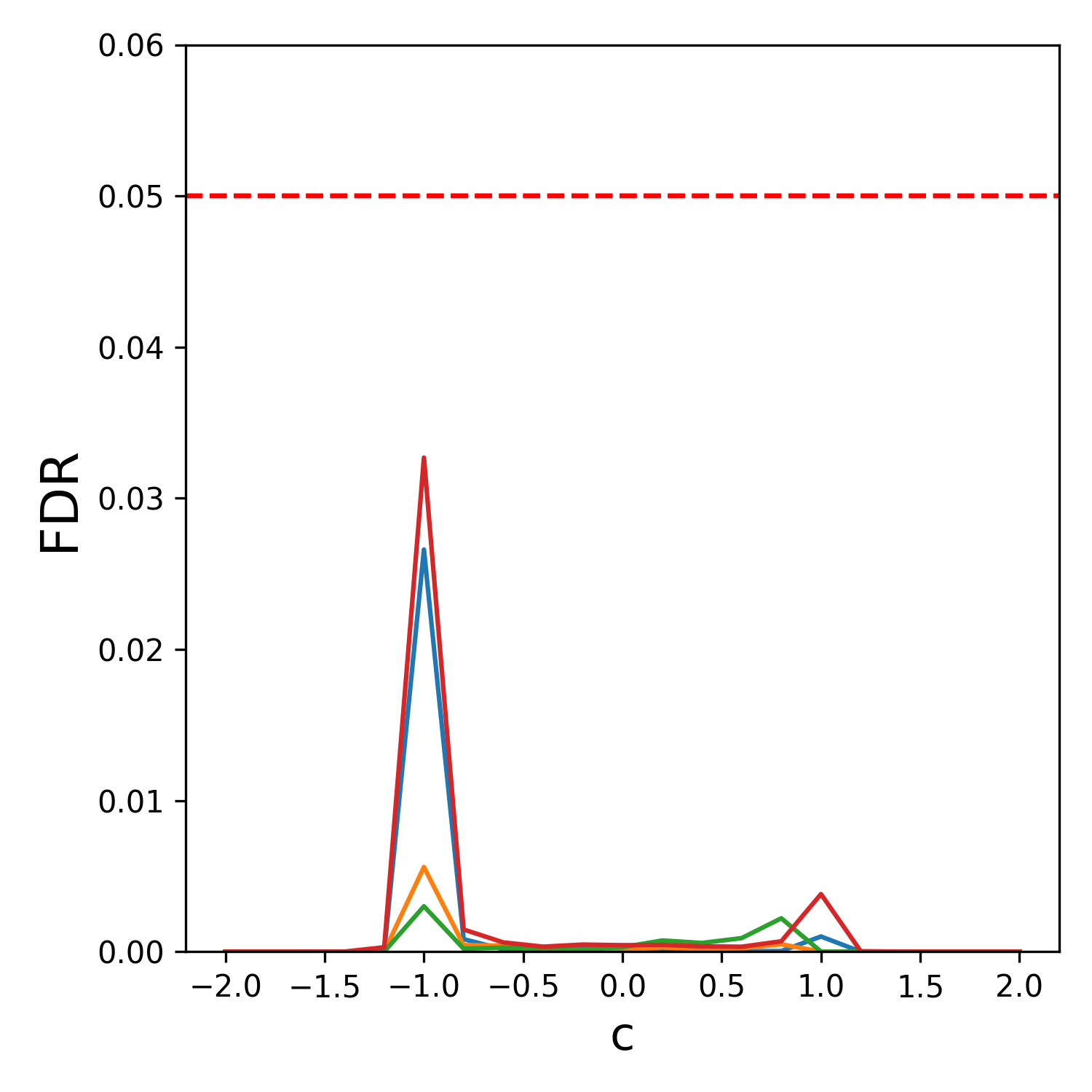}}
\newcommand{\simFDRtht}{\includegraphics[width=0.23\textwidth]{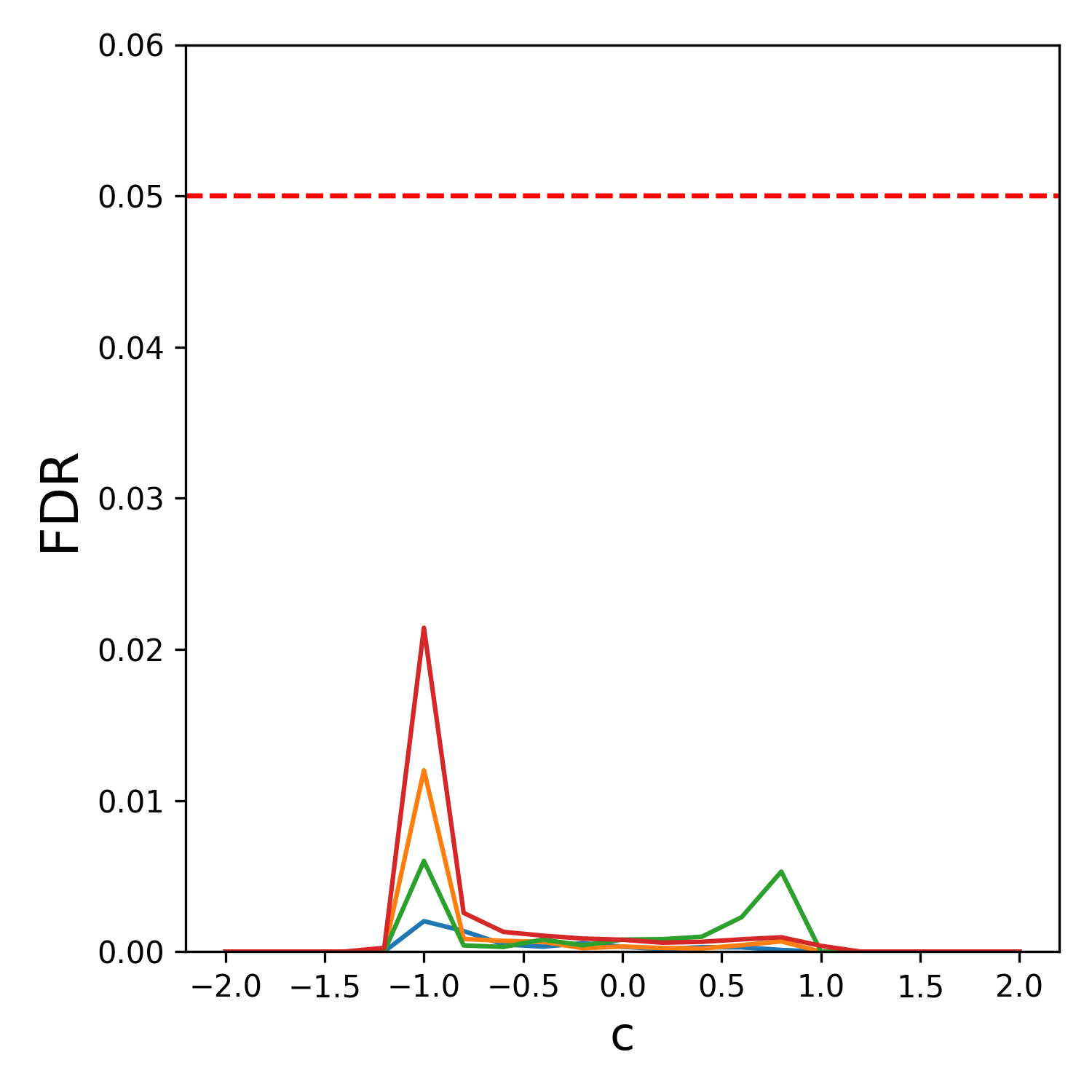}}
\newcommand{\simFDRthth}{\includegraphics[width=0.23\textwidth]{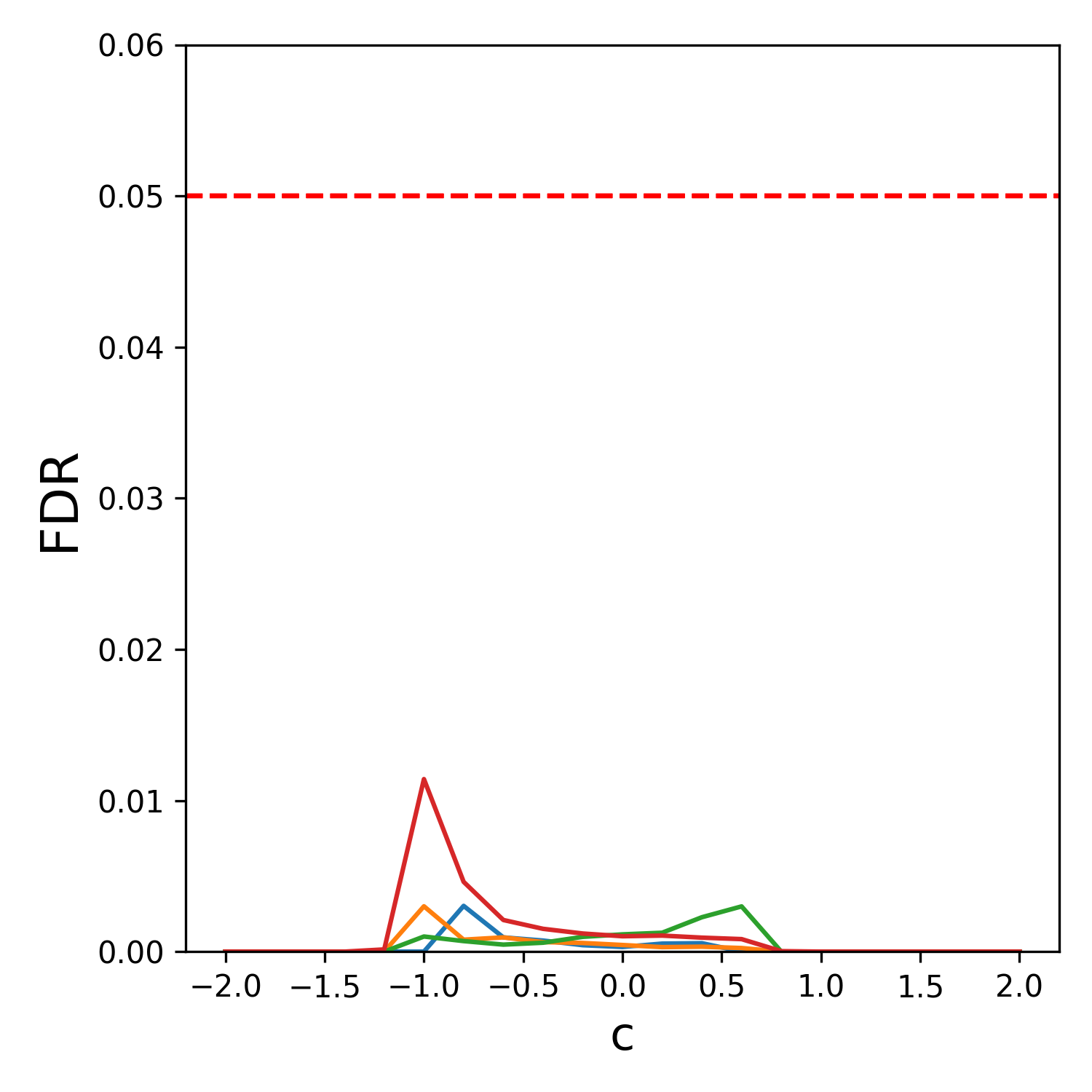}}

\newcolumntype{C}{>{\centering\arraybackslash}m{9em}}
\begin{figure}[H]\sffamily
\centering
\setlength{\tabcolsep}{0pt}
\begin{tabular}{CCCC}
\toprule
Signal  & $\text{FWHM}=5$ & $\text{FWHM}=10$ & $\text{FWHM}=15$  \\ 
\midrule
 \simFDRramp & \multicolumn{2}{c}{\simFDRoo}   & \simFDRot  \\ 
 \simFDRstep & \simFDRto & \simFDRtt & \simFDRtth \\ 
 \simFDRcircle & \simFDRtho & \simFDRtht & \simFDRthth \\ 
\bottomrule 
\end{tabular}
\caption{FDR simulation result for three signals (ramp, step and circle top to bottom) and three signal smoothing levels with noise smoothing at ${\rm FWHM}=5$. The $x$-axis denotes threshold $c$ ranging [-2, 2] with 0.2 increments. The $y$-axis denotes the empirical FDR. The red line denotes the joint method, the blue line the separate upper (BH), the yellow the separate lower (BH), and the green the separate lower (adaptive). The red dotted line signifies the nominal FDR level. Note here that the ramp signal has the same plots for different signal smoothing levels as they are unaffected by signal smoothing.}
\label{FDR-sim}
\end{figure}

Figure \ref{FDR-sim} shows the empirical FDR change per different $c$ levels for each possible combination of smoothing levels and signals. The results for the entire simulation are presented in the supplementary materials. The threshold $c$ ranges from -2 to 2, increasing by 0.2 on the $x$-axis. The red line shows the empirical FDR change for the joint method, the blue line for the separate upper (BH), the yellow line for the separate lower (BH), and the green line for the separate lower (adaptive). For the ramp signal, the signal smoothing is non-existent, thus showing the same result for three different levels of signal smoothing.

The simulation results suggest that in general, the empirical FDR is effectively controlled under the 0.05 level across four methods and for all values of $c$. This is with an exception of the peak at $c=-1$ in circle signal with ${\rm FWHM}=5$ from the joint method. Although maintained below the 0.05 level,  similar behaviour is observed at $c=-1$ and 1 for the step signal and $c=-1$ for the circle signal from the joint method. This is due to the fact that PRDS breaks down near the flat regions in the signals $c=1$ and  $c=-1$ when using the joint method. See discussion for further details.

The asymmetry between the peaks at $c=-1$ and $c=1$ in circle signal joint method simulation is explained by the asymmetry in the number of pixels included in the area below 1 (the background) and above -1 (the circle). More voxels are rejected when testing for the lower side $H_A^L(v):\mu(v) < c$ with $c=1$ than when testing for the upper side $H_A^U(v):\mu(v) > c$ with $c=-1$.  This leads to decreased denominator for the joint FDR at $c=-1$, while the other numbers involved in the calculation of joint FDR remain relatively insignificant, ultimately resulting in higher FDR at $c=-1$.

As predicted, the lower side adaptive method shows less conservative control of the FDR than the lower BH method in the circle signal images. This is in accordance with Blanchard and Roquain \cite{blanchard2009adaptive} which states that the proposed two-stage method works better when more tests are expected to be rejected which is equivalent to when there are more voxels below $c$ in the context of the negative one-sided test. This also explains why the lower side adaptive shows good control of the FDR when threshold is higher.

\subsection{FNDR Simulation Results}
\newcommand{\simFNDRoo}{\raisebox{1em}{\adjustbox{valign=m, max width=\dimexpr0.33\textwidth-2\tabcolsep}{\includegraphics{figs/sim/sim_legend.png}}}}
\newcommand{\simFNDRot}{\includegraphics[width=0.23\textwidth]{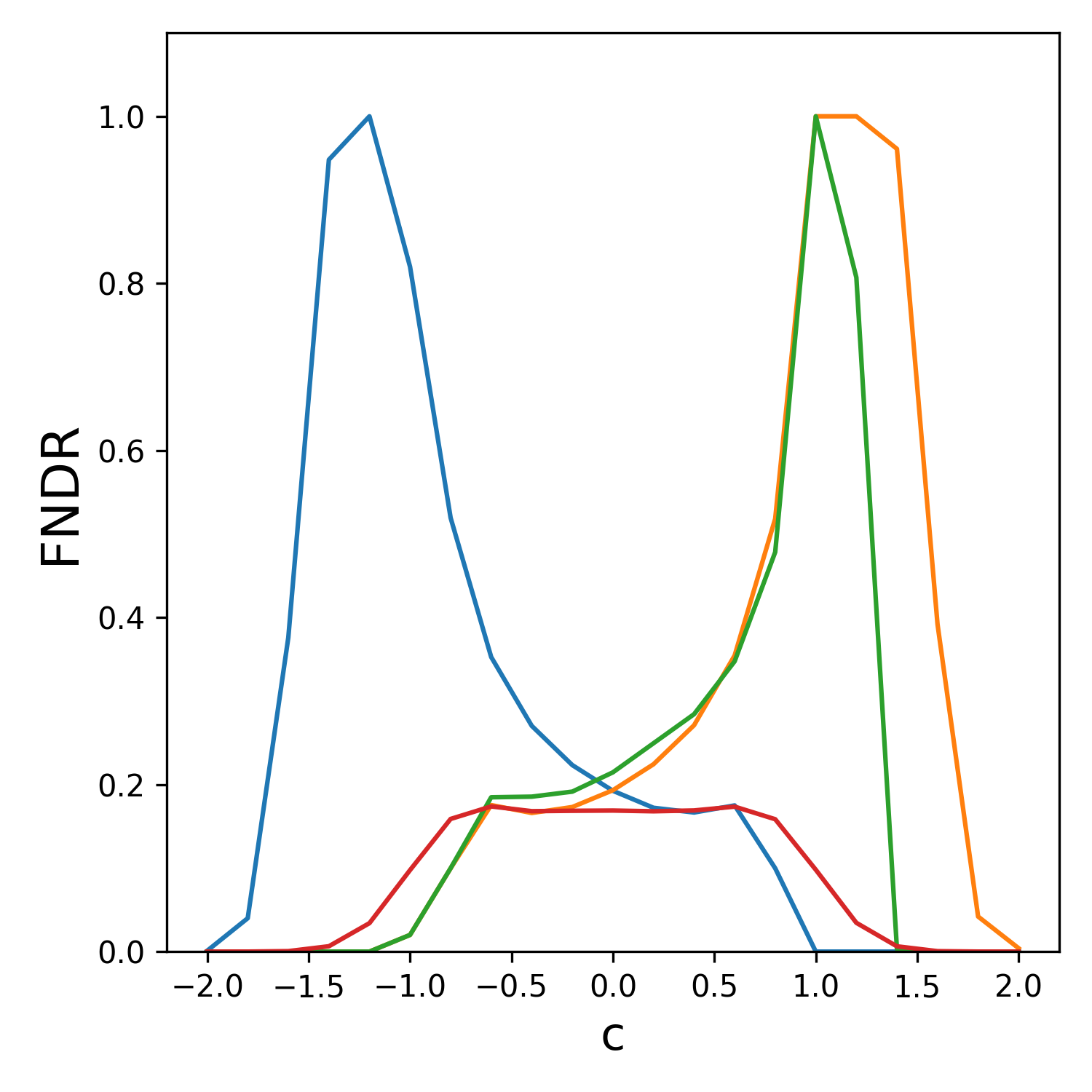}}
\newcommand{\simFNDRoth}{\includegraphics[width=0.23\textwidth]{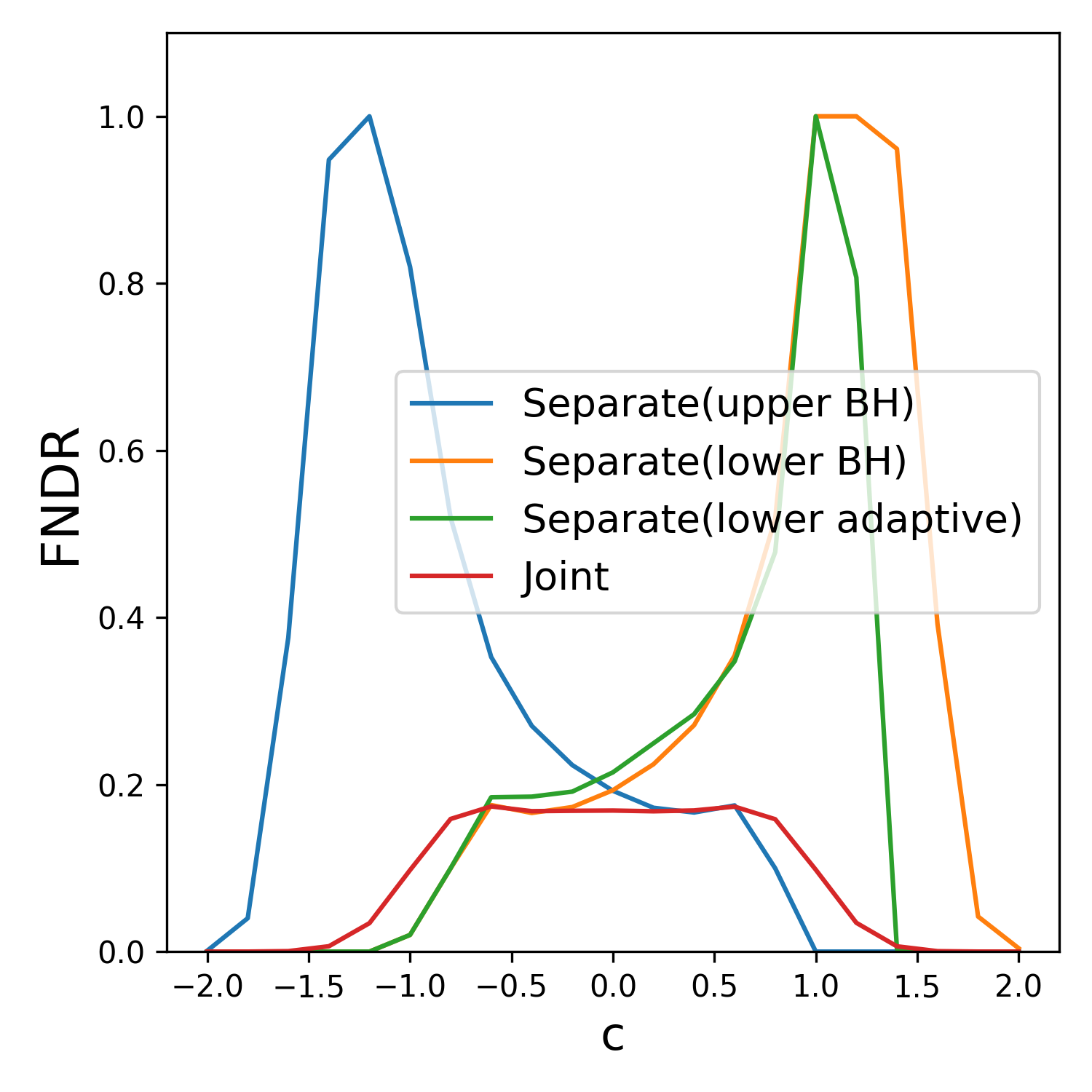}}
\newcommand{\simFNDRto}{\includegraphics[width=0.23\textwidth]{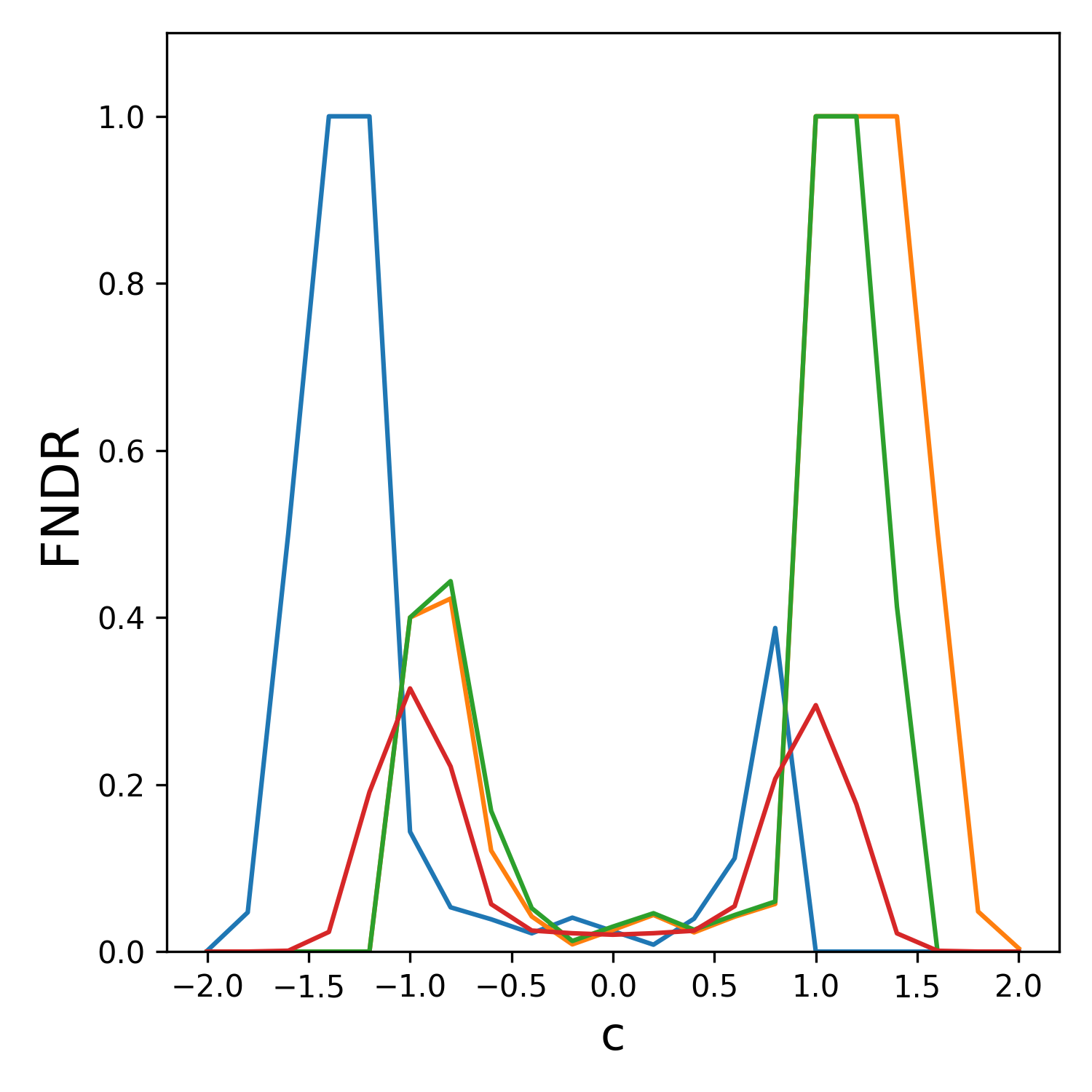}}
\newcommand{\simFNDRtt}{\includegraphics[width=0.23\textwidth]{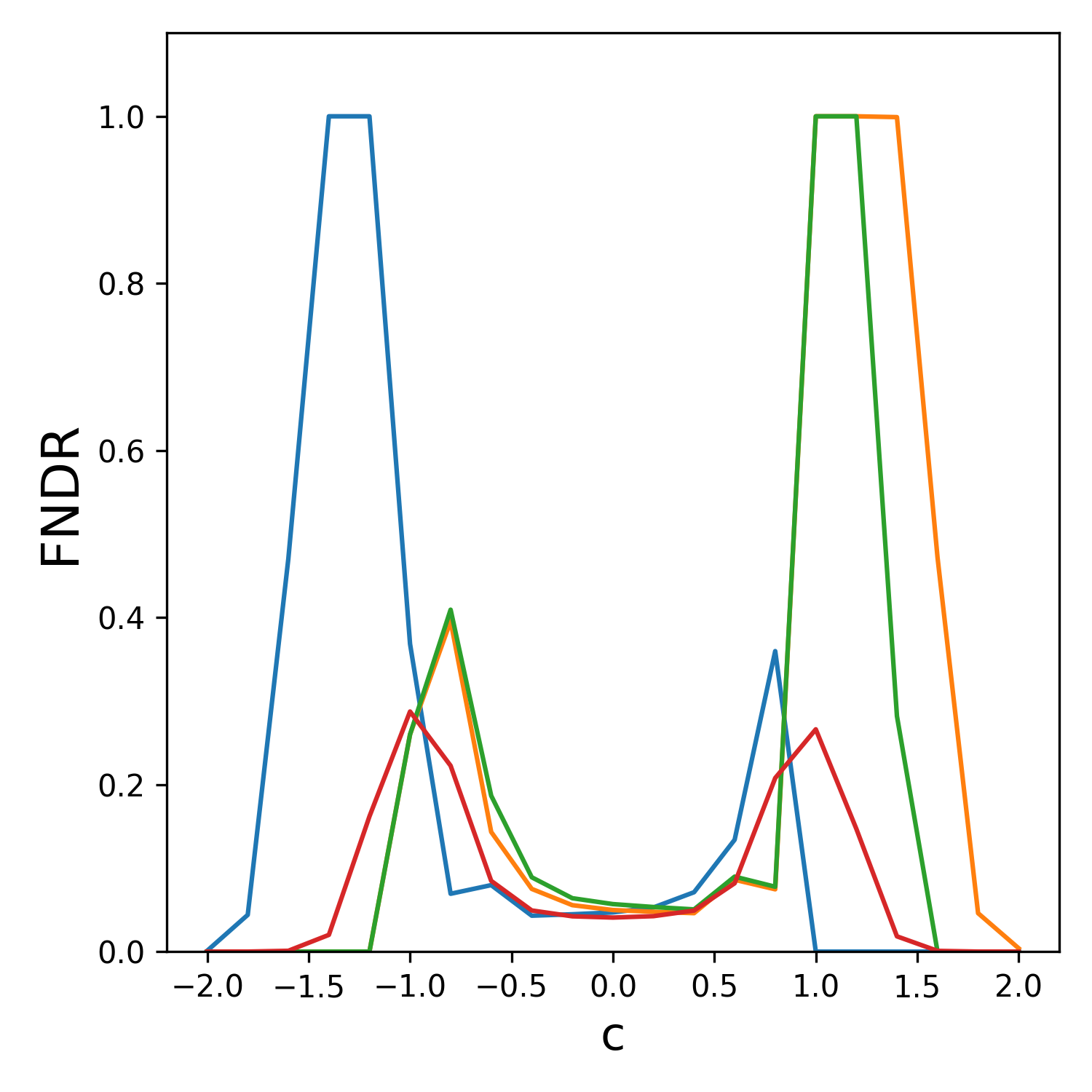}}
\newcommand{\simFNDRtth}{\includegraphics[width=0.23\textwidth]{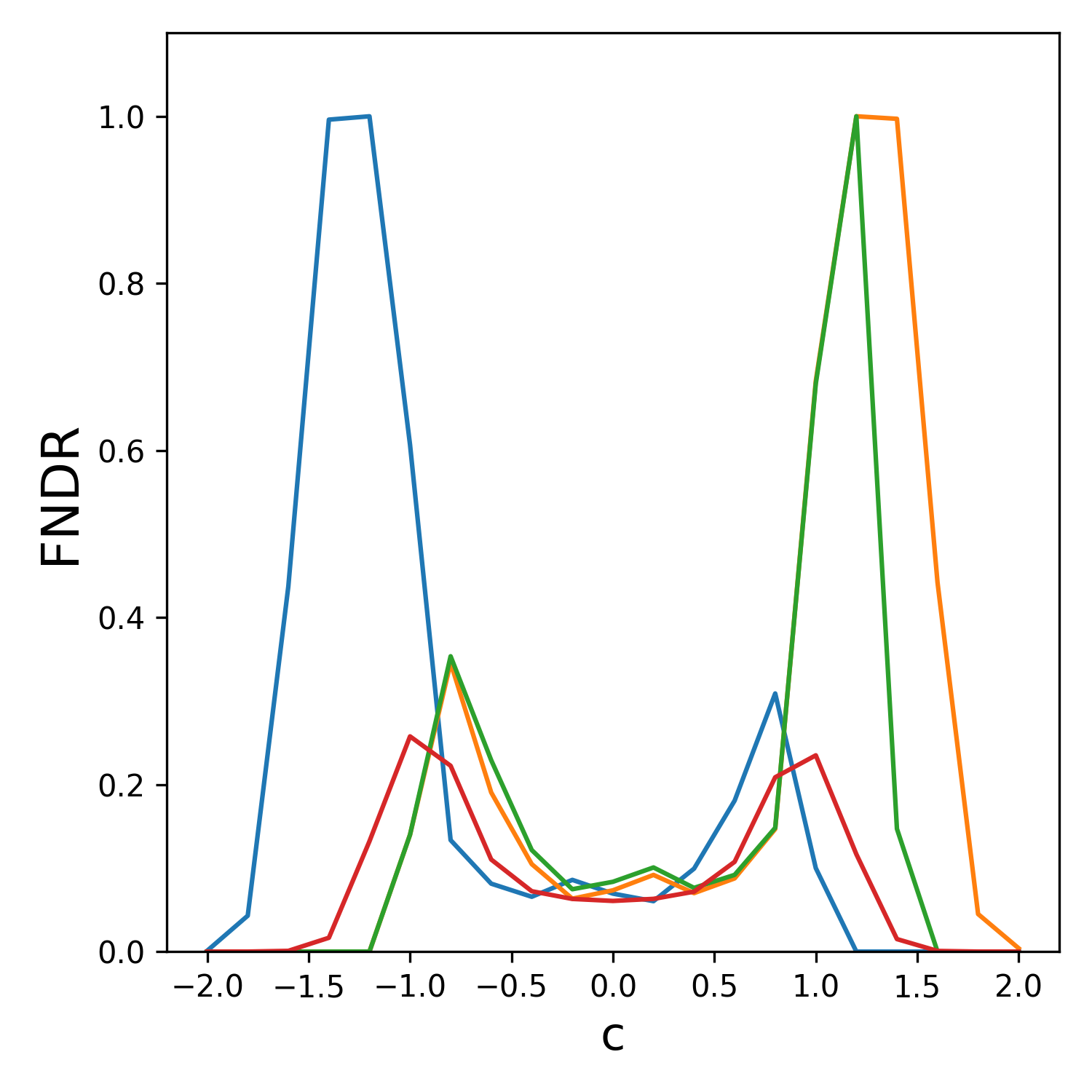}}
\newcommand{\simFNDRtho}{\includegraphics[width=0.23\textwidth]{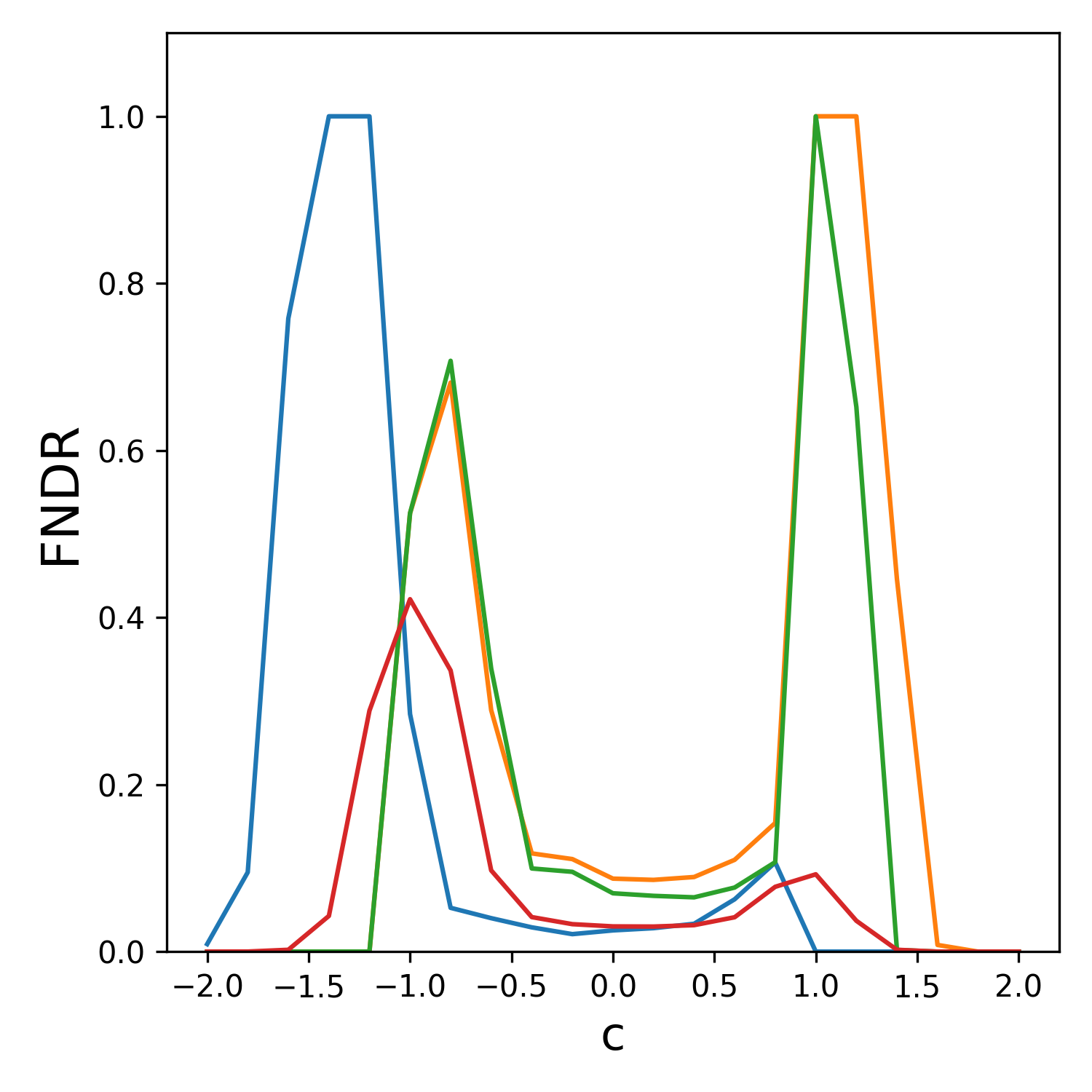}}
\newcommand{\simFNDRtht}{\includegraphics[width=0.23\textwidth]{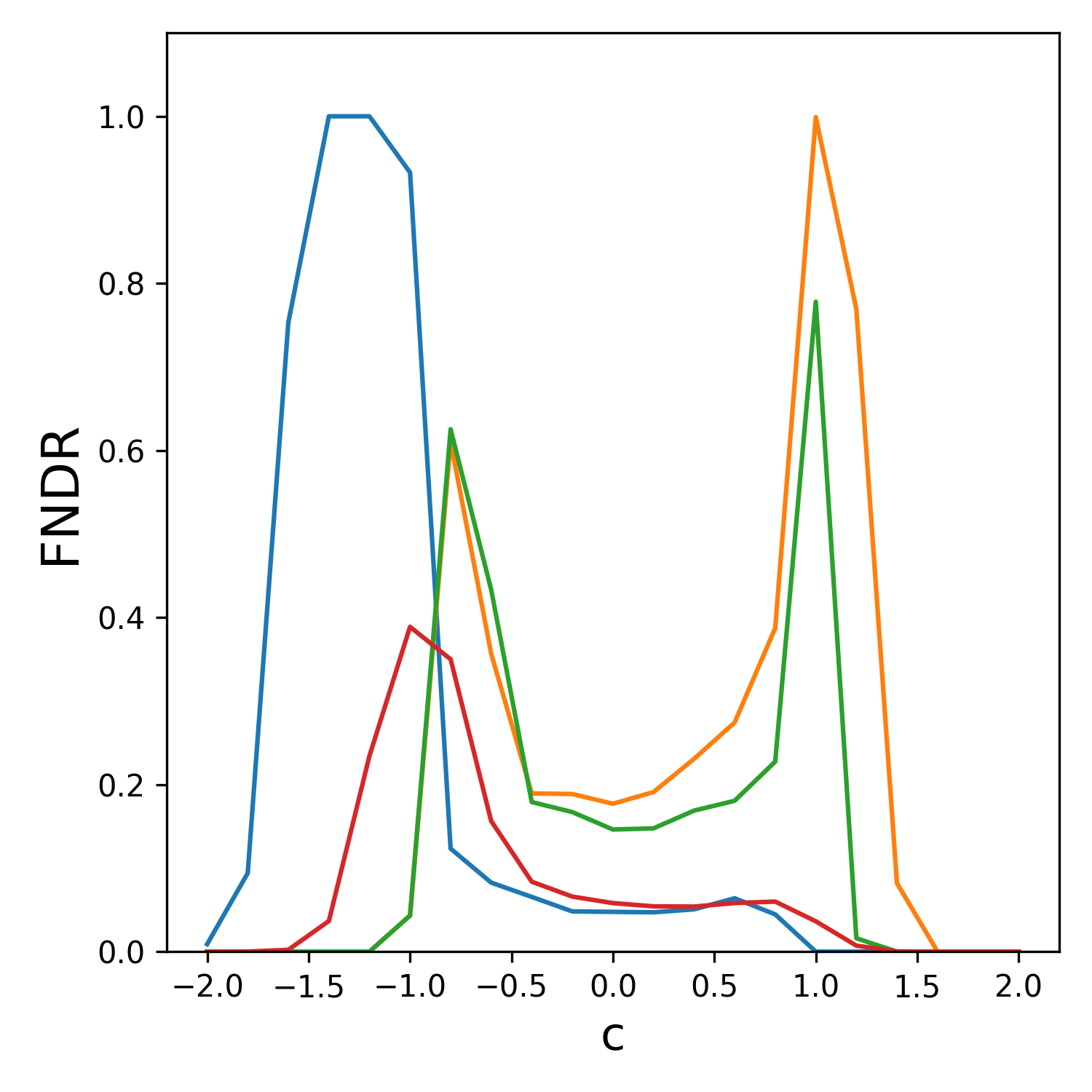}}
\newcommand{\simFNDRthth}{\includegraphics[width=0.23\textwidth]{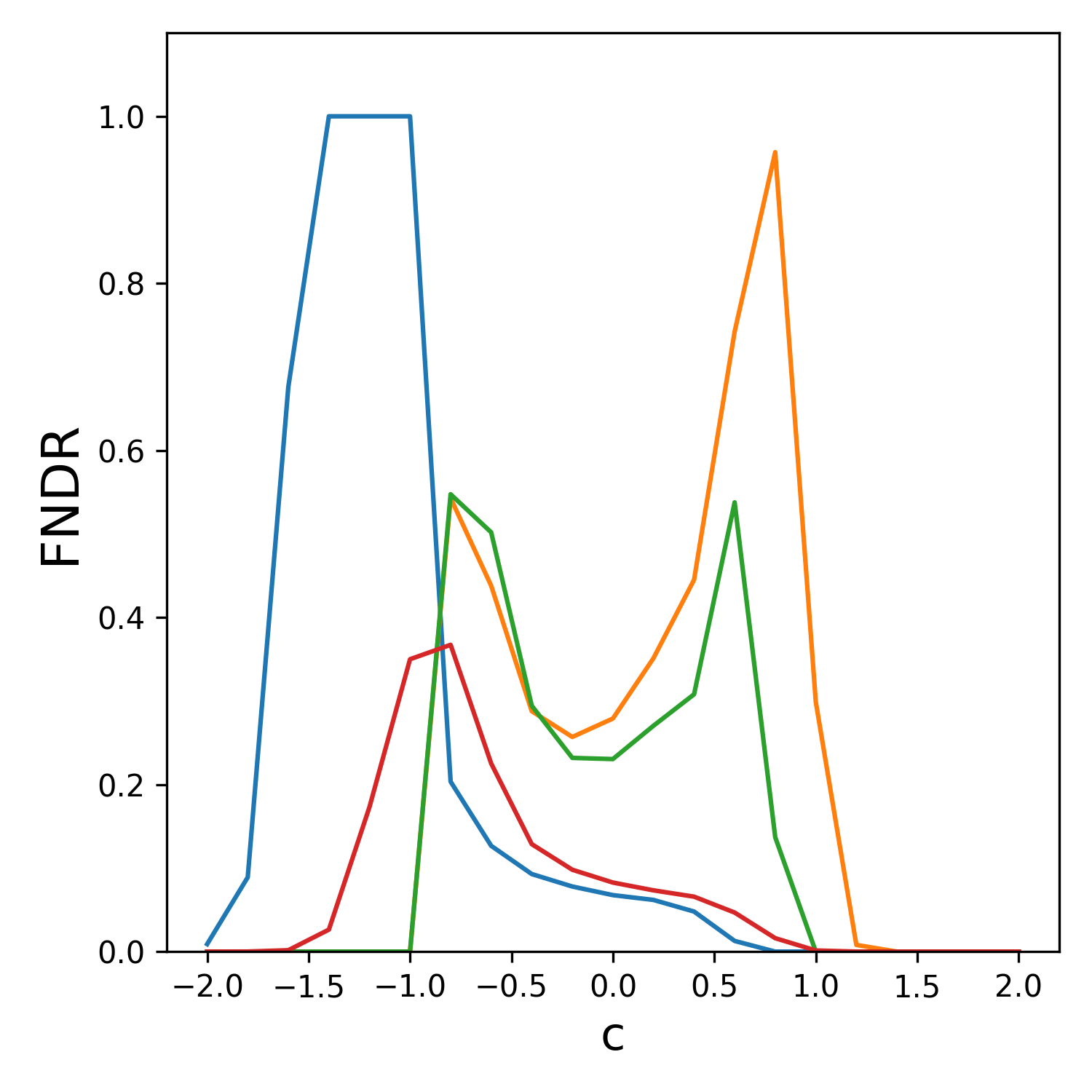}}

\newcolumntype{C}{>{\centering\arraybackslash}m{9em}}
\begin{figure}[H]\sffamily
\centering
\setlength{\tabcolsep}{0pt}
\begin{tabular}{CCCC}
\toprule
Signal & $\text{FWHM}=5$ & $\text{FWHM}=10$ & $\text{FWHM}=15$  \\ 
\midrule
 \simFDRramp & \multicolumn{2}{c}{\simFNDRoo} & \simFNDRot  \\ 
 \simFDRstep & \simFNDRto & \simFNDRtt & \simFNDRtth \\ 
 \simFDRcircle & \simFNDRtho & \simFNDRtht & \simFNDRthth \\ 
\bottomrule 
\end{tabular}
\caption{FNDR simulation result for three signals (ramp, step and circle top to bottom) and three signal smoothing levels with noise smoothing at ${\rm FWHM}=5$. The $x$-axis denotes threshold $c$ ranging [-2, 2] with 0.2 increments. The $y$-axis denotes the empirical FNDR. The red line denotes the joint method, the blue line the separate upper (BH), the yellow the separate lower (BH), and the green the separate lower (adaptive). Note here that the ramp signal has the same plots for different signal smoothing levels as they are unaffected by signal smoothing.}
\label{FNDR-sim}
\end{figure}

The simulation results for the empirical FNDR are presented in Figure \ref{FNDR-sim} with the same simulation settings as the FDR. 

Overall, the joint method shows consistently lower empirical FNDR compared to other methods. High peaks are displayed around the boundary values of $c \in (-2, -1)$ and $c \in (1, 2)$ in signals for separate methods. This is because once the threshold is set too extreme, i.e. below -1 for the positive test or above 1 for the negative test, essentially all the voxels are supposed to be rejected with $\mathcal{A}_c$ being $\mathcal{V}$. Therefore, any non-rejected voxels due to noise are all false non-rejections. Since the noise is distributed $N(0,1)$, if the threshold is set too high or low, i.e. past -2 or 2, rarely are voxels falsely not-rejected due to the effect of very big or very small noise.

Much like the FDR simulation, the asymmetric pattern in the circle signal for joint method stems from the asymmetry in the number of voxels in the circle signal and the background. Otherwise, the pattern of empirical FNDR demonstrated by the step and circle signals are similar.

Generally, lower BH shows higher FNDR than lower adaptive as the threshold gets higher, signifying that the two-stage adaptive procedure exhibits a higher power than the BH procedure when more tests are expected to be rejected. As can be seen from Figure \ref{FNDR-sim}, signal smoothing does not bring as much improvement in power detection with the empirical FNDR plots, showing similar range of empirical FNDR values across different signal smoothing levels.

\section{Human Connectome Project Data Application\label{dataapplication}}

\subsection{The Human Connectome Project Data Description}
The Human Connectome Project (HCP) is a five-year study which aims to map structural and functional human brain connectivity in young adults through multiple neuroimaging modalities \cite{van2012human}. In this section, we detail results of a real-data analysis performed on the HCP dataset with 77 participants using the FDCR methodologies proposed in this paper, as well as the method from SSS which refers to the bootstrap-based confidence region construction algorithm following Bowring (2019) \cite{BOWRING2019116187} and SSS \cite{SSS}. The FDCRs are built on the 77 subject-level contrast images from N-back working memory task comparing the 0-back vs. 2-back memory task. In the N-back working memory task, the participants are shown a sequence of pictures of objects and asked to remember them either immediately or 2 turns later. The image pre-processing details are presented in \cite{glasser2013minimal, barch2013function}.

\subsection{False Discovery Controlled Regions Application Results}
The $95\%$ FDCRs are constructed on fMRI scans from 77 subjects as a real data application of the proposed methods after applying additional smoothing with Gaussian kernel with FWHM $= 2.25$ to match the results shown in Bowring (2019) \cite{BOWRING2019116187}. FDCRs using 1) the joint method with $\alpha=0.1$, 2) the separate method with BH adjustment for upper and lower side each with $\alpha=0.05$, 3) the separate method with BH adjustment for upper side and two-stage adaptive procedure for lower with $\alpha=0.05$, and 4) SSS ($\alpha=0.05$) were compared with threshold level 1.0\%, 1.5\%, and 2.0\% Blood Oxygenation Level Dependent (BOLD) change. Joint control FDCRs are produced with $\alpha = 0.1$ instead of 0.05 for the reasons mentioned in Section 3.

\newcolumntype{C}{>{\centering\arraybackslash}m{8em}}
\begin{figure}
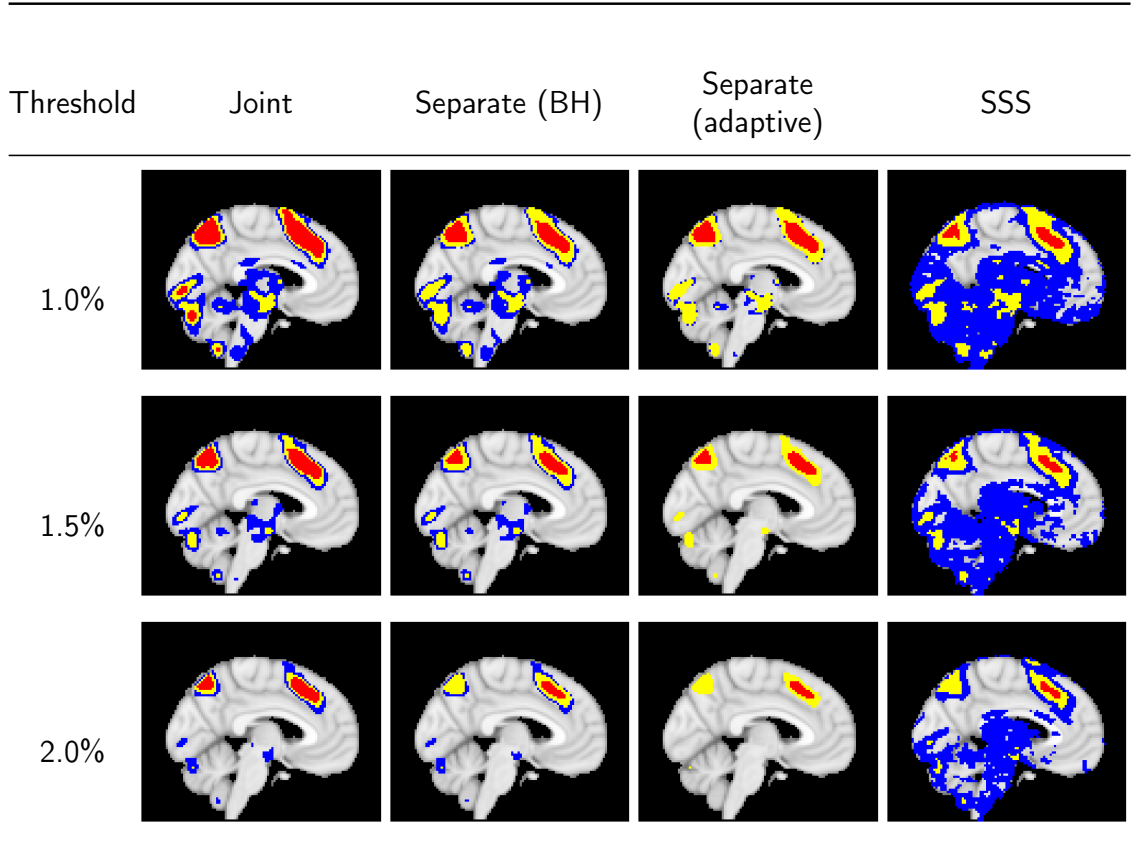
\sffamily
\setlength{\tabcolsep}{0pt}
\begin{tabular}{c*4{C}@{}}
\toprule
Threshold & Joint & Separate (BH) & Separate (adaptive) & SSS \\ 
\midrule
1.0\% & \dsrc & \dsrcc & \dsrccc & \dsrcccc \\ 
1.5\% & \dsrrc & \dsrrcc & \dsrrccc & \dsrrcccc \\ 
2.0\% & \dsrrrc & \dsrrrcc & \dsrrrccc & \dsrrrcccc \\ 
\bottomrule 
\end{tabular}

\caption{FDCR application to the HCP data with 77 subjects in the sagittal slice view (X=-4, MNI space). The red area denotes the upper FDCR $\hat{\mathcal{A}}_c^+$, the yellow area including the red area denotes the excursion set $\hat{\mathcal{A}}_c$, and the blue area including the yellow area denotes the lower FDCR $\hat{\mathcal{A}}_c^-$. The rows differ in the threshold level $c$ by which the FDCRs are constructed. The columns denote different methods. Overall, FDR controlling hypothesis testing methods which are presented in the first three columns, show tighter spatial inference result when compared to SSS, thus providing a higher degree of localization and interpretability to the identified regions. The FDCRs show activation in areas such as the precuneus cortex, paracingulate gyrus, middle frontal gyrus, and angular gyrus which matches the expectation from the literature as areas involved in working memory.}
\label{HCP-sagittal}
\end{figure}

\newcolumntype{C}{>{\centering\arraybackslash}m{8em}}
\begin{figure}\sffamily
\setlength{\tabcolsep}{0pt}
\begin{tabular}{c*4{C}@{}}
\toprule
Threshold & Joint & Separate (BH) & Separate (adaptive) & SSS \\ 
\midrule
1.0\% & \dcrc & \dcrcc & \dcrccc & \dcrcccc \\ 
1.5\% & \dcrrc & \dcrrcc & \dcrrccc & \dcrrcccc \\ 
2.0\% & \dcrrrc & \dcrrrcc & \dcrrrccc & \dcrrrcccc \\ 
\bottomrule 
\end{tabular}

\caption{FDCR application to the HCP data with 77 subjects in the coronal slice view (y=16, MNI space). The red area denotes the upper FDCR $\hat{\mathcal{A}}_c^+$, the yellow area including the red area denotes the excursion set $\hat{\mathcal{A}}_c$, and the blue area including the yellow area denotes the lower FDCR $\hat{\mathcal{A}}_c^-$. The rows differ in the threshold level $c$ by which the FDCRs are constructed. The columns denote different methods. Overall, FDR controlling hypothesis testing methods which are presented in the first three columns, show tighter spatial inference result when compared to SSS, thus providing a higher degree of localization and interpretability to the identified regions. The FDCRs show activation in areas such as the precuneus cortex, paracingulate gyrus, middle frontal gyrus, and angular gyrus which matches the expectation from the literature as areas involved in working memory.}
\label{HCP-coronal}
\end{figure}

\newcolumntype{C}{>{\centering\arraybackslash}m{8em}}
\begin{figure}\sffamily
\setlength{\tabcolsep}{0pt}
\begin{tabular}{c*4{C}@{}}
\toprule
Threshold & Joint & Separate (BH) & Separate (adaptive) & SSS \\ 
\midrule
1.0\% & \darc & \darcc & \darccc & \darcccc \\ 
1.5\% & \darrc & \darrcc & \darrccc & \darrcccc \\ 
2.0\% & \darrrc & \darrrcc & \darrrccc & \darrrcccc \\ 
\bottomrule 
\end{tabular}

\caption{FDCR application to the HCP data with 77 subjects in the axial slice view (Z=48, MNI space). The red area denotes the upper FDCR $\hat{\mathcal{A}}_c^+$, the yellow area including the red area denotes the excursion set $\hat{\mathcal{A}}_c$, and the blue area including the yellow area denotes the lower FDCR $\hat{\mathcal{A}}_c^-$. The rows differ in the threshold level $c$ by which the FDCRs are constructed. The columns denote different methods. Overall, FDR controlling hypothesis testing methods which are presented in the first three columns, show tighter spatial inference result when compared to SSS, thus providing a higher degree of localization and interpretability to the identified regions. The FDCRs show activation in areas such as the precuneus cortex, paracingulate gyrus, middle frontal gyrus, and angular gyrus which matches the expectation from the literature as areas involved in working memory.}
\label{HCP-axial}
\end{figure} 
\begin{figure}\sffamily
\setlength{\tabcolsep}{0pt}

\end{figure}

Figures \ref{HCP-sagittal}, \ref{HCP-coronal}, and \ref{HCP-axial} show the FDCR result for sagittal, coronal, and axial slice views respectively. For all slices, FDR controlling methods show tighter inference of both upper and lower regions compared to the SSS method. SSS shows smaller upper region and larger lower region which suggests more conservative inference compared to FDR controlling testing based methodologies. This is due to the fact that by controlling for FDR, the method allows for more false discoveries in exchange for more discoveries in general. The joint FDCRs are run at a nominal level of $0.1$, which as described in Section \ref{joint} corresponds to joint FDR control at the $0.05$ level, and the regions they produce are comparable in extent to those of the separate methods. Naturally, as the threshold $c$ goes up the area enclosed between the upper and lower FDCRs decreases, since fewer locations have signal exceeding the higher level and the excursion set being estimated is correspondingly smaller.

FDCRs with separate controls of FDR for lower and upper are presented in two forms for comparison: one with BH procedure for the lower FDCR, and the other one with the two-stage adaptive procedure for the lower FDCR. The upper FDCR remains the same as both methods uses BH procedure for the upper set FDR control. Lower FDCRs with adaptive method are smaller than lower sets with BH procedure which is to be expected as the two-stage adaptive procedure is less conservative when more voxels are thought to be rejected. In the context of negative one-sided testing, this is equivalent to when there are fewer voxels above $c$ than below $c$.

\subsection{Realistic Noise Simulation Results}
To investigate how our FDCRs control the FDR and FNDR with various sample sizes, a
realistic noise simulation was conducted. Following \cite{eklund2016cluster}, this
regresses fake task designs on resting state data in order to generate noise with the
spatial and distributional structure of real fMRI data. This procedure, which has come to be known as the Eklund test,
represents a gold standard for method validation in neuroimaging, and has been used to
establish the false positive rates of a number of procedures
\cite{lohmann2018, andreella2023, davenport2024surf}. As the noise is taken from real resting state
scans, it retains the spatial correlation, non-stationarity and non-Gaussianity of
fMRI data, none of which is captured by parametric simulation. We take this approach, using 3,500 noise fields created in the same setting as
\cite{davenport2024surf}, which were obtained by applying this procedure to resting
state data from the UK Biobank. There the first level design was a block
design consisting of randomly phased sequences of blocks of 20 time points, convolved with a
haemodynamic response function, with the resulting first level contrast maps
transformed to 2 mm MNI space. Applied to resting state data this provides very realistic null data \cite{eklund2016cluster}.

To obtain a setting in which the ground truth is known but the noise remains
realistic, these 3,500 noise fields, of dimension (91, 109, 91), were masked and added to the
sample mean of the HCP data (mean over the field $=1.03$, range $=[-43.9, 61.9]$).
This sample mean is then treated as the ground truth population mean against which the
FDR and FNDR are calculated, so that the excursion set $\mathcal{A}_c$ is known
exactly for every level $c$.

\begin{figure}[H]
    \centering
    \includegraphics[width=1\linewidth]{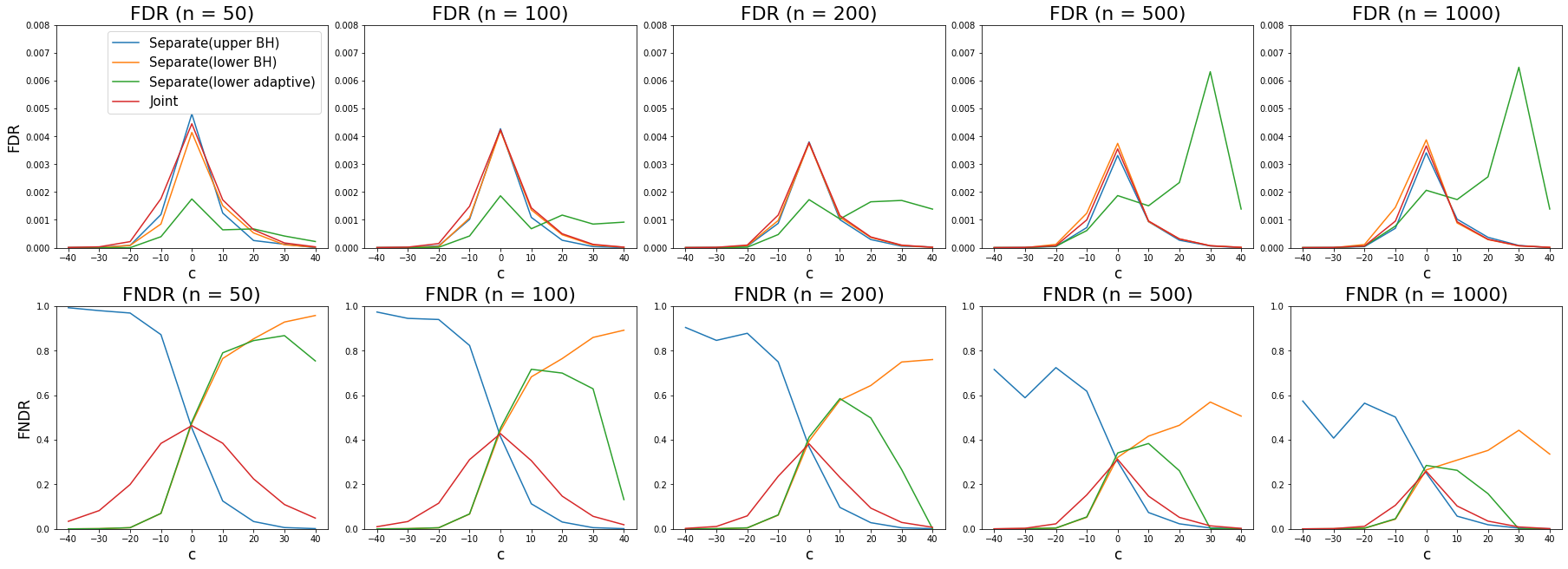}
    \caption{Realistic noise simulation results showing the empirical FDR (top) and FNDR (bottom) for samples sizes ranging from 100 to 2000.}
    \label{rn_sim}
\end{figure}

Given sample size $n \in \{50$, 100, 200, 500, $1000\}$, and threshold $c \in \{-40,$ $-$30, $-$20, $-$10, 0, 10, 20, 30, $40\}$, for each iteration of the simulation, $n$ number of images are sampled from the 3,500 synthetic images, and the FDR/FNDR are calculated for the threshold $c$. Simulation is repeated 500 times per given setting, and the averaged empirical FDR/FNDR is calculated and plotted in Figure \ref{rn_sim}.

Overall, all methods show FDR control well below the nominal 0.05 level. Separate lower adaptive shows lower empirical FNDR compared to separate lower BH which is consistent with the manuscript simulation results. Increase FNDR is observed at the extreme $c$ values for separate methods which is again consistent with the manuscript simulation results. As the sample size increases, empirical FNDR decreases below the 0.45 level, while the empirical FDR remain consistent across sample size increase.

\section{Conclusion and Discussion\label{conclusion}}
This paper developed upper and lower FDCRs for the excursion set $\mathcal{A}_c$, with spatial FDR control, based on directional hypothesis testing. 
These regions were constructed for separate upper and lower control and joint control over both directions. In order to boost power a two-stage adaptive approach was incorporated for the separate control and a natural adaptation was used for the joint approach. In the latter, the BH procedure is run at the nominal level $2\alpha$: because exactly one of the two hypotheses at each location is null, only half of the $2m$ joint hypotheses can be true nulls, and the resulting FDR is therefore controlled at the desired level $\alpha$. Error rate simulations across a variety of scenarios showed that the empirical FDR was almost always controlled to the nominal level $\alpha = 0.05$. Real data applications using fMRI data from the HCP showed that the FDCRs bounds were informative and particularly were more spatially precise than CRs obtained by existing methods \cite{BOWRING2019116187, SSS}.

For the joint FDCRs, we use two one-sided tests rather than naively using a two-sided testing approach in which the hypothesis being tested is of the form \[ H_0(v): \mu(v) = c \quad \mbox{vs.} \quad H_A(v): \mu(v) \neq c.\] At first glance naively testing against this null hypothesis and applying the BH might seem like a simpler approach for creating joint FDCRs. However, rejecting this null hypothesis does not provide directional error control, as the direction of the effect is not specified. Instead, testing against two one-sided null hypotheses (\ref{hypothesis-joint1} and \ref{hypothesis-joint2}) in the positive and the negative directions, as we do in our joint construction, allows us to provide directional inference.



For the FDCRs with separate error control, the FDR is controlled separately for the positive and negative directions. The joint FDCRs instead control the FDR simultaneously by considering both sets of directional nulls. Joint error control is theoretically more desirable than separate error control and the right choice of which to use may depend on the questions that users are trying to answer.

The two approaches are complementary and which to use depends on the error rate the analyst wishes to control. The separate control method is more informative, because it admits possibly different error rates in the upper and lower FDCRs. For example, when estimating a positive excursion set, the upper FDCR gives an indication of the minimal extent of the excursion set, whereas the lower FDCR gives an indication of the maximal extent of the excursion set, and controlling the error rate in one may be more important than in the other. The joint method is more concise in that it controls the FDR over the entire image, but it does not distinguish whether the errors occur in the upper FDCR or the lower FDCR. If the question of interest is one-directional in nature, using the separate method provides more informative FDCRs in terms of interpretability and error control.

In our simulations we observed that when the underlying signals in the data are thought to be relatively evenly distributed below and above the threshold $c$, the joint method usually provide less conservative FDCRs under the nominal $\alpha$ level (see e.g. the ramp and step signal simulation in Figure \ref{FDR-sim} first and second rows). Meanwhile, with the presence of signals with unbalanced distribution of areas below and above $c$, the lower FDCR with two-stage adaptive procedure provides better results (as signified by the circle signal simulation; c.f. Figure \ref{FDR-sim} third row). Most fMRI data would fall under this latter case.

A further observation of note is the presence of the peaks in the empirical FDR plots at $c= -1, 1$ for the circle and step signals in Figure \ref{FDR-sim} . This can be attributed to the fact that PRDS, which is a condition on the dependence structure of the $p$-values that is sufficient for the validity of the BH procedure \cite{benjamini2001control, roquain2008two}, breaks down around $c=-1$ and 1 due to stronger negative dependency among the $p$-values for the lower and upper directions, near the boundary - see the discussion in Section S6.4 of the supplementary material. The peaks of the FDR in Figure \ref{FDR-sim} are reduced in height with the presence of stronger signal smoothing as the gradients of the signal become less sharp.

Compared to SSS, the FDR-controlling method we propose in this work provides less conservative, thus tighter inference results. Aside from the power gain, the FDR-controlling framework based on hypothesis testing allows error control in finite samples, without requiring asymptotic assumptions. 

As a closing note, we discuss potential extensions of this work. More immediate applications include the construction of the FDCRs for Cohen's $d$ or $R^2$ in place of the mean values in the image \cite{BOWRING2021117477, algina1999comparison}. FDCRs obtained via test inversion extend easily to other dependence structures, since the multiple testing step is modular: the BH procedure used in Algorithms 1 to 3 may be replaced by any other FDR controlling procedure valid under the dependence at hand, such as those of \cite{benjamini2006adaptive, benjamini2001control, benjamini2000adaptive}. FDCRs could also be constructed in order to control other measures of error such as the false discovery proportion (FDP) \cite{rosenblatt2018all,davenport2022fdp,blain2022notip} or in the context of selective inference on the hierarchies of hypotheses, such as the whole brain, regions, and individual voxels \cite{benjamini2014selective, blanchard2020post, rosenblatt2018all}. The framework of \cite{benjamini2014selective} is particularly relevant in this regard, as it controls the FDR across families of hypotheses while accounting for the selection of the families themselves, which is precisely the situation that arises when regions of the brain are selected for further examination on the basis of the same data. The directional nature of the inference performed here also connects to the literature on directional error control discussed in the introduction, and in particular to \cite{winkler2024false}, who consider directional FDR control in neuroimaging.




\section{Funding}
This work was partially supported by the National Institutes of Health [R01MH128923 to H.R., A.S., and S.D., R01EB026859 to T.M.S., A.S., and S.D.].

\section{Acknowledgements}
Data were provided in part by the Human Connectome Project, WU-Minn Consortium (Principal Investigators: David Van Essen and Kamil Ugurbil; 1U54MH091657) funded by the 16 NIH Institutes and Centers that support the NIH Blueprint for Neuroscience Research; and by the McDonnell Center for Systems Neuroscience at Washington University.

The research was partly supported under UK BioBank application \#34077, with bulk image data shared within (with UK BioBank permission) from application 8107.

\section{Code Availability}
The code used for simulations and data analysis can be found in the \href{https://github.com/HowonRyu/ConfidenceSet}{github repository}.

\section{Conflict of Interest}
None declared.

\newpage

\printbibliography
\section{Supplementary Materials}

\subsection{Simulation Results}

\begin{figure}[H]
    \centering
    \includegraphics[scale=0.45]{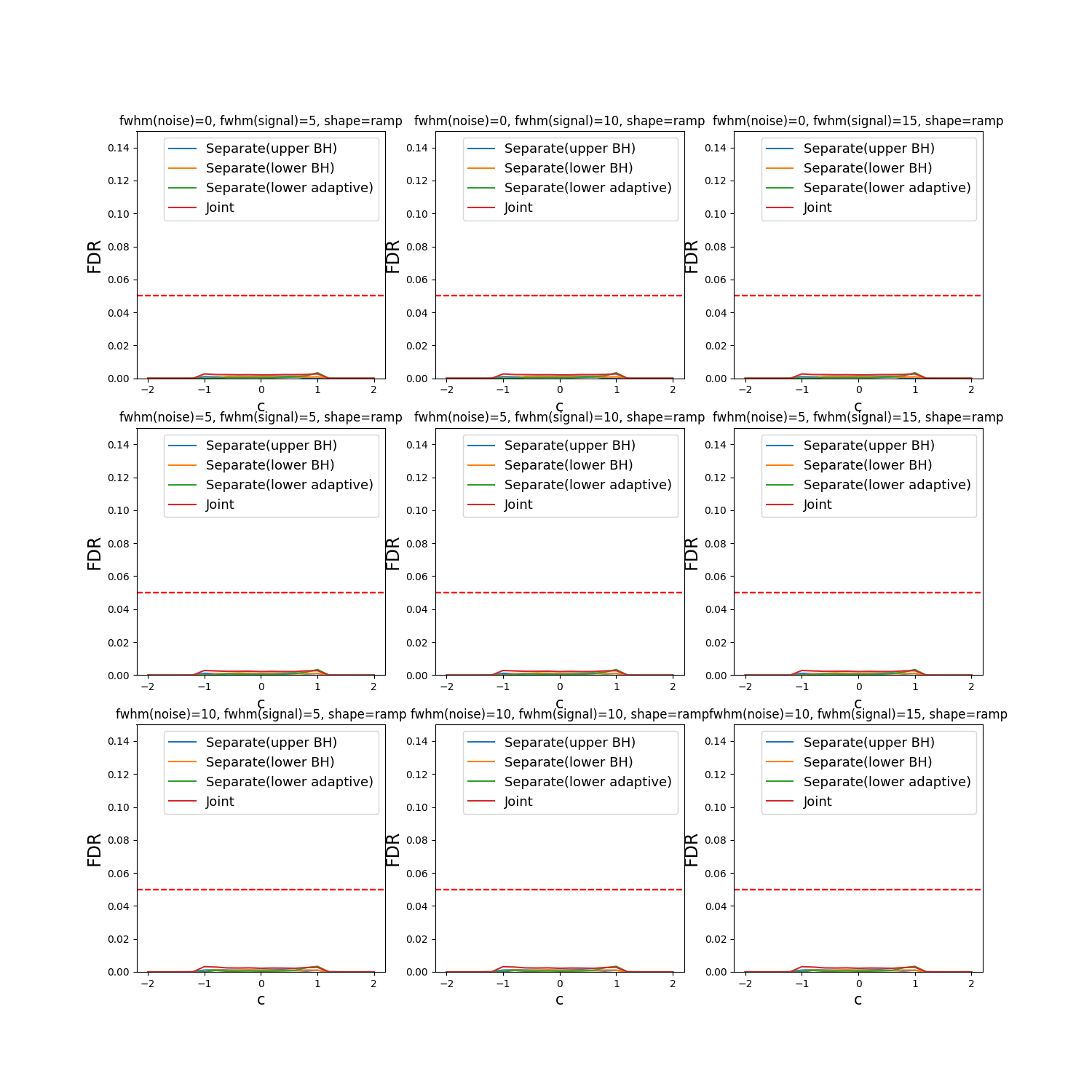}
    \label{supp-FDR_ramp}
    \caption{FDR simulation result for the ramp signal. The rows differ in noise smoothing (FWHM 0, 5, and 10) and the columns differ in signal smoothing levels (FWHM 5, 10, and 15). The $x$-axis denotes threshold c ranging [-2, 2] with 0.2 increments. The $y$-axis denotes the empirical FDR. The red line denotes the joint method, the blue line the separate upper (BH), the yellow the separate lower (BH), and the green the separate lower (adaptive). The red dotted line signifies the nominal FDR level.}
\end{figure}

\begin{figure}[H]
    \centering
    \includegraphics[scale=0.45]{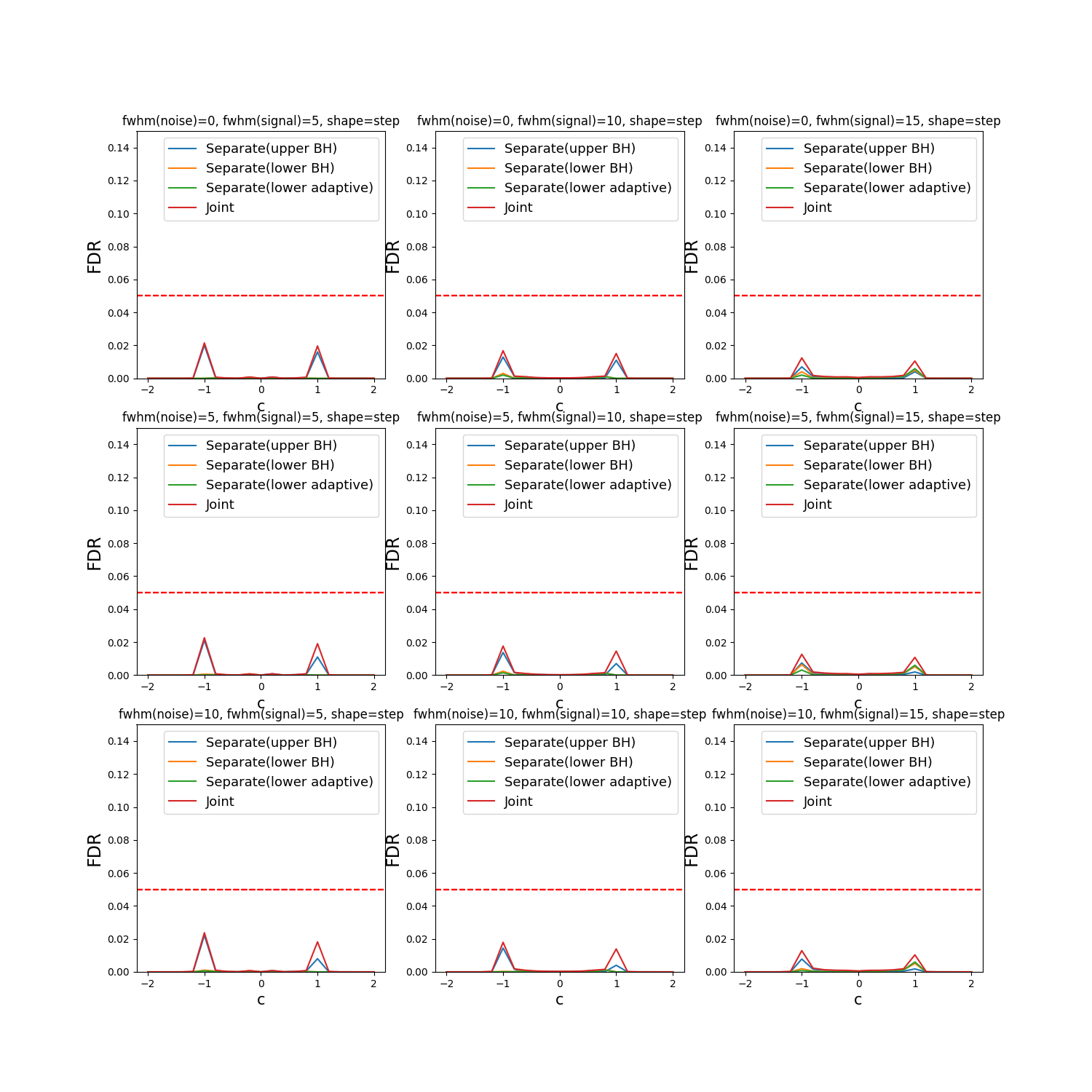}
    \label{supp-FDR_step}
        \caption{FDR simulation result for the step signal. The rows differ in noise smoothing (FWHM 0, 5, and 10) and the columns differ in signal smoothing levels (FWHM 5, 10, and 15). The $x$-axis denotes threshold c ranging [-2, 2] with 0.2 increments. The $y$-axis denotes the empirical FDR. The red line denotes the joint method, the blue line the separate upper (BH), the yellow the separate lower (BH), and the green the separate lower (adaptive). The red dotted line signifies the nominal FDR level.}
\end{figure}

\begin{figure}[H]
    \centering
    \includegraphics[scale=0.45]{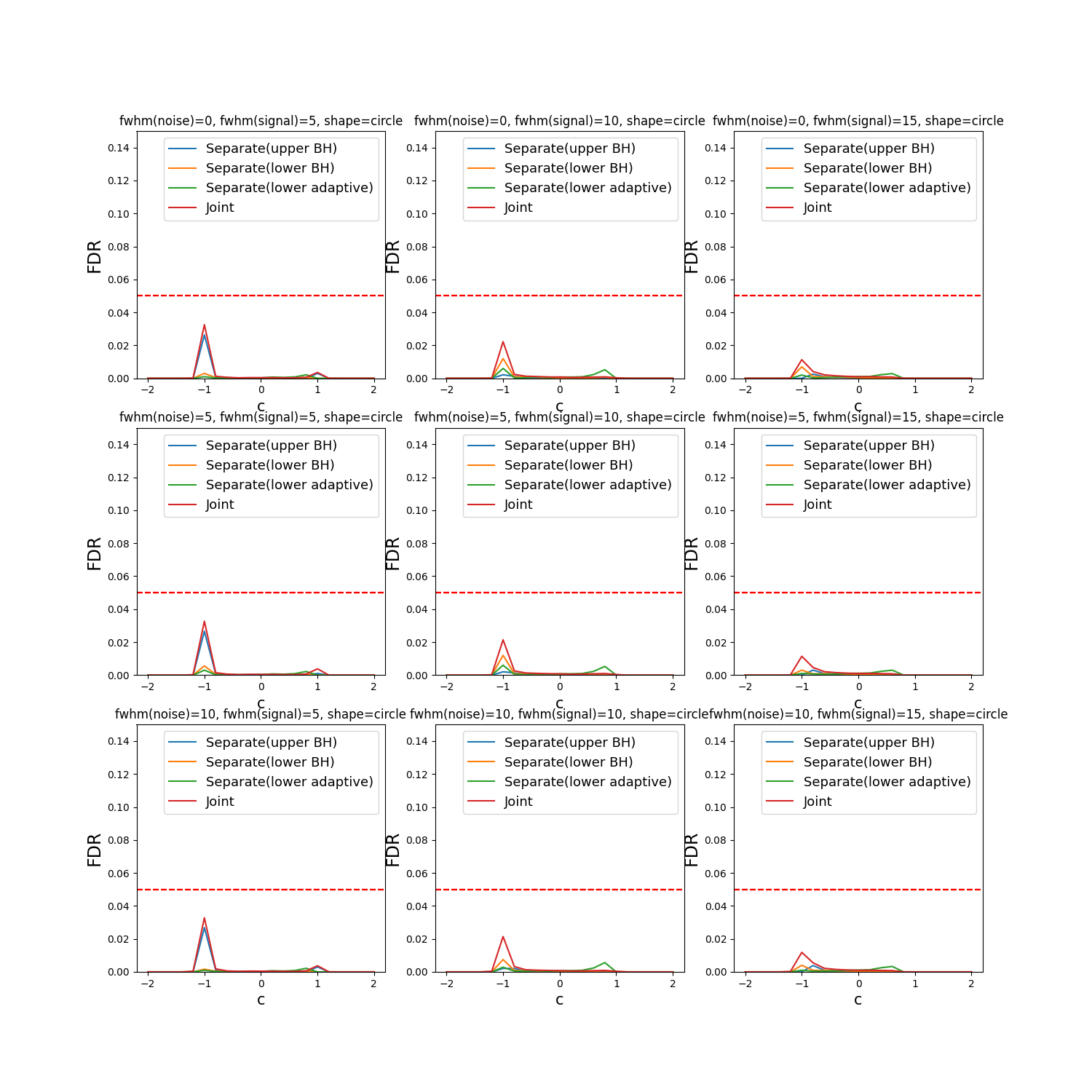}
    \label{supp-FDR_circle}
        \caption{FDR simulation result for the circle signal. The rows differ in noise smoothing (FWHM 0, 5, and 10) and the columns differ in signal smoothing levels (FWHM 5, 10, and 15). The $x$-axis denotes threshold c ranging [-2, 2] with 0.2 increments. The $y$-axis denotes the empirical FDR. The red line denotes the joint method, the blue line the separate upper (BH), the yellow the separate lower (BH), and the green the separate lower (adaptive). The red dotted line signifies the nominal FDR level.}
\end{figure}

\begin{figure}[H]
    \centering
    \includegraphics[scale=0.45]{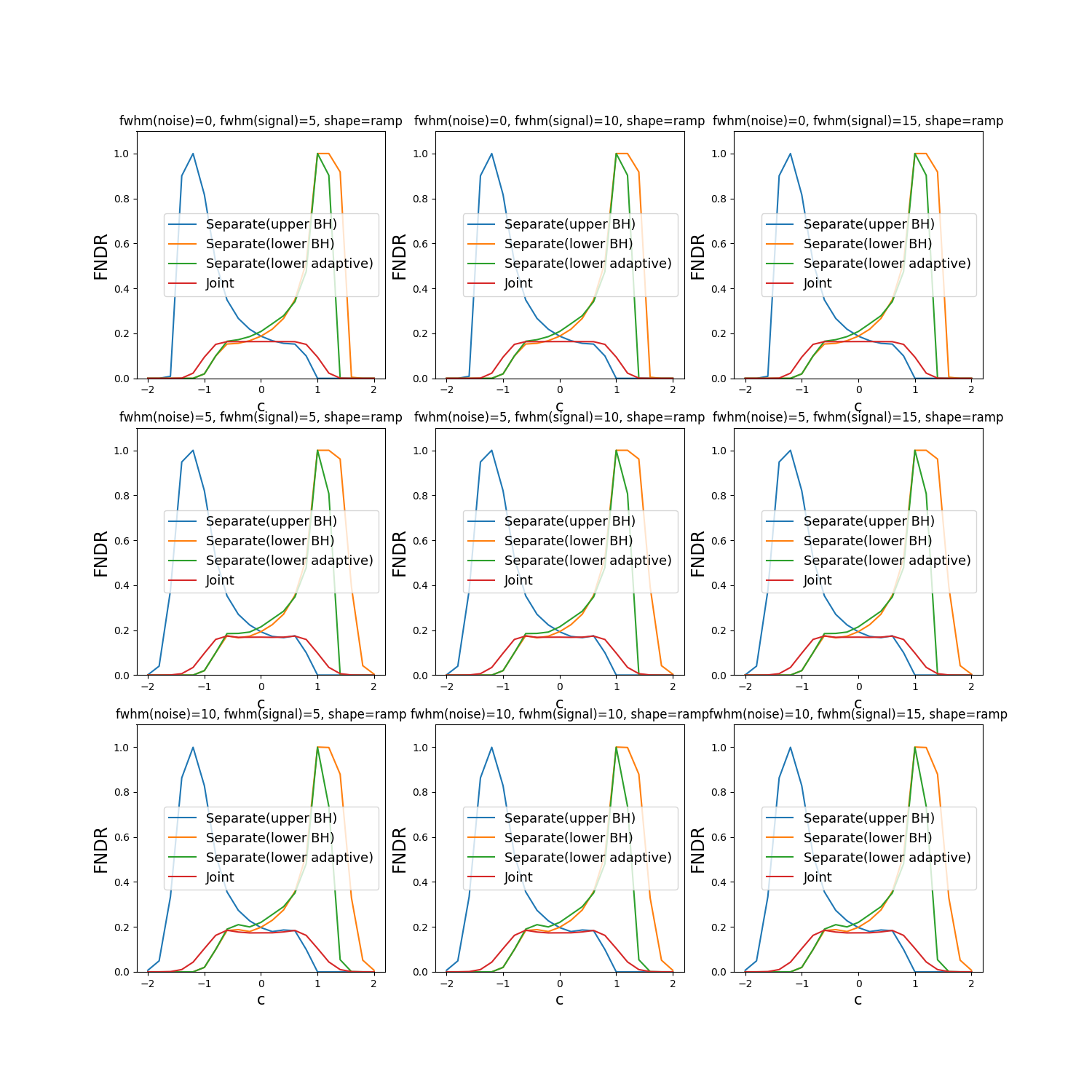}
    \label{supp-FNDR_ramp}
    \caption{FNDR simulation result for the ramp signal. The rows differ in noise smoothing (FWHM 0, 5, and 10) and the columns differ in signal smoothing levels (FWHM 5, 10, and 15). The $x$-axis denotes threshold c ranging [-2, 2] with 0.2 increments. The $y$-axis denotes the empirical FNDR. The red line denotes the joint method, the blue line the separate upper (BH), the yellow the separate lower (BH), and the green the separate lower (adaptive).}
\end{figure}

\begin{figure}[H]
    \centering
    \includegraphics[scale=0.45]{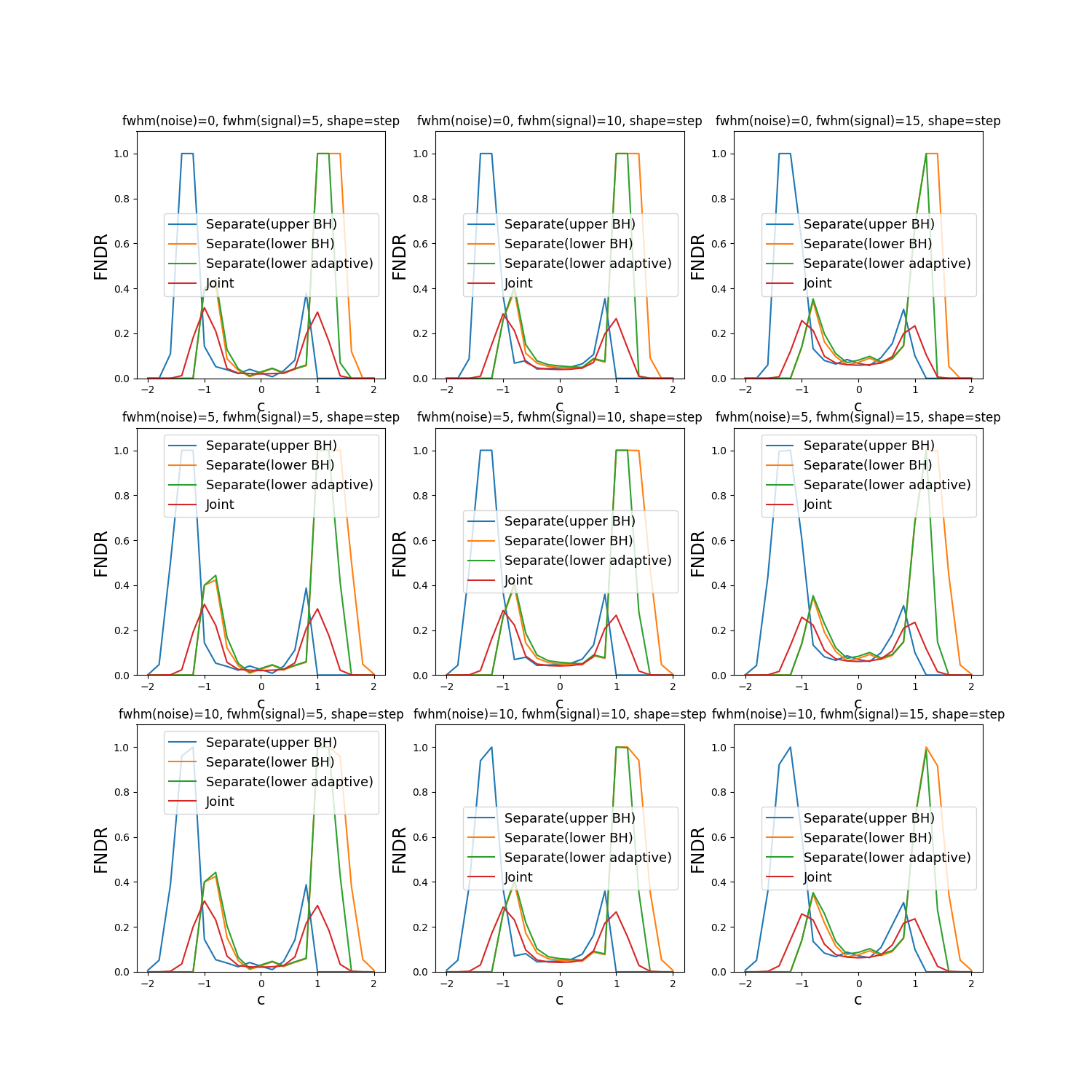}
    \label{supp-FNDR_step}
    \caption{FNDR simulation result for the step signal. The rows differ in noise smoothing (FWHM 0, 5, and 10) and the columns differ in signal smoothing levels (FWHM 5, 10, and 15). The $x$-axis denotes threshold c ranging [-2, 2] with 0.2 increments. The $y$-axis denotes the empirical FNDR. The red line denotes the joint method, the blue line the separate upper (BH), the yellow the separate lower (BH), and the green the separate lower (adaptive).}
\end{figure}

\begin{figure}[H]
    \centering
    \includegraphics[scale=0.45]{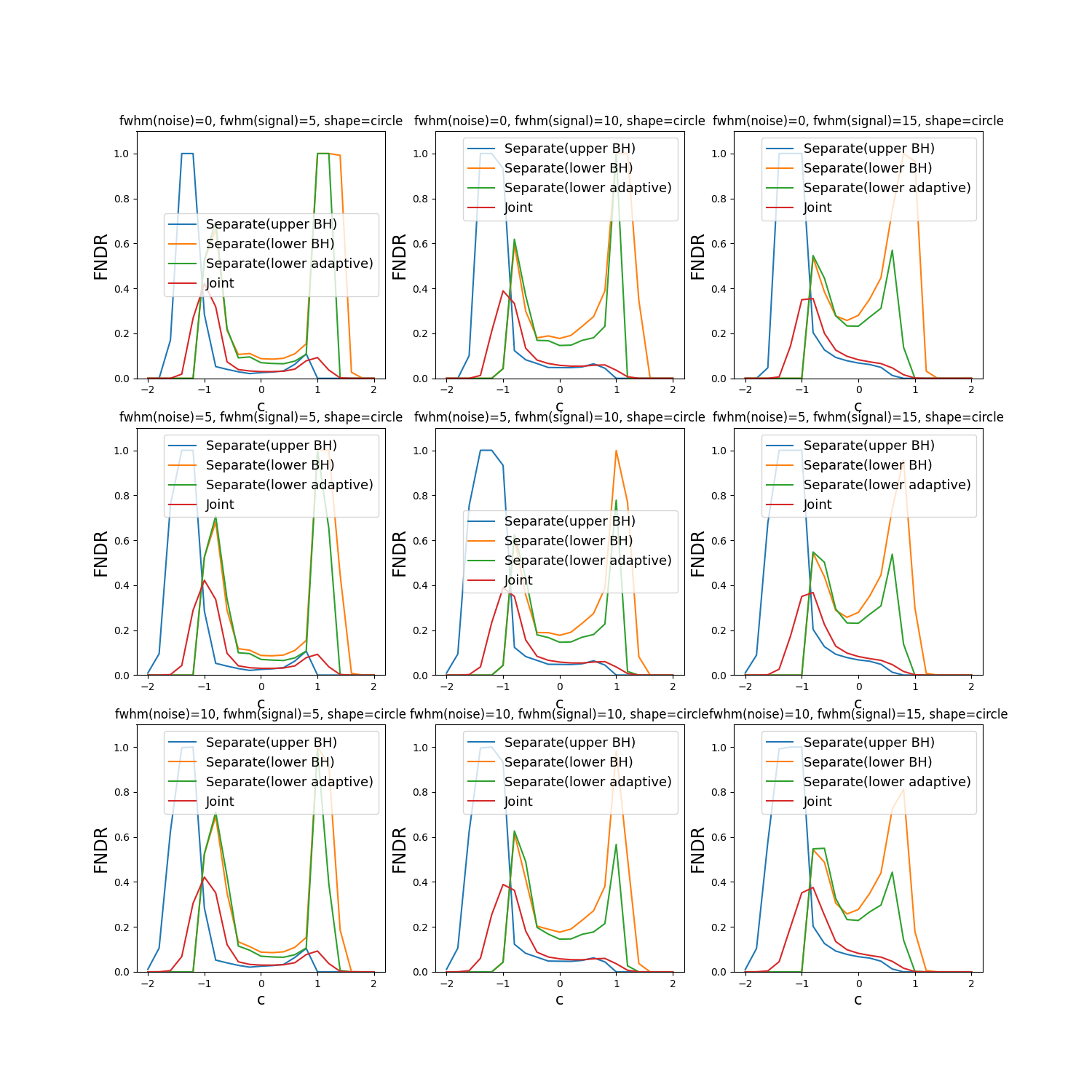}
    \label{supp-FNDR_circle}
    \caption{FNDR simulation result for the circle signal. The rows differ in noise smoothing (FWHM 0, 5, and 10) and the columns differ in signal smoothing levels (FWHM 5, 10, and 15). The $x$-axis denotes threshold c ranging [-2, 2] with 0.2 increments. The $y$-axis denotes the empirical FNDR. The red line denotes the joint method, the blue line the separate upper (BH), the yellow the separate lower (BH), and the green the separate lower (adaptive).}
\end{figure}

\subsection{HCP Applications}
\begin{figure}[H]
    \centering
    \includegraphics[scale=0.235]{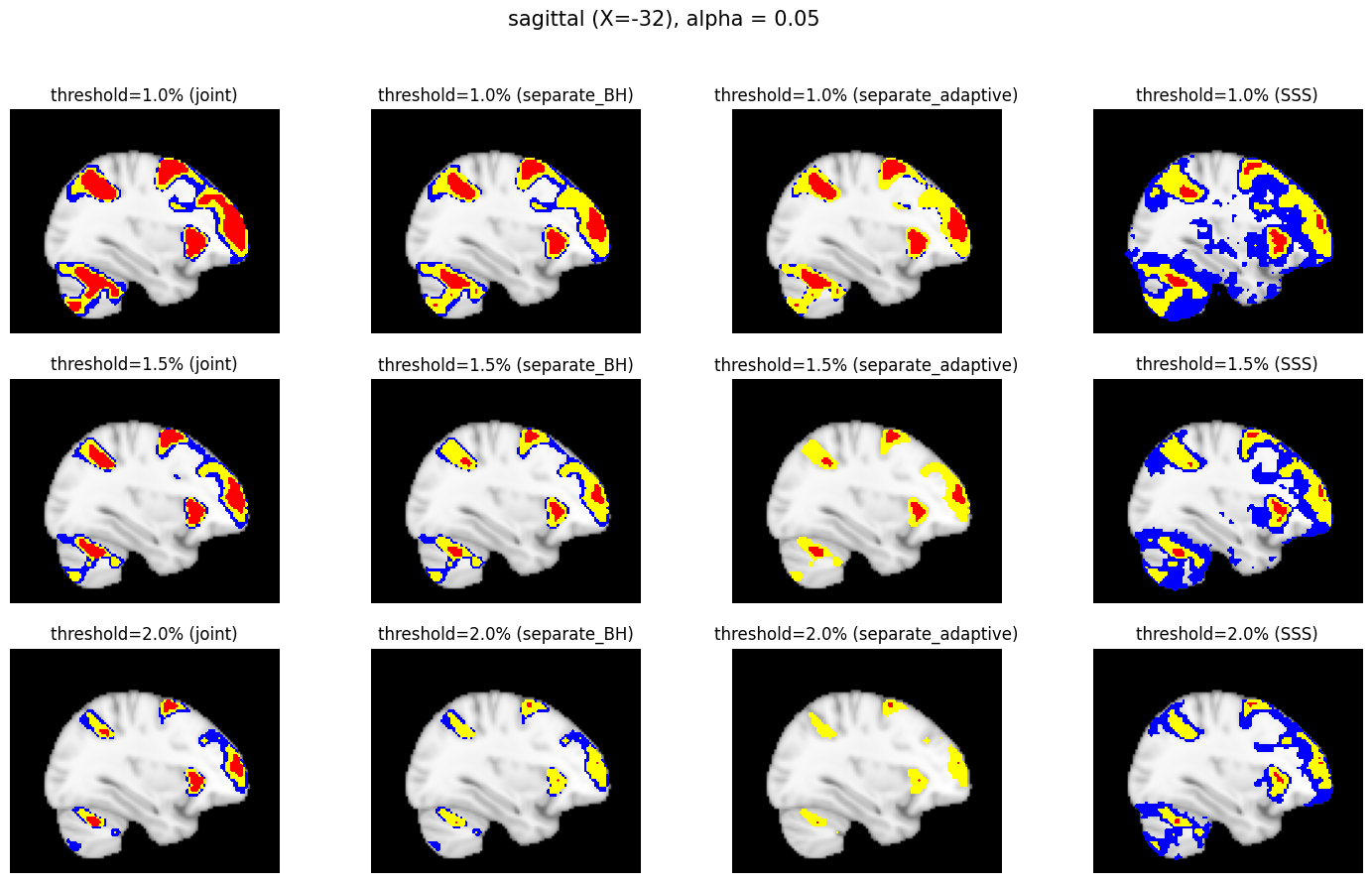}
    \includegraphics[scale=0.235]{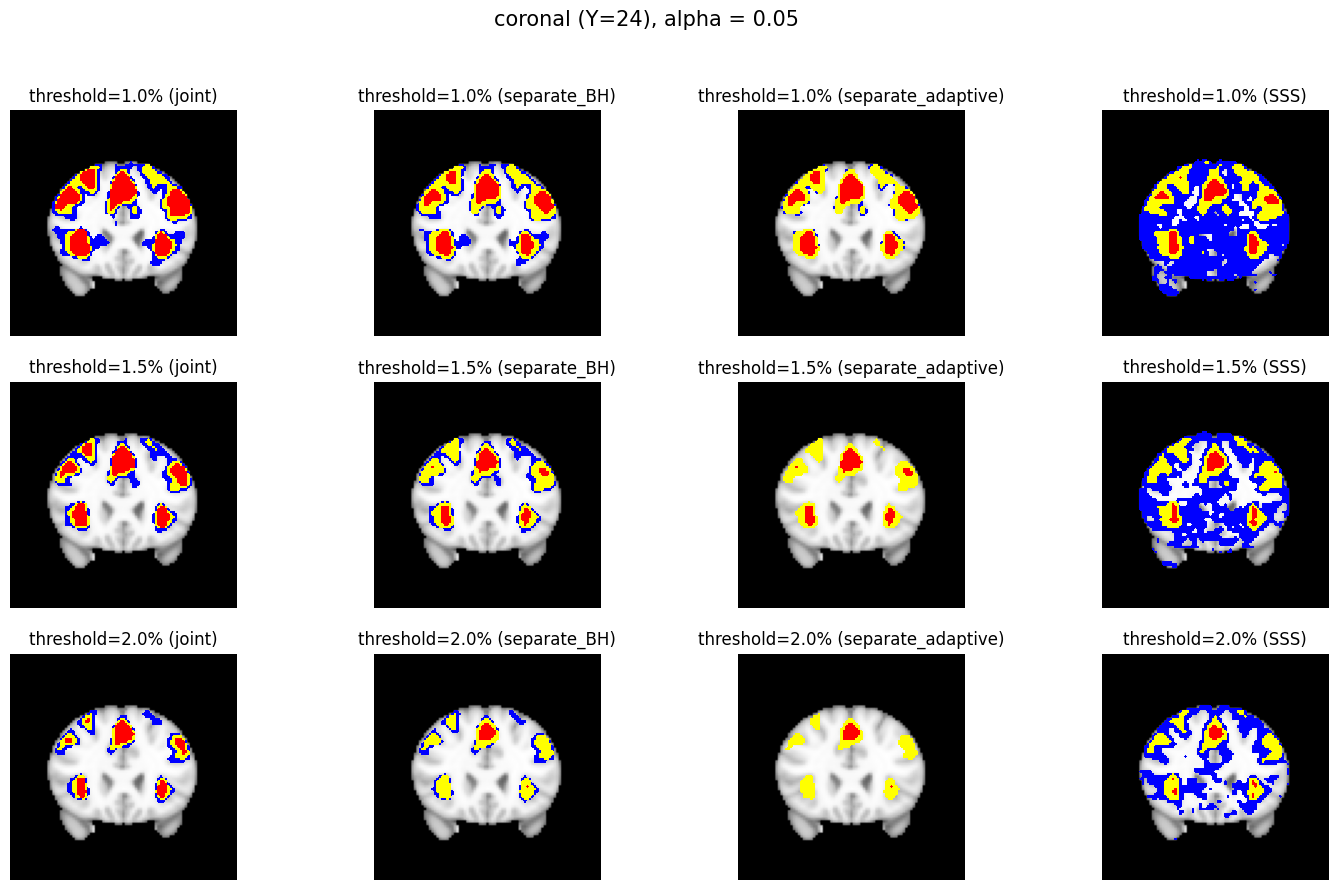}
    \includegraphics[scale=0.235]{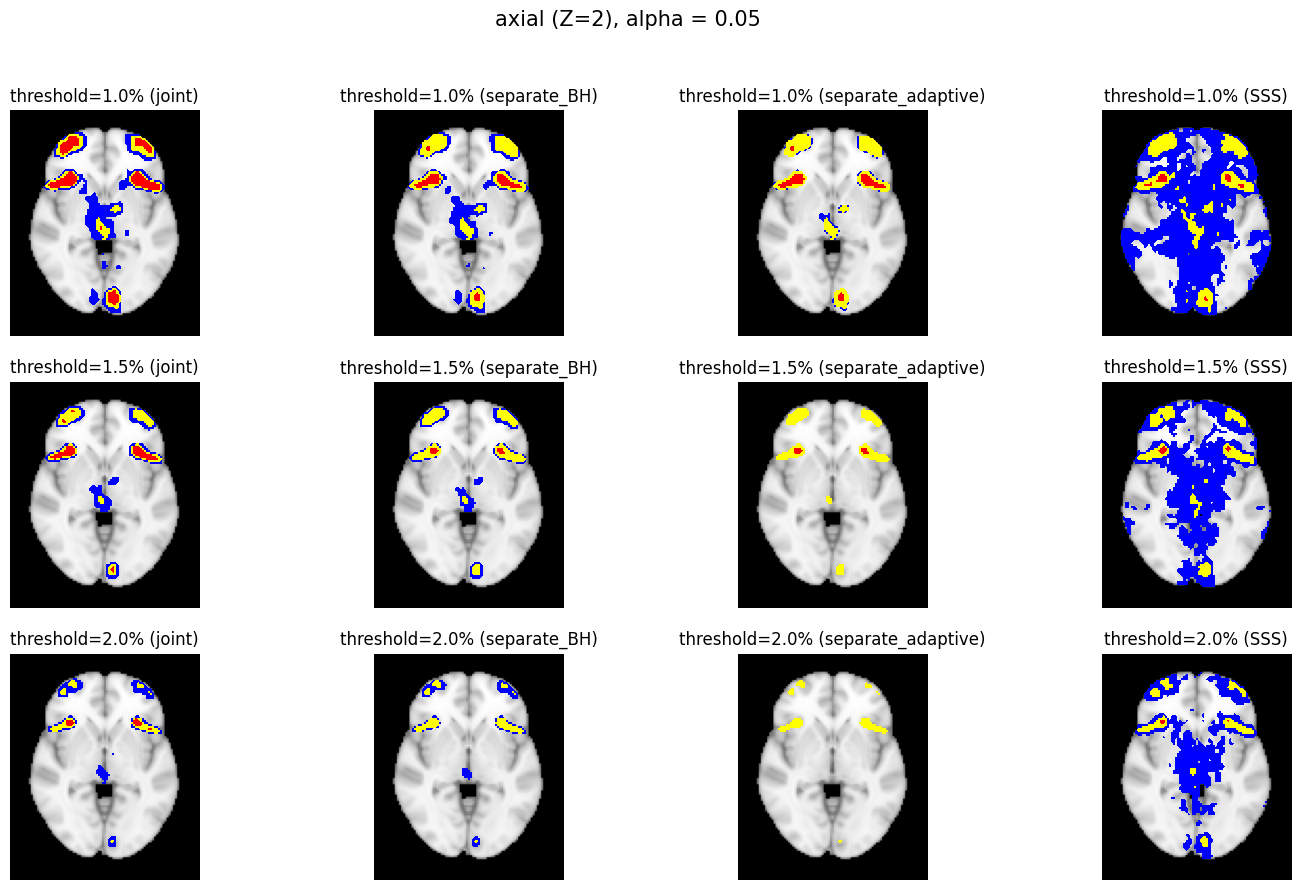}
    \label{supp-HCP}
\caption{FDCR application to the HCP data with 77 subjects in sagittal (X=-32), coronal (Y=24), and axial (Z=-2) from top to bottom. The red area denotes the upper FDCR $\hat{\mathcal{A}}_c^+$, the yellow area including the red area denotes the excursion set $\hat{\mathcal{A}}_c$, and the blue area including the yellow area denotes the lower FDCR $\hat{\mathcal{A}}_c^-$. The rows differ in the threshold level $c$ by which the FDCRs are constructed. The columns denote different methods. Overall, FDR controlling hypothesis testing methods which are presented in the first three columns, show tighter spatial inference result when compared to SSS, thus providing a higher degree of localization and interpretability to the identified regions. The FDCRs show activation in areas such as the precuneus cortex, paracingulate gyrus, middle frontal gyrus, and angular gyrus which matches the expectation from the literature as areas involved in working memory.}
\end{figure}

\end{document}